\documentclass[aps, prd, amsmath, floats, floatfix, twocolumn,superscriptaddress, 
showpacs]{revtex4}
\allowdisplaybreaks

\usepackage{psfrag}
\usepackage[dvips]{graphicx}
\usepackage{amsmath}
\usepackage{amsfonts}
\usepackage{bm}
\usepackage{amsmath}
\usepackage{multirow}
\usepackage[usenames]{color}

\usepackage{ulem}
\newcommand{\hide}[1]{{}}

\def\laq{\raise 0.4ex\hbox{$<$}\kern -0.8em\lower 0.62ex\hbox{$\sim$}}
\def\gaq{\raise 0.4ex\hbox{$>$}\kern -0.7em\lower 0.62ex\hbox{$\sim$}}

\newcommand{\nn}{\nonumber}

\newcommand{\vS}{\mbox{\boldmath${S}$}}

\newcommand{\vL}{\mbox{\boldmath${L}$}}

\newcommand{\Maryland}{\affiliation{Maryland Center for Fundamental
    Physics \& Joint Space-Science Institute, Department of Physics, University of Maryland, College
    Park, MD 20742}}
\newcommand{\Raman}{\affiliation{Raman Research Institute, Bangalore 560 080, India}}
\newcommand{\Osaka}{\affiliation{Department of Earth and Space Science, Graduate School of Science, Osaka University, 
Toyonaka 560 0043, Japan}}

\begin{document}

\title{Post-Newtonian factorized multipolar waveforms for spinning, non-precessing black-hole binaries}

\author{Yi Pan} \Maryland %
\author{Alessandra Buonanno} \Maryland %
\author{Ryuichi Fujita} \Raman %
\author{Etienne Racine} \Maryland %
\author{Hideyuki Tagoshi} \Osaka %

\begin{abstract}
We generalize the {\it factorized} resummation of multipolar waveforms 
introduced by Damour, Iyer and Nagar to spinning black holes. 
For a nonspinning test-particle spiraling a Kerr black hole in the equatorial plane, 
we find that factorized multipolar amplitudes which replace the residual relativistic amplitude 
$f_{\ell m}$ with its $\ell$-th root, $\rho_{\ell m} = f_{\ell m}^{1/\ell}$,  
agree quite well with the numerical amplitudes up to the Kerr-spin value 
$q \leq 0.95$ for orbital velocities $v \leq 0.4$. The numerical 
amplitudes are computed solving the Teukolsky equation with a spectral code.  
The agreement for prograde orbits and large spin values of the Kerr black hole can be 
further improved at high velocities by properly factoring out the lower-order post-Newtonian contributions 
in $\rho_{\ell m}$. The resummation procedure results in a 
better and systematic agreement between numerical and analytical amplitudes (and energy fluxes) than standard Taylor-expanded post-Newtonian approximants. This is particularly true for higher-order modes, such as (2,1), (3,3), (3,2), and (4,4) for which less spin post-Newtonian terms are known. We also extend the factorized resummation of 
multipolar amplitudes to generic mass-ratio, non-precessing, spinning black holes. 
Lastly, in our study we employ new, recently computed, higher-order 
post-Newtonian terms in several subdominant modes, 
and compute explicit expressions for the half and one-and-half post-Newtonian contributions to the 
odd-parity (current) and even-parity (odd) multipoles, respectively. Those results can be used to 
build more accurate templates for ground-based and space-based gravitational-wave detectors.
\end{abstract}
\date{\today}

\pacs{04.25.Nx, 04.30.Db}

\maketitle

\section{Introduction}
\label{sec:introduction}

An international network of kilometer-scale laser-interferometric
gravitational-wave detectors, consisting of the Laser-Interferometer
Gravitational-wave Observatory (LIGO) \cite{Waldman:2006} and 
Virgo~\cite{Acernese:2008zz} are currently operating at the best 
sensitivity ever in the frequency range $10 \mbox{--} 10^3$ Hz. 
We expect that in the next decade the 
Laser Interferometer Space Antenna (LISA)~\cite{schutz_lisa_science} 
will be also operating, but in the frequency range $10^{-4} \mbox{--} 10^{-1}$ Hz.

Binary black holes are among the most promising sources for those
detectors. During the last thirty years, the search for gravitational waves from coalescing 
black-hole binaries with LIGO, Virgo and LISA has prompted the development of highly-accurate, analytical 
template families to be employed in matched-filtering analysis. Those template 
families are based on the post-Newtonian (PN) approximation of the two-body dynamics and gravitational 
radiation~\cite{Sasaki:2003xr,Blanchet2006}. In PN theory the multipolar waveforms are derived as 
a Taylor-expansion in $v/c$ ($v$ being the binary characteristic velocity and $c$ the speed of light).   
More recently, Damour, Iyer and Nagar~\cite{DINresum}~\footnote{The factorized waveform 
for the (2,2) mode appeared first in Ref.~\cite{Damour2007}.} have proposed 
a resummation of the multipolar waveforms in which the Taylor-expanded multipolar waveforms computed in PN theory 
are re-written in a factorized, resummed form as 
\begin{equation}
h_{\ell m}=h_{\ell m}^{(N,\epsilon_p)}\,
\hat{S}_{\rm eff}^{(\epsilon_p)}\,T_{\ell m}\,e^{i\delta_{\ell m}}\,f_{\ell m}\,.
\end{equation}
The several factors in the above $h_{\ell m}$ have the following meaning. The factor 
$h_{\ell m}^{(N,\epsilon_p)}$ is the leading Newtonian term; 
$\hat{S}_{\rm eff}^{(\epsilon_p)}$ is the relativistic conserved energy or angular momentum 
of the effective moving source; $T_{\ell m}$ 
resums an infinite number of leading logarithms entering the tail effects; $e^{i\delta_{\ell m}}$ 
is a supplementary phase which contains phase effects which are not contained in the complex 
$T_{\ell m}$, and, finally, $f_{\ell m}$ contain residual terms which can be carefully resummed 
to improve its behaviour as function of $\ell$. The better agreement of the factorized multipolar 
waveforms to the exact numerical results suggests that the factors entering the $h_{\ell m}$'s can 
capture effects, such as the presence of a pole in the effective source for quasi-circular orbits and 
the inclusion of all leading logarithms in tail terms, 
that are missed when expanded in a PN series and truncated 
at a certain PN order. 

In Refs.~\cite{Damour2007,DINresum}, the factorized waveforms 
for a test particle orbiting around a Schwarzschild black hole were computed, including also the case of 
comparable-mass nonspinning black holes. It was found that factorized waveforms 
agree better with numerical (exact) results than Taylor-expanded waveforms. In particular, 
in the test-particle limit, Ref.~\cite{DINresum} compared the analytical factorized $(l,m)$ modes and 
gravitational-wave energy flux to the numerical results obtained by Berti~\cite{Yunes:2008tw},  
solving the Teukolsky equation. The factorized waveforms have been also 
employed in the effective-one-body formalism and compared to waveforms 
computed in numerical-relativity simulations~\cite{Damour2007,Damour2009a,Buonanno2009a}.  
Also in this case, the agreement of the factorized waveforms to the numerical waveforms  
is better than the one of the Taylor-expanded waveforms, especially during the last 
stages of inspiral and plunge, and close to merger.

\begin{table*}[t]\label{tab:PNorder}
\begin{tabular}{|c|cc|cc|cc|cc|cc|cc|cc|}
  \hline
  & \multicolumn{2}{|c|}{$C_{2m}$}  
  & \multicolumn{2}{|c|}{$C_{3m}$} & \multicolumn{2}{|c|}{$C_{4m}$} & \multicolumn{2}{|c|}{$C_{5m}$} & \multicolumn{2}{|c|}{$C_{6m}$} & \multicolumn{2}{|c|}{$C_{7m}$} & \multicolumn{2}{|c|}{$C_{8m}$} \\
  & $0$ & $1$ & $0$ & $1$ & $0$ & $1$ & $0$ & $1$ & $0$ & $1$ & $0$ & $1$ & $0$ & $1$ \\
  \hline\hline
    PN orders beyond $C^{(N,0)}_{22}$ (nonspin)
  & 5.5 & 6 & 5 & 5.5 & 4.5 & 5 & 4 & 4.5 & 3.5 & 4 & 3 & 3 & 3 & 3.5 \\
  PN orders beyond $C^{(N,0)}_{22}$ (spin) 
  & 4 & 4 & 4 & 4 & 4 & 4 & 4 & 4 & 4 & 4 & 4 & 4 & 4 & 4 \\
  \hline
  PN orders beyond $C^{(N,\epsilon_p)}_{\ell m}$ (nonspin)
  & 5.5 & 5.5 & 4.5 & 4.5 & 3.5 & 3.5 & 2.5 & 2.5 & 1.5 & 1.5 & 0.5 & 0.5 & 0 & 0 \\
  PN orders beyond $C^{(N,\epsilon_p)}_{\ell m}$ (spin) 
  & 4 & 3.5 & 3.5 & 3 & 3 & 2.5 & 2.5 & 2 & 2 & 1.5 & 1.5 & 1 & 1 & 0.5 \\\hline \hline
  PN orders beyond $C^{(N,\epsilon_p)}_{\ell m}$ needed for nonspin 5.5PN-flux  
  & 5.5 & 4.5 & 4.5 & 3.5 & 3.5 & 2.5 & 2.5 & 1.5 & 1.5 & 0.5 & 0.5 & 0 & 0 & 0 \\ 
  PN orders beyond $C^{(N,\epsilon_p)}_{\ell m}$ needed for spin 4PN-flux 
  & 4 & 3 & 3 & 2 & 2 & 1 & 1 & 0 & 0 & 0 & 0 & 0 & 0 & 0 \\
  \hline
\end{tabular}
\caption{PN orders currently available in the multipolar waveforms $C_{\ell m}$ (in the adiabatic 
approximation $C_{\ell m}=-m^2\,\Omega^2\,h_{\ell m}$). In the first two rows, we 
list the nonspin and spin PN orders beyond the leading order Newtonian term $C^{(N,0)}_{22}$. 
In the next two rows, we list the nonspin and spin PN orders beyond the leading-order term for each mode 
$C^{(N,\epsilon_p)}_{\ell m}$. In the last two rows, we list the PN orders beyond the leading-order term for each mode 
$C^{(N,\epsilon_p)}_{\ell m}$ that are needed to compute the nonspin 5.5PN--energy-flux and the spin 4PN--energy-flux.
For each $C_{\ell m}$, the two columns refer to the parity of the multipolar waveform $\epsilon_p = 0$ and $=1$.}
\end{table*}

In this paper we extend the factorized multipolar waveforms to the case of a test-particle 
orbiting around a Kerr black hole on the equatorial plane. In the case of a test particle
orbiting around a Schwarzschild black hole, the Taylor-expanded 
multipolar waveforms were derived through the PN order needed to 
compute the 5.5PN energy flux ~\cite{TTS55PN},
although their explicit formulas were not available in the literature. 
In the case of a test particle orbiting
around a Kerr black hole on the equatorial plane, spin terms in the
Taylor-expanded multipolar waveforms were derived through the PN order needed to 
compute the 4PN energy flux~\cite{TSTS1996}, but their explicit formulas 
were not published. Motivated by this work, 
Tagoshi and Fujita ~\cite{TagoshiFujita2010} have recently computed the spinning and nonspinning 
Taylor-expanded multipolar waveforms up to 4PN and 5.5PN order (see Table~\ref{tab:PNorder} for a summary), respectively. Also, recently, Fujita and Iyer~\cite{FujitaIyer2010} have independently computed the 
nonspinning Taylor-expanded multipolar waveforms up to 5.5PN order.

Since, as said above, explicit expressions of the Taylor-expanded multipolar waveforms are 
not available in the literature, even at lower PN orders~\cite{TTS55PN, TSTS1996},
 we write those expressions explicitly in this paper (see Appendix~\ref{AppendixA}) 
decomposing them in -2 spin-weighted spheroidal harmonics. Then,
we apply the transformation from -2 spin-weighted spheroidal harmonics
to -2 spin-weighted spherical harmonics, and build the factorized
multipolar waveforms decomposed in -2 spin-weighted spherical
harmonics.  The latter decomposition is the one commonly used in the
fields of numerical relativity and gravitational-wave data analysis.
Finally, we compare the factorized waveforms to numerical (exact)
waveforms for a test particle orbiting around a Kerr black hole, on
the equatorial plane, solving the Teukolsky equation 
\cite{FujitaTagoshi04,FujitaTagoshi05,FujitaHikidaTagoshi09}. Finally,  we derive 
the factorized multipolar waveforms for spinning, nonprecessing 
black holes of comparable masses. Those factorized waveforms 
were recently used in the spinning effective-one-body model
of Ref.~\cite{EOBNRS} and compared to numerical-relativity simulations
of spinning, nonprecessing equal-mass black holes from the Caltech-Cornell-CITA 
collaboration.

This paper is organized as followed. In Sec.~\ref{sec:tmlresum} we work out the 
factorized waveforms decomposed in -2 spin-weighted spherical harmonics for a test-particle orbiting a Kerr 
black hole, on the equatorial plane. In Sec.~\ref{sec:tmlcomparison}, we compare 
the gravitational-wave energy flux and the $(l,m)$ modes 
of analytical factorized waveforms to numerical waveforms. The numerical results are obtained solving 
the Teukolsky equation \cite{FujitaTagoshi04,FujitaTagoshi05,FujitaHikidaTagoshi09}. 
In Sec.~\ref{sec:gmrresum} we derive the factorized waveforms for 
generic mass-ratio spinning, non-precessing black holes. Section~\ref{sec:conclusions} 
summarizes our main conclusions. 
In Appendix~\ref{AppendixA} we write the Taylor-expanded multipolar waveforms in the test-particle limit through the PN order currently known. In Appendices~\ref{AppendixE}, \ref{AppendixB}, 
\ref{AppendixC}, and \ref{AppendixF} we give the complete expressions of the 
$f_{\ell m}$'s, $C_{\ell m}$'s, $\rho_{\ell m}$'s and $\delta_{\ell m}$'s for $4< l \leq 8$.  
Finally, in Appendix~\ref{AppendixD} we compute the $l$ and $m$ dependence of the spin terms 
in the mass and current multipole moments at 0.5PN order 
and 1.5PN order, respectively. 

\section{Factorized multipolar waveforms for a test particle orbiting around a Kerr black hole}
\label{sec:tmlresum}

We consider a nonspinning test-particle orbiting around a Kerr black hole, and extend the factorized 
waveforms of Ref.~\cite{DINresum} to the case where the motion is quasi-circular and 
confined to the equatorial plane, that is the spinning, 
nonprecessing case. The factorized multipolar waveforms are built as the production of a leading order term $h_{\ell m}^{(N,\epsilon_p)}$ and a higher order correction term $\hat{h}_{\ell m}$ consisting four factors~\cite{DINresum}
\begin{equation}\label{hlm}
h_{\ell m}=h_{\ell m}^{(N,\epsilon_p)}\,\hat{h}_{\ell m}=h_{\ell m}^{(N,\epsilon_p)}\,
\hat{S}_{\rm eff}^{(\epsilon_p)}\,T_{\ell m}\,e^{i\delta_{\ell m}}\,f_{\ell m}\,.
\end{equation}
where $\epsilon_p$ denotes the parity of the multipolar waveform. In
the quasi-circular case, $\epsilon_p$ is the parity of $\ell+m$:
$\epsilon_p=\pi(\ell+m)$. The leading term $h_{\ell
  m}^{(N,\epsilon_p)}$ in Eq.~(\ref{hlm}) is the Newtonian order waveform 
\begin{equation}\label{hlmNewt}
h_{\ell m}^{(N,\epsilon_p)}=\frac{G\,M\,\nu}{c^2\,R}\,n_{\ell m}^{(\epsilon_p)}\,c_{\ell+\epsilon_p}(\nu)\,v^{(\ell+\epsilon_p)}\,Y^{\ell-\epsilon_p,-m}\,\left(\frac{\pi}{2},\phi\right)\,,
\end{equation}
where $v$ is the orbital velocity, $Y^{\ell m}(\theta,\phi)$ are the scalar spherical
harmonics, $n_{\ell m}^{(\epsilon_p)}$ are 
\begin{subequations}
\begin{eqnarray}
n^{(0)}_{\ell m}&=&(im)^\ell\frac{8\pi}{(2\ell+1)!!}\sqrt{\frac{(\ell+1)(\ell+2)}{\ell(\ell-1)}}\,, \nonumber \\ \label{nlmeven} \\
n^{(1)}_{\ell m}&=&-(im)^\ell\frac{16\pi i}{(2\ell+1)!!}\sqrt{\frac{(2\ell+1)(\ell+2)(\ell^2-m^2)}{(2\ell-1)(\ell+1)\ell(\ell-1)}}\,,
\nonumber \\ \label{nlmodd} 
\end{eqnarray}
\end{subequations}
and $c_{\ell+\epsilon_p}(\nu)$ are functions of the symmetric mass ratio $\nu\equiv m_1\,m_2/M^2$, with
$M = m_1+m_2$:
\begin{eqnarray}
\label{cleps}
c_{\ell+\epsilon}(\nu)&=&\left(\frac{1}{2}-\frac{1}{2}\sqrt{1-4\nu}\right)^{\ell+\epsilon-1}\nonumber \\
&& +(-)^{\ell+\epsilon}\left(\frac{1}{2}+\frac{1}{2}\sqrt{1-4\nu}\right)^{\ell+\epsilon-1}\,.
\end{eqnarray}
Although in this section we consider the test particle limit $m_1 \equiv M \gg m_2  \equiv \mu$, 
that is $\nu \rightarrow 0$, the above relations will be used for generic $\nu$ in Sec.~\ref{sec:gmrresum} and Appendix~\ref{AppendixD}. 

We shall define the source factor $\hat{S}_{\rm
    eff}^{(\epsilon_p)}$ and the tail factors $T_{\ell m}$ in
  Sec.~\ref{sec:tmltailsource}. In Secs.~\ref{sec:tmltailsource}, 
~\ref{sec:tmlexpand}, we compute the imaginary and real PN spin
  effects in the $e^{i\delta_{\ell m}}$'s and $f_{\ell m}$'s
  respectively. We shall obtain those quantities by requiring that when we Taylor expand the factorized
  waveforms~\eqref{hlm} the results coincide through 4PN order, for the spin
  terms, and 5.5PN order, for the nonspinning terms, with the Taylor-expanded
  waveforms given in Sec.~\ref{sec:tmltaylor} and
  Appendix~\ref{AppendixA}.  

\subsection{Taylor-expanded multipolar waveforms}
\label{sec:tmltaylor}

The Newman-Penrose scalar $\Psi_4=-(\ddot{h}_+ - i\,\ddot{h}_\times)$ can be 
decomposed in either -2 spin-weighted spherical harmonics 
$_{-2}Y_{\ell m}(\theta,\phi)\equiv _{-2}P_{\ell
  m}(\theta)\,e^{im\phi}$, or -2 spin-weighted spheroidal harmonics 
$_{-2}S^{a\omega_0}_{\ell m}(\theta)\,e^{im\phi}$ as
\begin{eqnarray}
r\Psi_4&=&\sum_{\ell}\sum_{m=-\ell}^\ell C_{\ell m}\;_{-2}Y_{\ell m}(\theta,\phi)\,e^{i\omega_0(r^*-t)}\,, \nonumber \\
	&=& \sum_{\ell}\sum_{m=-\ell}^\ell C_{\ell m}\;\frac{_{-2}P_{\ell m}(\theta)}{\sqrt{2\pi}}\,e^{i\omega_0(r^*-t)+im\phi}\,,
\nonumber\\                 
       &=&\sum_{\ell}\sum_{m=-\ell}^\ell Z_{\ell m\omega_0}\,\frac{_{-2}S^{a\omega_0}_{\ell m}(\theta)}{\sqrt{2\pi}}\,e^{i\omega_0(r^*-t)+im\phi}\,,
\end{eqnarray}
where $a$ is the spin of the Kerr black hole, having the dimension of
length (while we also define $q \equiv a/M$), and $\omega_0=m\Omega$
is a multiple of the orbital frequency $\Omega$. Since -2
  spin-weighted spheroidal harmonics are eigenfunctions of the
  Teukolsky equation, it is natural to expand its solution in the
  spheroidal basis. In the fields of numerical relativity and
  gravitational-wave data analysis, however, the -2 spin-weighted
  spherical harmonics are commonly used, because they do not 
  depend on the spin and the frequency, as the -2 spin-weighted
  spheroidal harmonics do.

The -2 spin-weighted spherical and spheroidal harmonic bases are related by 
\begin{eqnarray}
\label{StoY}
_{-2}S^{a\omega_0}_{\ell m}(\theta)&=&_{-2}P_{\ell m}(\theta) + a\,\omega_0\,\sum_{\ell'}c^{\ell'}_{\ell m}\,_{-2}P_{\ell'm}(\theta) \nonumber \\
&& +(a\,\omega_0)^2\,\sum_{\ell'}d^{\ell'}_{\ell m}\,_{-2}P_{\ell'm}(\theta)+{\cal O}(a\,\omega_0)^3\,.\nonumber \\
\end{eqnarray}
The coefficients $c^{\ell'}_{\ell m}$ and $d^{\ell'}_{\ell m}$ are given in Ref.~\cite{TSTS1996} as
\begin{eqnarray}
c^{\ell'}_{\ell m}&=&\begin{cases} 
\frac{2}{(\ell+1)^2}\sqrt{\frac{(\ell+3)(\ell-1)(\ell+m+1)(\ell-m+1)}{(2\ell+1)(2\ell+3)}}\,, & \ell'=\ell+1 \\
-\frac{2}{\ell^2}\sqrt{\frac{(\ell+2)(\ell-2)(\ell+m)(\ell-m)}{(2\ell+1)(2\ell-1)}}\,, & \ell'=\ell-1 \\
0\,, & \text{otherwise} \end{cases} \nonumber\\
\end{eqnarray}
and if $\ell'=\ell$ we have
\begin{equation}
d^{\ell'}_{\ell m}=
-\frac{1}{2} \left[ \left(c^{\ell+1}_{\ell m}\right)^2 + \left(c^{\ell-1}_{\ell m}\right)^2 \right]\,, 
\end{equation}
while if $\ell' \neq \ell$ we have
\begin{widetext}
\begin{eqnarray}
d^{\ell'}_{\ell m}&=&\frac{1}{\lambda_0(\ell)-\lambda_0(\ell')}\,\left\{ -[2m+\lambda_1(\ell,m)]\,
[\delta_{\ell' \ell+1}c^{\ell+1}_{\ell m} +  \delta_{\ell' \ell-1}c^{\ell-1}_{\ell m}-\delta_{\ell' \ell}\lambda_2(\ell,m)] \right. \nonumber \\ 
&& \left. - 4c^{\ell+1}_{\ell m}\sqrt{\frac{2\ell+3}{2\ell'+1}}\langle \ell+1,m,1,0|\ell',m \rangle\langle \ell+1,2,1,0|\ell',2 \rangle - 4c^{\ell-1}_{\ell m}\sqrt{\frac{2\ell-1}{2\ell'+1}}\langle \ell-1,m,1,0|\ell',m \rangle\langle \ell-1,2,1,0|\ell',2 \rangle \right. \nonumber \\
&& \left. +\frac{2}{3}\left[\delta_{\ell' \ell}-\sqrt{\frac{2\ell+1}{2\ell'+1}}\langle \ell,m,2,0|\ell',m \rangle\langle \ell,2,2,0|\ell',2 \rangle\right] 
\right\} \,,
\end{eqnarray}
\end{widetext}
where $\langle j_1,m_1,j_2,m_2|J,M \rangle$ is a Clebsch-Gordan coefficient and
\begin{eqnarray}
\lambda_0(\ell)&=& (\ell-1)(\ell+2) \,,\\
\lambda_1(\ell,m)&=& -2\,m\,(\ell^2+\ell+4)/(\ell^2+\ell) \,,\\
\lambda_2(\ell,m)&=& -2(\ell+1)\left(c^{\ell+1}_{\ell m}\right)^2+2\ell\left(c^{\ell-1}_{\ell m}\right)^2+\frac{2}{3} \nonumber\\
&& -\frac{2}{3}\frac{(\ell+4)(\ell-3)(\ell^2+\ell-3m^2)}{\ell(\ell+1)(2\ell+3)(2\ell-1)} \,.
\end{eqnarray}

\ 

In the nonspinning case, $_{-2}S^{a\omega_0}_{\ell m}(\theta)$ reduces to $_{-2}P_{\ell
m}(\theta)$ which has the closed expression
\begin{widetext}
\begin{equation}
_{-2}P_{\ell m}(\theta)=(-1)^m\sqrt{\frac{(l+m)!(l-m)!(2l+1)}{2(l+2)!(l-2)!}}\sin^{2l}\left(\frac{\theta}{2}\right)\,
\sum_{r=0}^{l+2}{{l+2}\choose{r}}{{l-2}\choose{r-2-m}}(-1)^{l-r+2}\cot^{2r-2-m}\left(\frac{\theta}{2}\right)\,.
\end{equation}
\end{widetext}
The -2 spin-weighted spherical harmonic basis is orthonormal in the sense that
\begin{equation}
\int_0^\pi\,_{-2}P_{\ell m}(\theta)_{-2}\,P_{\ell'm'}(\theta)\,\sin\theta\,d\theta=\delta_{\ell\ell'}\,\delta_{mm'}\,.
\end{equation}
To explicitly write the modes $C_{\ell m}$ and $Z_{\ell m\omega_0}$ expanded in $v$, 
we find it convenient to introduce the following notation
\begin{eqnarray}
C_{\ell m}&=&C^{(N,\epsilon_p)}_{\ell m}\,\hat{C}_{\ell m}\,, \\
Z_{\ell m\omega_0}&=&Z^{(N,\epsilon_p)}_{\ell m\omega_0}\,\hat{Z}_{\ell m\omega_0}\,, 
\label{hatZlmw0}
\end{eqnarray}
where $C^{(N,\epsilon_p)}_{\ell m}$ and $Z^{(N,\epsilon_p)}_{\ell
  m\omega_0}$ represent the Newtonian contributions and, as said above, 
$\epsilon_p$ denotes the parity of the multipolar waveform. In the adiabatic limit,
$C_{\ell m}=-m^2\,\Omega^2\,h_{\ell m}$. Therefore, whereas the Newtonian
contribution to $C_{\ell m}$ and $h_{\ell m}$ differ by a factor of
$-m^2\,\Omega^2$, the PN corrections are the same, i.e., $\hat{C}_{\ell
  m}=\hat{h}_{\ell m}$. The Newtonian contributions
in the $C_{\ell m}$'s or $Z_{\ell m\omega_0}$'s are [see Eq.~\eqref{hlmNewt}],

\ 

\ 

\begin{eqnarray}
\label{NewtZlmw}
C^{(N,\epsilon_p)}_{\ell m}=Z^{(N,\epsilon_p)}_{\ell m\omega_0}
&=&-m^2\nu\,n^{(\epsilon_p)}_{\ell m}\,c_{\ell+\epsilon_p}(\nu)\,
v^{(\ell+\epsilon_p+6)}
\nonumber\\
&&\qquad\times Y^{\ell-\epsilon_p,-m}(\pi/2,\phi)\,,
\end{eqnarray}
where we define $v=(M\,\Omega)^{1/3}$. In Refs.~\cite{TTS55PN,TSTS1996}, the 
Taylor-expanded multipolar waveforms were calculated at the PN order 
needed to compute the nonspinning 5.5PN--energy-flux and spin 4PN--energy-flux, respectively. 
For the purpose of the present paper, Tagoshi and Fujita \cite{TagoshiFujita2010} 
extended the computation of the multipolar waveforms at higher PN order. Although those new PN corrections 
are not sufficient for computing the energy flux at the next order (6PN and 4.5PN order 
in the nonspinning and spinning cases, respectively), they do improve our knowledge of 
the multipolar waveforms, as we shall discuss below. In Table~\ref{tab:PNorder}, we
list the PN orders available to us in each multipolar waveform $C_{\ell m}$, 
while the explicit Taylor-expanded waveforms, $\hat{Z}_{\ell m\omega_0}$'s, 
are given in Appendix~\ref{AppendixA}.

We compute the $\hat{C}_{\ell m}$'s from the $\hat{Z}_{\ell m\omega_0}$'s by applying Eq.~\eqref{StoY} 
and the orthogonality condition of the -2 spin-weighted spherical harmonics
\begin{widetext}
\begin{eqnarray}
C_{\ell m} &=& \int_{S^2}\,d\Omega\,r\Psi_4\,_{-2}Y^*_{\ell m}e^{-i\omega_0(r^*-t)}
= \int_{S^2}\,d\Omega \sum_{\ell'}\sum_{m'=-\ell'}^{\ell'} Z_{\ell' m'\omega_0}
\frac{_{-2}S^{a\omega_0}_{\ell' m'}\,_{-2}P^*_{\ell m}}{2\pi}\,e^{i\,(m'-m)\,\phi}\,,
\nonumber\\
&=& \int_0^{\pi}\,\sin\theta\,d\theta \sum_{\ell'} Z_{\ell' m\omega_0}\left[ _{-2}P_{\ell' m} + a\omega_0\sum_{\ell''}c^{\ell''}_{\ell' m}\,_{-2}P_{\ell'' m} + (a\,\omega_0)^2\sum_{\ell''}d^{\ell''}
_{\ell' m}\,_{-2}P_{\ell'' m} \right]\,_{-2}P^*_{\ell m}\,,
\nonumber\\
&=& Z_{\ell m\omega0} + a\,\omega_0\,\sum_{\ell'}c^{\ell}_{\ell' m}\,Z_{\ell' m\omega0} + 
(a\,\omega_0)^2\sum_{\ell'}d^{\ell}_{\ell' m}\,Z_{\ell' m\omega0} + {\cal O}(a\,\omega_0)^3\,.
\end{eqnarray}
\end{widetext}
We notice that the mixing of spheroidal waveforms happens among modes
with the same $m$ and different $\ell$.  

The $C_{\ell m}$ modes are computed in perturbation
  theory~~\cite{Poisson:1993vp,Tagoshi94,TSTS1996,TTS55PN} using a coordinate system  
  different from the one used in PN calculations~\cite{Kidder2008,BFIS}.  
  When expressing both modes in terms of the orbital frequency they should 
  coincide. However, the presence of tail terms in both calculations 
  demands a careful treatment. In PN calculations, the tail
  terms contain a freely specifiable constant $r_0$ that
  corresponds to the difference in the origins of the retarded time in
  radiative coordinates and in harmonic coordinates in which the
  equations of motion are given (see e.g., Eq.~(3.16) in Ref.~\cite{BFIS}). 
  This constant can be absorbed into the phase of the PN modes (see e.g., Eq.~(8.8) 
in Ref.~\cite{BFIS})  once it is traded with $x_0$ (or $v_0$)~\cite{Kidder2008} 
as
\begin{equation}
\log x_0\equiv 2 \log v_0\equiv\frac{11}{18}-\frac{2}{3}\gamma_E-\frac{4}{3}\log 2-\frac{2}{3}\log (r_0/M)\,,
\label{x0}
\end{equation}
where $\gamma_E = 0.577\,215 \dots$ is the Euler's constant, and
throughout the paper, we use ``$\log$'' to denote the natural
logarithm. In perturbation theory calculations, Schwarzschild or Boyer-Lindquist 
coordinates are used. The waveforms at infinity are naturally expressed with the tortoise coordinate,
and the relation between the Schwarzshild or Boyer-Lindquist coordinate and 
the tortoise coordinate has an arbitrary constant which 
in Refs.~\cite{Poisson:1993vp,Tagoshi94,TTS55PN,TSTS1996} is fixed to $-2M\,\log(2M)$.~\footnote{  
We note that in perturbation-theory 
calculations this constant cancels out in the combination $h_+ - i h_\times$~\cite{Tagoshi94}.}

We find that to recover the PN results we need to express some of 
the $\gamma_E$'s in the perturbation-theory modes~\cite{TagoshiFujita2010} 
in terms of $x_0$ and $r_0$ using Eq.~(\ref{x0}) and 
set $r_0= 2M/\sqrt{e}$. More specifically, we replace some of the $\gamma_E$'s using 
the following equation~\cite{Kidder2008}
\begin{equation}
\log v_0\equiv\frac{11}{36}-\frac{1}{3}\gamma_E-\frac{2}{3}\log 2-\frac{1}{3}\log (2 e^{-1/2})\,.
\end{equation}
We notice that the constant $r_0$ will appear later in our definition of the tail 
term $T_{\ell m}$ of the factorized resummed waveforms. In fact, since the $T_{\ell m}$ term
resums all tail integrals that contain $r_0$ at known orders, it is
the only term in the resummed waveforms that depends on $r_0$. Finally, 
to ease the notation, we follow Ref.~\cite{DINresum} and introduce
${\rm eulerlog}_m(v^2)=\gamma_E + \log 2 + \log m + 1/2\log v^2$ into
our $C_{\ell m}$ expressions.

Below we list the $\hat{C}_{\ell m}$'s through $l=4$, and give the expressions for $4< \ell < 8$ 
in Appendix~\ref{AppendixE}. The differences between the $\hat{C}_{\ell m}$'s 
and $\hat{Z}_{\ell m\omega_0}$'s concern only spin terms. We obtain 
\begin{widetext}
\begin{subequations}
\begin{eqnarray}\label{hatC2m}
\hat{C}_{22}&=&\hat{Z}_{22\omega_0} -\frac{20\,q}{189}\,v^5 + \frac{40\,q^2}{567}\,v^6
			+ \frac{386\,q}{1\,701}\,v^7 + \left[ -\frac{40\,\pi\,q}{189} 
			+ \frac{7\,720\,q^2}{83\,349} +\frac{20}{63}\,i\,q-\frac{80}{63}\,i\,q\log\left(\frac{v}{v_0}\right)\right]\,v^8,\\
\hat{C}_{21}&=&\hat{Z}_{21\omega_0} -\frac{q}{63}\,v^3 + \frac{8\,q}{189}\,v^5 
			- \left[ \frac{\pi\,q}{63} -
			   \frac{271\,q^2}{4\,536} -
			   \left(\frac{1}{45}+\frac{2}{63}\log 2\right)\,i\,q+\frac{2}{21}\,i\,q\log\left(\frac{v}{v_0}\right) \right]\,v^6-\left(\frac{607\,q}{12\,474}+\frac{q^3}{378}\right)\,v^7,\nonumber\\
\end{eqnarray}
\begin{eqnarray}\label{hatC3m}
\hat{C}_{33} &=& \hat{Z}_{33\omega_0} - \frac{3\,q}{20}\,v^5 + \frac{3\,q^2}{32}\,v^6 + \frac{117\,q}{220}\,v^7,\\
\hat{C}_{32} &=& \hat{Z}_{32\omega_0} + \frac{4\,q}{3}\,v - \frac{31\,q}{9}\,v^3 
			+ \left[ \frac{8\,\pi\,q}{3} - \frac{46\,q^2}{27}+16\,i\,q\log\left(\frac{v}{v_0}\right) \right]\,v^4 + \left( -\frac{7\,694\,q}{4\,455}+\frac{4\,q^3}{3} \right)\,v^5 
\nonumber\\
&&+ \left[ -\frac{2\,683\,q^2}{810}-\frac{62\,\pi\,q}{9}+\frac{i\,q}{5}-\frac{1\,24}{3}\,i\,q \log \left(\frac{v}{v_0}\right) \right]\,v^6,\\
\hat{C}_{31} &=& \hat{Z}_{31\omega_0} + \frac{32\,q}{9}\,v^3 -
 \frac{16\,q^2}{3}\,v^4 - \frac{79\,q}{36}\,v^5 +
 \left[\frac{32\,\pi\,q}{9} - \frac{4\,349\,q^2}{2\,592} -
  \frac{16}{9}(1+4\log 2)\,i\,q +
  \frac{64}{3}\,i\,q\log\left(\frac{v}{v_0}\right) \right]\,v^6 
\nonumber\\
&&+ \left[ \frac{286\,q^3}{27}-\frac{16\,\pi\,q^2}{3}+\frac{8\,i\,q^2}{3}+\frac{32}{3}\,i\,q^2 \log  2 -\frac{3\,935\,q}{3\,564}-32\,i\,q^2 \log \left(\frac{v}{v_0}\right) \right]\,v^7,
\end{eqnarray}
\begin{eqnarray}\label{hatC4m}
\hat{C}_{44}&=&\hat{Z}_{44\omega_0} - \frac{224\,q}{1\,375}\,v^5 + \frac{672\,q^2}{6\,875}\,v^6,\\
\hat{C}_{43}&=&\hat{Z}_{43\omega_0} + \frac{5\,q}{4}\,v-\frac{1\,396\,q}{275}\,v^3+ \left[-\frac{17\,q^2}{8}+\frac{15\,\pi\,q}{4}-\frac{21\,i\,q}{4}+\frac{15}{2}\,i\,q \log\left(\frac{3}{2}\right)+\frac{45}{2}\,i\,q \log\left(\frac{v}{v_0}\right)\right]\,v^4
\nonumber\\
&&+ \left(\frac{15\,q^3}{8}+\frac{51\,567\,q}{28\,600}\right)\,v^5,\\
\hat{C}_{42}&=&\hat{Z}_{42\omega_0} + 3\,q\,v^3 -\frac{6\,q^2}{7}\,v^4-\frac{17\,953\,q}{2\,750}\,v^5 + \left[-\frac{3\,562\,709\,q^2}{673\,750}+6\,\pi\,q-9\,i\,q+36\,i\,q \log \left(\frac{v}{v_0}\right)\right]\,v^6,\\
\hat{C}_{41}&=&\hat{Z}_{41\omega_0} +\frac{5\,q}{4}\,v-\frac{919\,q}{275}\,v^3+ \left[-\frac{191\,q^2}{56}+\frac{5\,\pi\,q}{4}-\frac{7\,i\,q}{4}-\frac{5}{2}\,i\,q \log  2 +\frac{15}{2}\,i\,q \log\left(\frac{v}{v_0}\right)\right]\,v^4
\nonumber\\
&&+ \left(\frac{110\,711\,q}{28\,600}-\frac{95\,q^3}{56}\right)\,v^5,
\label{hatC41}
\end{eqnarray}
\end{subequations}
\end{widetext}
where the $\hat{Z}_{\ell m}$'s can be found in Appendix~\ref{AppendixA}. 
We notice that whereas the $\hat{Z}_{\ell m}$'s contain 0.5PN spin terms 
(relative to $Z^{N}_{\ell m}$'s), the $\hat{C}_{\ell m}$'s do not, except 
for $\hat{C}_{2 1}$.
The spin terms in the multipolar waveforms (\ref{hatC2m})--(\ref{hatC41}) 
agree with the currently known spin terms computed in Ref.~\cite{ABFO}, and with 
the 0.5PN and 1.5PN spin terms in the odd and even-parity modes computed in 
Appendix~\ref{AppendixD}.

\subsection{Source and tail terms}
\label{sec:tmltailsource}

In the limit of a nonspinning test particle of mass $\mu$ orbiting around a Kerr
black hole of mass $M$ in a quasi-circular equatorial orbit, the energy and
orbital angular momentum, in Boyer-Lindquist coordinates, read~\cite{Bardeen1972}
\begin{eqnarray}
\label{Etpl}
\frac{E(r)}{\mu}&=&\frac{1-2M/r+a\,M^{1/2}/r^{3/2}}{\sqrt{1-3M/r+2a\,M^{1/2}/r^{3/2}}}\,, \\
\frac{L(r)}{\mu M}&=&\sqrt{\frac{r}{M}}\,\frac{1-2a\,M^{1/2}/r^{3/2}+a^2/r^2}{\sqrt{1-3M/r+2 a\,M^{1/2}/r^{3/2}}}\,,
\label{Ltpl}
\end{eqnarray}
where $r=(1-a\,v^3)^{2/3}/v^2$. The source term in the factorized waveform (\ref{hlm}) is 
\begin{equation}
\hat{S}_{\rm eff}^{(\epsilon)}=\begin{cases} \frac{E(r)}{\mu}\,, & \epsilon=0\,, \\ \frac{L(r)}{(\mu\,M/v)}\,, 
& \epsilon=1\,, \end{cases}
\end{equation}
where $\mu M/v$ is the Newtonian angular momentum. We use the resummed tail factor $T_{\ell m}$ given in 
Eq. (19) of Ref.~\cite{DINresum}
\begin{equation}
T_{\ell m}=\frac{\Gamma(\ell+1-2i\hat{\hat{k}})}{\Gamma(\ell+1)}\,e^{\pi\hat{\hat{k}}}\,e^{2i\hat{\hat{k}}\,\log(2kr_0)}\,,
\end{equation}
where $k=m\,\Omega$, $\hat{\hat{k}}=H_{\rm real}\,k$,
and the real Hamiltonian in the test-particle limit reduces to $H_{\rm
  real}=M$. Once again, we emphasize that the constant $r_0$ must take 
a fixed numerical value, $2M/\sqrt{e}$~\footnote{Note that in Ref.~\cite{DINresum} the authors 
chose $r_0 = 2M$.}, to reproduce the correct 
test-particle limit waveforms. We notice that there is no spin contribution to $T_{\ell m
}$ since the latter resums the corrections to the waveform when
traveling through a long-range Coulomb-type potential generated by the
mass $M$~\cite{Khriplovich:1997ms,Asada:1997zu}. Spin effects generate
a short-range potential, thus they do not contribute to $T_{\ell
  m}$. 

We compute the phase correction factors $e^{i\delta_{\ell m}}$ in
  Eq.~(\ref{hlm}) by Taylor expanding the factorized waveforms
  $h_{\ell m}$ given in Eq.~(\ref{hlm}), comparing the result with the
  $C_{\ell m}$ waveforms derived in Sec.~\ref{sec:tmltaylor} (in the
  circular-orbit, adiabatic approximation $C_{\ell m} = - (m
  \Omega)^2\,h_{\ell m}$), and collecting all imaginary terms into
  $\delta_{\ell m}$. We obtain
\begin{subequations}
\begin{eqnarray}
\label{delta2m}
\delta_{22} &=& \frac{7}{3}\,v^3+\left(\frac{428\,\pi}{105}-\frac{4\,q}{3}\right)\,v^6+\frac{20\,q}{63}\,v^8
\nonumber\\
&&+\left(\frac{1\,712 \pi^2}{315}-\frac{2\,203}{81}\right)\,v^9\,, \\
\delta_{21} &=& \frac{2}{3}\,v^3+\left(\frac{107\,\pi}{105}-\frac{17\,q}{35}\right)\,v^6+\frac{3\,q^2}{140}\,v^7
\nonumber\\
&&+\left(\frac{214 \pi^2}{315}-\frac{272}{81}\right)\,v^9\,,
\end{eqnarray}
\begin{eqnarray}
\delta_{33} &=& \frac{13}{10}\,v^3 + \left(\frac{39\,\pi}{7}-\frac{81\,q}{20}\right)\,v^6
\nonumber\\
&&+\left(\frac{78\,\pi^2}{7}-\frac{227\,827}{3\,000}\right)\,v^9\,, \\
\delta_{32} &=& \frac{2}{3}\,v^3+4\,q\,v^4+\left(\frac{52\,\pi}{21}-\frac{136\,q}{45}\right)\,v^6
\nonumber\\
&&+\left(\frac{208\,\pi^2}{63}-\frac{9\,112}{405}\right)\,v^9\,, \\
\delta_{31} &=& \frac{13}{30}\,v^3+\left(\frac{61\,q}{20}+\frac{13\,\pi}{21}\right)\,v^6-\frac{24\,q^2}{5}\,v^7
\nonumber\\
&&+\left(\frac{26\,\pi^2}{63}-\frac{227\,827}{81\,000}\right)\,v^9\,,
\end{eqnarray}
\begin{eqnarray}
\delta_{44} &=& \frac{14}{15}\,v^3 + \left(\frac{25\,136\,\pi}{3\,465}-\frac{464\,q}{75}\right)\,v^6\,, \\
\delta_{43} &=& \frac{3}{5}\,v^3 + \frac{11\,q}{4}\,v^4+\frac{1\,571 \pi}{385}\,v^6\,, \\
\delta_{42} &=& \frac{7}{15}\,v^3 + \left(\frac{212\,q}{75}+\frac{6\,284\,\pi}{3\,465}\right)\,v^6\,, \\
\delta_{41} &=& \frac{1}{5}\,v^3 + \frac{11\,q}{12}\,v^4+\frac{1\,571 \pi}{3\,465}\,v^6\,.
\label{delta4m}
\end{eqnarray}
\end{subequations}
Notice that the nonspinning terms in the $\delta_{\ell m}$ already  
appeared in Ref.~\cite{DINresum}, except for the terms at 3PN order ($v^6$) 
[Ref.~\cite{DINresum} did compute $\delta_{22}$ at 3PN order]. 
We find that those 3PN-order terms in the $\delta_{\ell m}$ 
are necessary to obtain full agreement between the factorized 
waveforms and the nonspinning $\hat{C}_{\ell m}$ waveforms through 3PN order. 
We note that the nonspinning terms at 3PN order in the $\delta_{\ell m}$'s are the same as 
the 3PN phase terms in $Z_{\ell m\omega_0}$ in Ref.~\cite{Tagoshi94}. This happens because
in the test-particle limit the PN expansion of $T_{\ell m}$ 
does not contain imaginary terms at 3PN. Thus, for $q=0$, the phase corrections $\delta_{\ell m}$ 
at 3PN order do not contain any additional terms other than the 3PN phase terms in $Z_{\ell m\omega_0}$. 
We further note that some of the above $\delta$'s can be obtained directly in the standard PN and test-particle
limit calculations. For example, the terms proportional to $\pi v^6$ and $\pi^2 v^9$ (for $q=0$)
are the same as the phase factors in the asymptotic amplitude in the test-particle limit
calculations (e.g., Eqs.~(30)--(32) in Ref.~\cite{Tagoshi94} and Eq.~(4.17) in Ref.~\cite{MSSTT1997}).

\subsection{Taylor-expanded residual terms and their resummation}
\label{sec:tmlexpand}

In the circular-orbit, adiabatic approximation $C_{\ell m} = - (m \Omega)^2\,h_{\ell m}$. By Taylor expanding 
the factorized waveforms $h_{\ell m}$ given in Eq.~(\ref{hlm}), comparing the result with the $C_{\ell m}$ 
waveforms derived in Sec.~\ref{sec:tmltaylor}, and factoring out the imaginary terms 
in the $\delta_{\ell m}$ of Eqs.~(\ref{delta2m})--(\ref{delta4m}), 
we derive the $f_{\ell m}$'s in Eq.~(\ref{hlm}). 
We notice that in the case of even-parity modes, the determination of the $f_{\ell m}$ is unique. 
In the case of odd-parity modes, it depends on the choice of the source which, as explained 
above, can be either the energy or the angular momentum. We denote with $f_{\ell m}^L$ and 
$f_{\ell m}^H$ the odd-parity modes computed with the energy and angular-momentum sources, respectively. 
[Since in both cases the source is a real quantity, the phases $\delta_{\ell m}$'s remain the same.] 
We obtain through $\ell = 4$ (see the Appendix~\ref{AppendixB} for modes with $4 < \ell \leq 8$)
\begin{widetext}
\begin{subequations}
\begin{eqnarray}
\label{f22}
f_{22}&=&1-\frac{43}{21}\,v^2-\frac{4\,q}{3}\,v^3+\left(q^2-\frac{536}{189}\right)\,v^4-\frac{118\,q}{63}\,v^5+\left(\frac{8\,q^2}{63}-\frac{856\,\text{eulerlog}_2(v^2)}{105}+\frac{21\,428\,357}{727\,650}\right)\,v^6+\frac{1\,562\,q}{189}\,v^7
\nonumber\\
&&+\left(\frac{232\,q^2}{189}+\frac{36\,808\,\text{eulerlog}_2(v^2)}{2\,205}-\frac{5\,391\,582\,359}{198\,648\,450}\right)\,v^8+ \left(\frac{458\,816\,\text{eulerlog}_2(v^2)}{19\,845}-\frac{93\,684\,531\,406}{893\,918\,025}\right)\,v^{10} \,, \nonumber \\ \\
f^L_{21}&=&1-\frac{3\,q}{2} v-\frac{59}{28}\,v^2+\frac{61\,q}{12}\,v^3+\left(-3\,q^2-\frac{5}{9}\right)\,v^4+\frac{3}{16}\,q\,\left(4\,q^2-27\right)\,v^5
\nonumber\\
&&+ \left(\frac{4\,163\,q^2}{252}-\frac{214\,\text{eulerlog}_1(v^2)}{105}+\frac{88\,404\,893}{11\,642\,400}\right)\,v^6+ \left(-\frac{2\,593\,q^3}{168}+\frac{107}{35}\,q\,\text{eulerlog}_1(v^2)-\frac{11\,847\,887\,q}{1\,058\,400}\right)\,v^7
\nonumber\\
&&+ \left(\frac{6\,313\,\text{eulerlog}_1(v^2)}{1470}-\frac{33\,998\,136\,553}{4\,237\,833\,600}\right)\,v^8 + \left(\frac{214\,\text{eulerlog}_1(v^2)}{189}-\frac{214\,752\,050\,459}{21\,794\,572\,800}\right)\,v^{10} \,,
\end{eqnarray}
\begin{eqnarray}
f_{33}&=&1-\frac{7}{2}\,v^2-2\,q\,v^3+\left(\frac{3\,q^2}{2}-\frac{443}{440}\right)\,v^4+\frac{2\,q}{3}\,v^5+ \left(-\frac{7\,q^2}{4}-\frac{78\,\text{eulerlog}_3(v^2)}{7}+\frac{147\,471\,561}{2\,802\,800}\right)\,v^6
\nonumber\\
&&+\left(\frac{6\,187\,q}{330}-q^3\right)\,v^7+ \left(39\,\text{eulerlog}_3(v^2)-\frac{53\,641\,811}{457\,600}\right)\,v^8 \,,\\
f^L_{32}&=&1-\frac{164}{45}\,v^2+\frac{2\,q}{3}\,v^3+\left(q^2+\frac{854}{495}\right)\,v^4-\frac{1\,148\,q}{135}\,v^5+ \left(\frac{4\,q^2}{3}-\frac{104\,\text{eulerlog}_2(v^2)}{21}+\frac{110\,842\,222}{4\,729\,725}\right)\,v^6
\nonumber\\
&&+ \left(\frac{17\,056\,\text{eulerlog}_2(v^2)}{945}-\frac{97\,490\,306}{1\,702\,701}\right)\,v^8 \,,\\
f_{31}&=&1-\frac{13}{6}\,v^2-2\,q\,v^3+\left(\frac{1273}{792}-\frac{5\,q^2}{2}\right)\,v^4+\frac{38\,q}{9}\,v^5+ \left(\frac{43\,q^2}{12}-\frac{26\,\text{eulerlog}_1(v^2)}{21}+\frac{400\,427\,563}{75\,675\,600}\right)\,v^6
\nonumber\\
&&+\left(\frac{11\,q^3}{3}-\frac{2\,657\,q}{594}\right)\,v^7+ \left(\frac{169\,\text{eulerlog}_1(v^2)}{63}-\frac{12\,064\,573\,043}{1\,816\,214\,400}\right)\,v^8 \,,
\end{eqnarray}
\begin{eqnarray}
f_{44}&=&1-\frac{269}{55}\,v^2-\frac{8\,q}{3}\,v^3+\left(2\,q^2+\frac{63\,002}{25\,025}\right)\,v^4+\frac{262\,q}{55}\,v^5
\nonumber\\
&&\qquad\qquad- \left(\frac{2\,203\,q^2}{495}+\frac{50\,272\,\text{eulerlog}_4(v^2)}{3465}-\frac{11\,985\,502\,766}{156\,080\,925}\right)\,v^6 \,, \\
f^L_{43}&=&1-\frac{111}{22}\,v^2+\left(\frac{3\,q^2}{2}+\frac{225\,543}{40\,040}\right)\,v^4-\frac{12\,113\,q}{1\,540}\,v^5
+ \left(\frac{11\,337\,315\,611}{277\,477\,200}-\frac{3142\,\text{eulerlog}_3(v^2)}{385}\right)\,v^6 \,,\\
f_{42}&=&1-\frac{191}{55}\,v^2-\frac{8\,q}{3}\,v^3+\left(2\,q^2+\frac{76\,918}{25\,025}\right)\,v^4+\frac{368\,q}{55}\,v^5
- \left(\frac{97\,q^2}{495}+\frac{12\,568\,\text{eulerlog}_2(v^2)}{3\,465}-\frac{5\,180\,369\,659}{312\,161\,850}\right)\,v^6 \,, \nonumber\\ \\
f^L_{41}&=&1-\frac{301}{66}\,v^2+\left(\frac{3\,q^2}{2}+\frac{760\,181}{120\,120}\right)\,v^4-\left(\frac{10\,q^3}{3}+\frac{20\,033\,q}{13\,860}\right)\,v^5 + \left(\frac{4\,735\,160\,051}{2\,497\,294\,800}-\frac{3\,142\,\text{eulerlog}_1(v^2)}{3\,465}\right)\,v^6 \,,\nonumber\\ 
\label{f41}
\end{eqnarray}
\end{subequations}
\end{widetext}
\begin{figure*}
\begin{center}
\begin{tabular}{ccc}
  \includegraphics[width=0.315\linewidth]{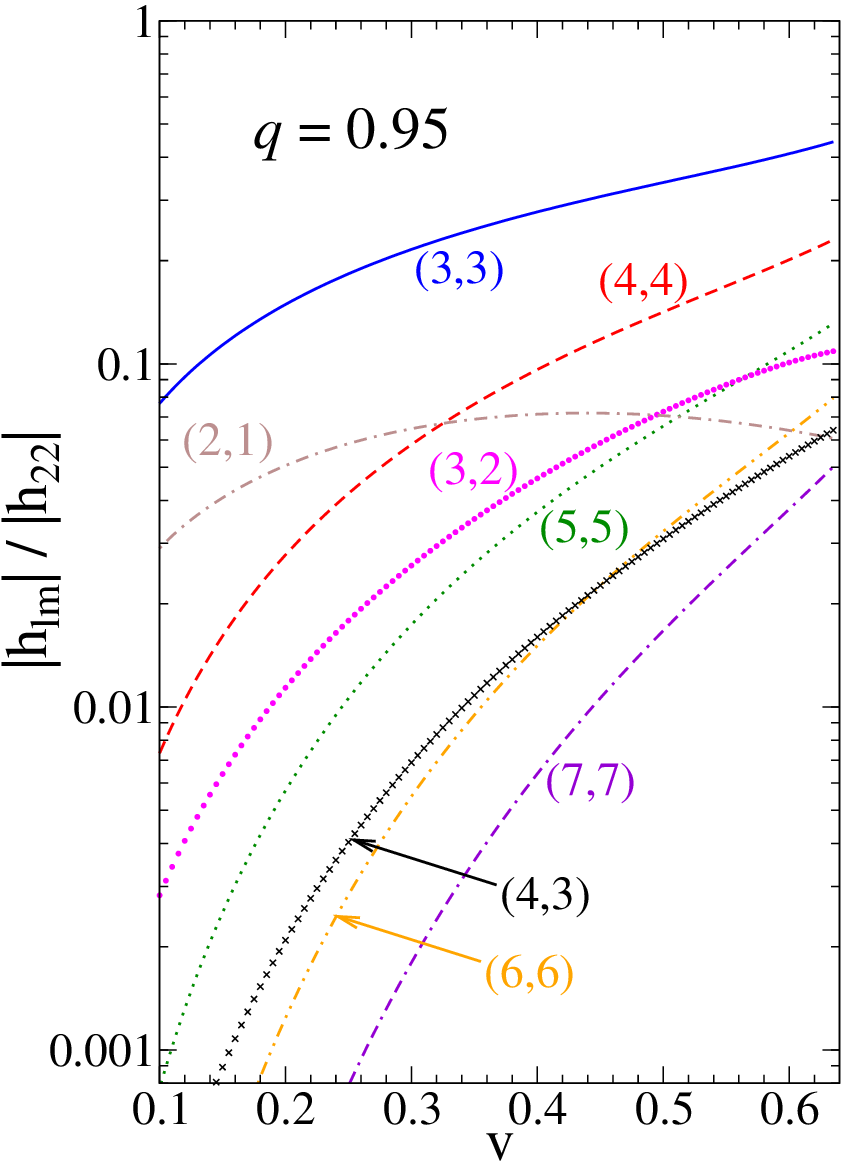} \quad
  \includegraphics[width=0.316\linewidth]{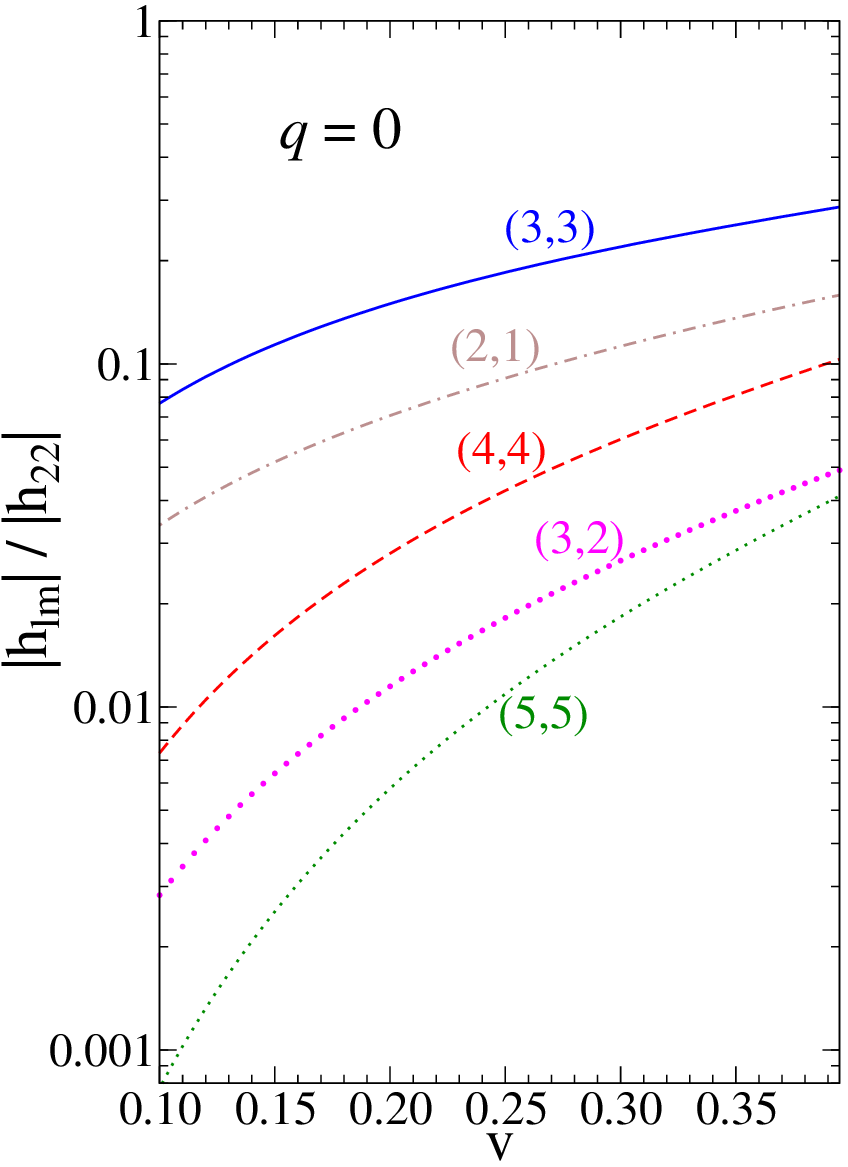} \quad
  \includegraphics[width=0.315\linewidth]{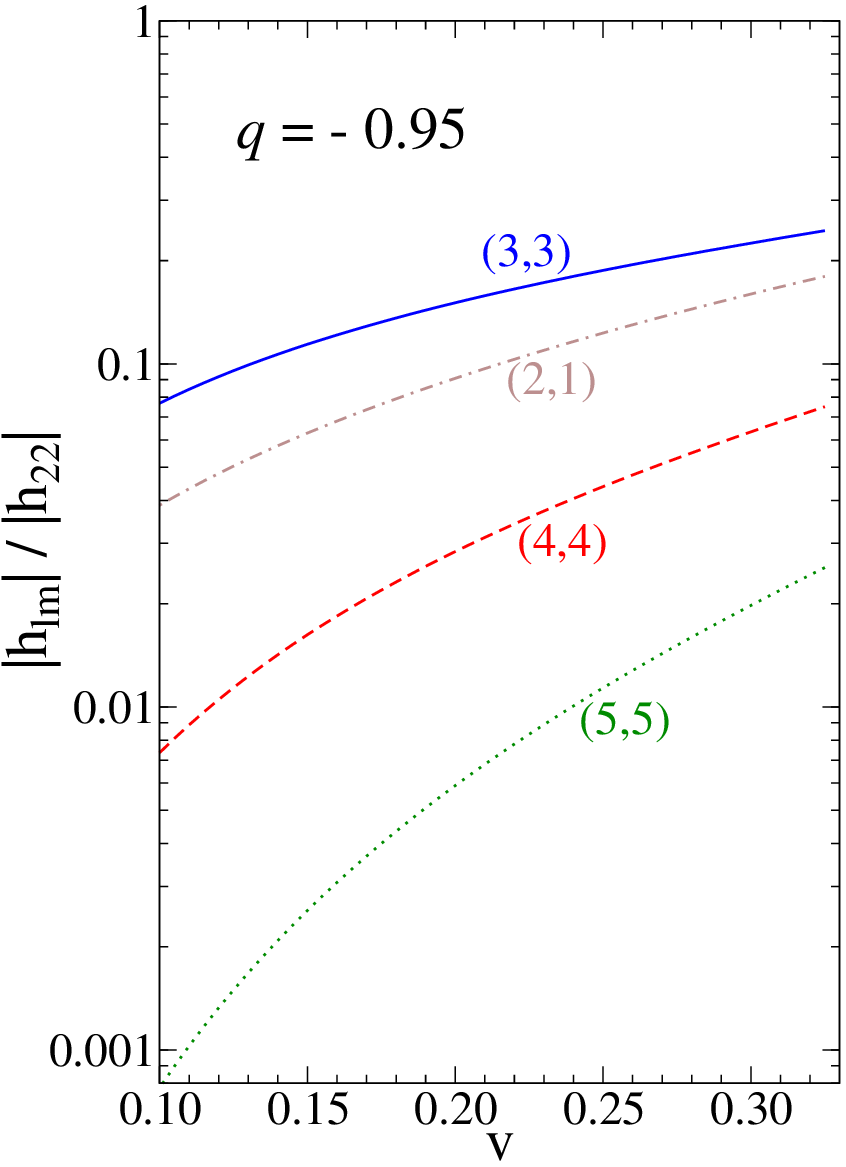} 
\end{tabular}
\end{center}
  \caption{Hierarchy of the numerically-computed modes $h_{\ell m}$ relative to that of the dominant 
$h_{22}$ mode. The spin values in the three panels from left to right are $q=0.95$, $0$, $-0.95$, respectively. 
The $x$-axis ranges between $0$ and $v_{\rm LSO}(q)$.}
\label{fig:modeamp}
\end{figure*}
where, as introduced above, we have defined ${\rm  eulerlog}_m(v^2)=\gamma_E+\log 2 + \log m + 1/2 \log v^2$ 
with $m = 1,2,3, \dots$. Note that all the nonspinning terms in Eqs.~(\ref{f22})--(\ref{f41}) 
appear at even powers of $v$, and the spin terms at odd powers of $v$. Moreover, except for 
the (2,1) odd-parity mode, all the other odd-parity modes do not have 
a spin contribution at 0.5PN order. This is consistent with the results of Appendix~\ref{AppendixD}.

As emphasized in Ref.~\cite{DINresum}, the decomposition of the Taylor-expanded 
multipolar waveform into several factors [see Eq.~(\ref{hlm})] is in itself a resummation 
procedure. In fact, the factorization of $T_{\ell m}$ has absorbed powers of $m \pi$, 
which introduce large coefficients in the Taylor-expanded waveform. Moreover, in the 
quasi-circular case assumed here, the factorization of the energy or angular-momentum 
sources, has extracted the pole located at the light-ring position 
$v = \sqrt{M/r_{\rm lr}}$ with $r_{\rm lr} = 2M \,\left [1 + \cos[2/3 \arccos(\mp a/M)]\right ]$ 
(where $\mp$ refers to prograde and retrograde orbits, respectively) which causes 
the coefficient of $v^{2n}$ in any PN-expanded quantity to grow as $r_{\rm lr}^n$ as 
$n \rightarrow \infty$. As we shall see in Sec.~\ref{sec:tmlcomparison}, despite those improvements, the $f_{\ell m}$'s above are not close enough to the exact results for large velocities.  

As we shall discuss in detail in Sec.~\ref{sec:tmlcomparison}, the
$f_{\ell m}$'s in the form of Taylor expanded power series
$\displaystyle f_{\ell m}=\sum_{k=0}^{N_{\ell m}} f_{\ell m}^{(k)}
v^k$ can be further improved by applying the Pad\'e-summation
  and/or the $\rho$-resummation ~\cite{DINresum}. In the
  Pad\'e-summation, we replace $f_{\ell m}$ with its Pad\'e
  approximant, i.e. with the rational function $({\sum\limits_{k=0}^M a_k\,v^k})\Big/({\sum\limits_{k=0}^N b_k\,v^k})$, 
with $a_0= b_0=1$ and $M+N=N_{\ell m}$. The $\rho$-resummation consists in finding the polynomial 
function $\displaystyle \rho_{\ell m}=\sum_{k=0}^{N_{\ell m}} \rho_{\ell m}^{(k)} v^k$ such that the 
Taylor expanded power series of its $\ell$-th power $(\rho_{\ell m})^\ell$ agrees with $f_{\ell m}$ 
through order $N_{\ell m}$.

The motivation for the $\rho$-resummation is to reduce the magnitude of the 1PN-order 
nonspinning coefficient $f_{\ell m}^{(2)}$ of $f_{\ell m}$, which grows linearly
with $\ell$ (see Sec. IID of Ref.~\cite{DINresum}). In the nonspinning case, since $\rho_{\ell m}^{(1)}=f_{\ell m}^{(1)}=0$, we have $\rho_{\ell m}^{(2)}=f_{\ell m}^{(2)}/\ell$ and the linear dependence of $\ell$ is removed from $\rho_{\ell m}^{(2)}$. 
We find that such dependence
on $\ell$ does also affects the 1.5PN spin terms in the even-parity modes computed 
as function of $\ell$ and $m$ in Appendix~\ref{AppendixD}. In fact, we find 
that $h_{\ell m}^{\rm even} = - 2\ell\,q\,v^3/3$, and so
$f_{\ell m}^{\rm even} = - 2\ell\,q\,v^3/3$. Thus, we apply the $\rho$-resummation  
also to the spin terms, and find  (see Appendix~\ref{AppendixC} for modes with $ 4< \ell < 8$)
\begin{figure*}
\begin{center}
\begin{tabular}{ccc}
  \includegraphics[width=0.316\linewidth]{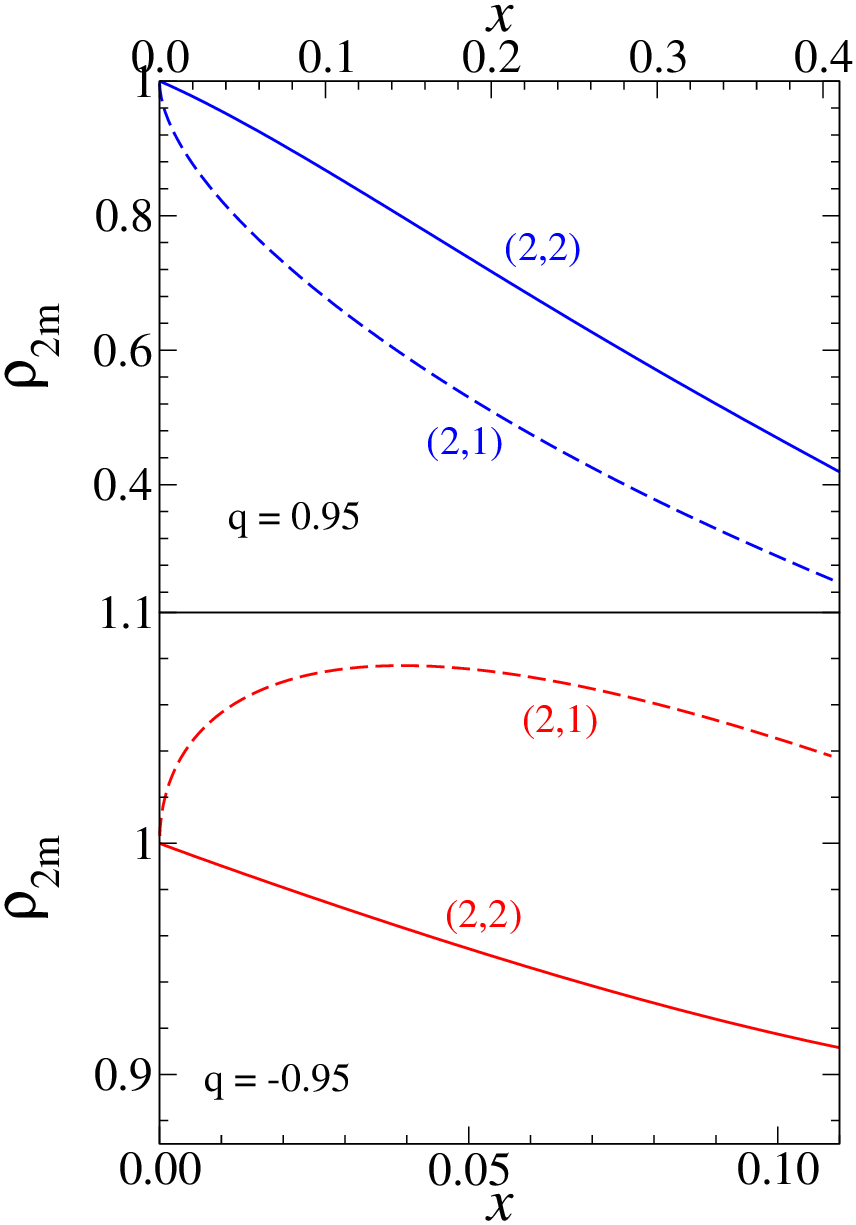} \quad
  \includegraphics[width=0.314\linewidth]{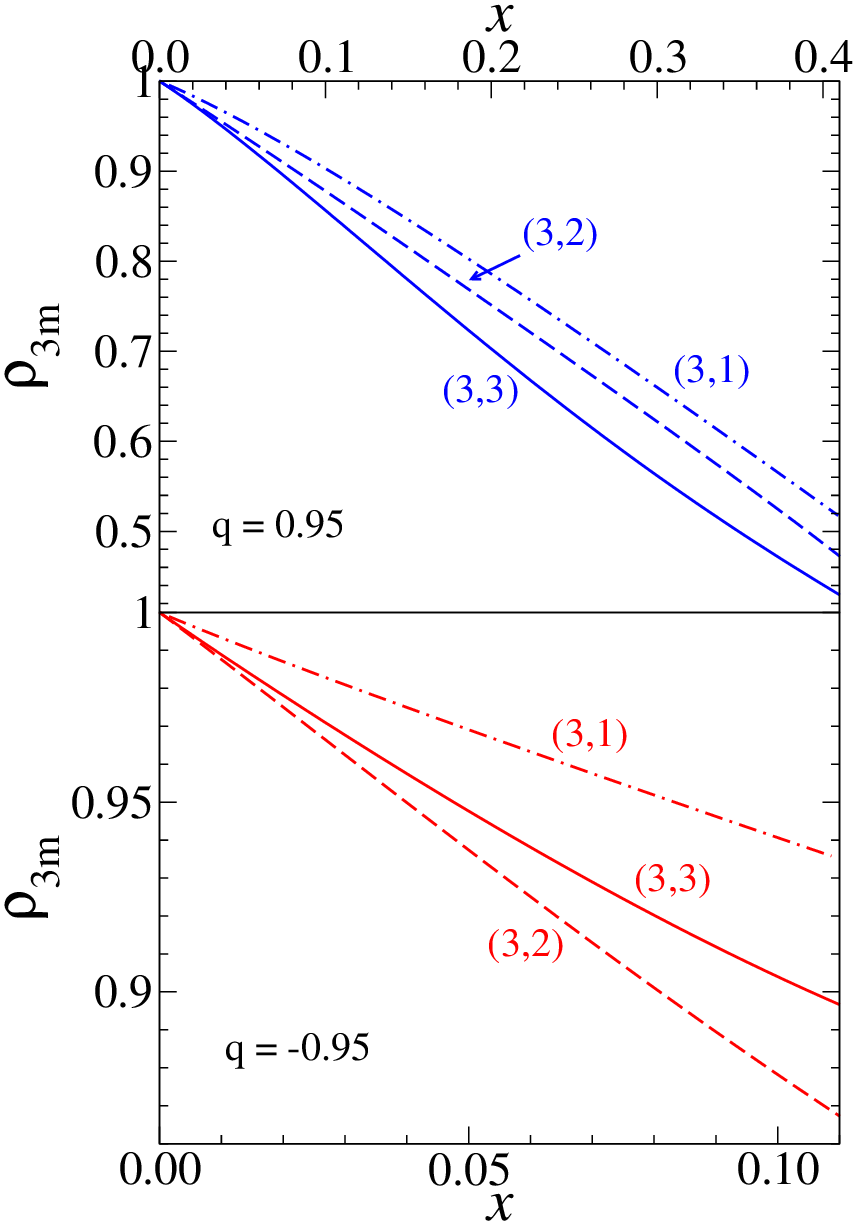} \quad
  \includegraphics[width=0.315\linewidth]{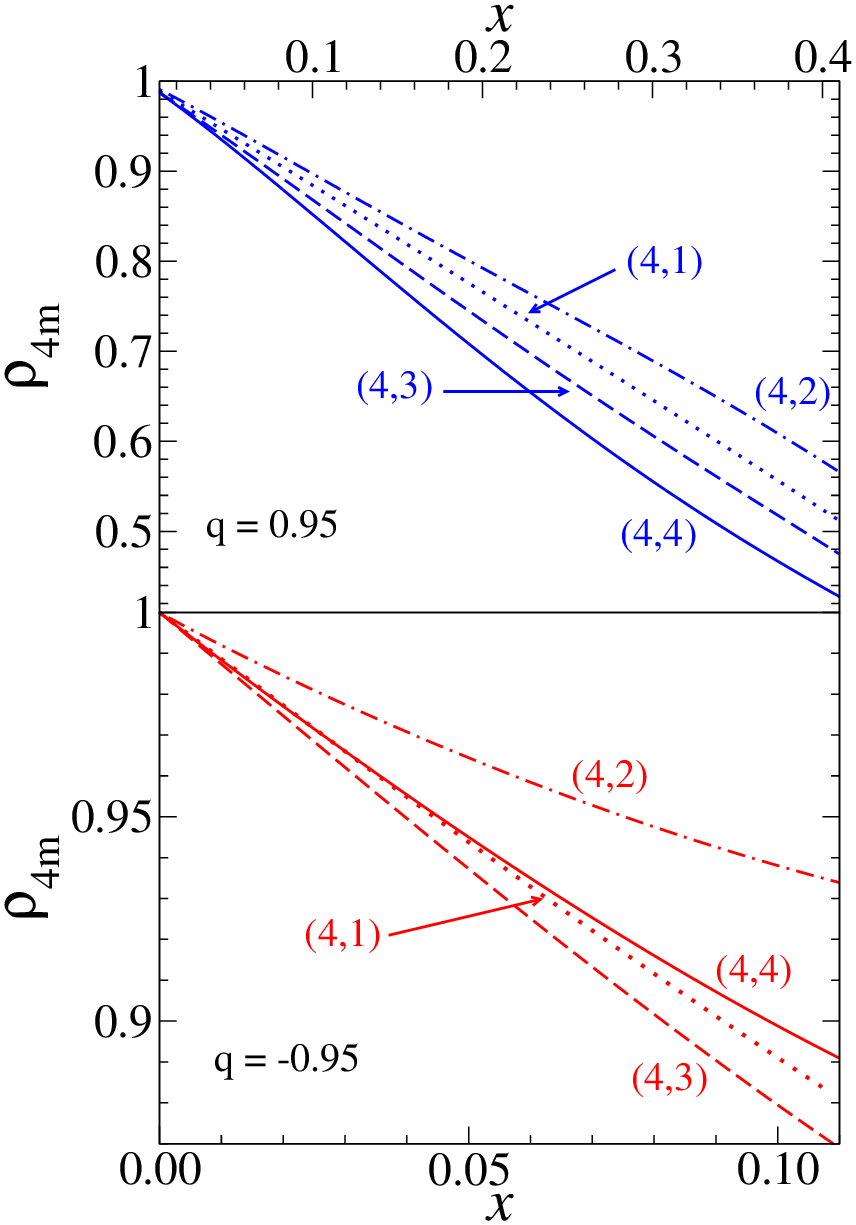} 
\end{tabular}
\end{center}
\caption{We plot the $\rho_{\ell m}$'s extracted from the numerical data as function of $x \equiv v^2$. 
The upper panels (blue colors) 
refer to $q = 0.95$, the lower panels (red colors) to $q =-0.95$. The variable $x$ ranges between 
$ 0 < x < x_{\rm LSO}(a)$.}
\label{fig:rholm234}
\end{figure*}

\begin{widetext}
\begin{subequations}\label{rholmtpl}
\begin{eqnarray}
\rho_{22}&=&1-\frac{43}{42} v^2-\frac{2\,q}{3}\,v^3+\left(\frac{q^2}{2}-\frac{20\,555}{10\,584}\right) v^4-\frac{34\,q}{21}\,v^5+ \left(\frac{89\,q^2}{252}-\frac{428\,\text{eulerlog}_2(v^2)}{105}+\frac{1\,556\,919\,113}{122\,245\,200}\right)\,v^6
\nonumber\\
&&+\left(\frac{q^3}{3}+\frac{18\,733\,q}{15\,876}\right) v^7+ \left(-\frac{q^4}{8}+\frac{18\,353\,q^2}{21\,168}+\frac{9\,202\,\text{eulerlog}_2(v^2)}{2\,205}-\frac{387\,216\,563\,023}{160\,190\,110\,080}\right)\,v^8
\nonumber\\
&&+ \left(\frac{439\,877\,\text{eulerlog}_2(v^2)}{55\,566}-\frac{16\,094\,530\,514\,677}{533\,967\,033\,600}\right)\,v^{10} \,,\\
\rho^L_{21}&=&1-\frac{3\,q}{4}\,v+\left(-\frac{9\,q^2}{32}-\frac{59}{56}\right) v^2+\left(\frac{1\,177\,q}{672}-\frac{27\,q^3}{128}\right)\,v^3-\left(\frac{47\,009}{56\,448}+\frac{865\,q^2}{1\,792}+\frac{405\,q^4}{2\,048}\right) v^4
\nonumber\\
&&-\left(\frac{98\,635\,q}{75\,264}-\frac{2\,031\,q^3}{7\,168}+\frac{1\,701\,q^5}{8\,192}\right)\,v^5+ \left(-\frac{15\,309\,q^6}{65\,536}+\frac{3\,897\,q^4}{16\,384}+\frac{9\,032\,393\,q^2}{1\,806\,336}-\frac{107\,\text{eulerlog}_1(v^2)}{105}\right.
\nonumber\\
&&\left.+\frac{7\,613\,184\,941}{2\,607\,897\,600}\right)\,v^6+ \left(-\frac{72\,171\,q^7}{262\,144}+\frac{18\,603\,q^5}{65\,536}-\frac{55\,169\,q^3}{16\,384}+\frac{107}{140}\,q\,\text{eulerlog}_1(v^2)-\frac{3\,859\,374\,457\,q}{1\,159\,065\,600}\right)\,v^7
\nonumber\\
&&+ \left(\frac{6\,313\,\text{eulerlog}_1(v^2)}{5\,880}-\frac{1\,168\,617\,463\,883}{911\,303\,737\,344}\right)\,v^8+ \left(\frac{5\,029\,963\,\text{eulerlog}_1(v^2)}{5\,927\,040}-\frac{63\,735\,873\,771\,463}{16\,569\,158\,860\,800}\right)\,v^{10} \,, \nonumber \\
\end{eqnarray}
\begin{eqnarray}
\rho_{33}&=&1-\frac{7}{6} v^2-\frac{2\,q}{3}\,v^3+\left(\frac{q^2}{2}-\frac{6\,719}{3\,960}\right) v^4-\frac{4\,q}{3}\,v^5+ \left(\frac{5\,q^2}{36}-\frac{26\,\text{eulerlog}_3(v^2)}{7}+\frac{3\,203\,101\,567}{227\,026\,800}\right)\,v^6
\nonumber\\
&&+\left(\frac{q^3}{3}+\frac{5\,297\,q}{2\,970}\right) v^7+ \left(\frac{13\,\text{eulerlog}_3(v^2)}{3}-\frac{57\,566\,572\,157}{8\,562\,153\,600}\right)\,v^8 \,,\\
\rho^L_{32}&=&1-\frac{164}{135} v^2+\frac{2\,q}{9}\,v^3+\left(\frac{q^2}{3}-\frac{180\,566}{200\,475}\right) v^4-\frac{2\,788\,q}{1\,215}\,v^5+ \left(\frac{488\,q^2}{405}-\frac{104\,\text{eulerlog}_2(v^2)}{63}+\frac{5\,849\,948\,554}{940\,355\,325}\right)\,v^6
\nonumber\\
&&+ \left(\frac{17\,056\,\text{eulerlog}_2(v^2)}{8\,505}-\frac{10\,607\,269\,449\,358}{3\,072\,140\,846\,775}\right)\,v^8 \,,\\
\rho_{31}&=&1-\frac{13}{18} v^2-\frac{2\,q}{3}\,v^3+\left(\frac{101}{7\,128}-\frac{5\,q^2}{6}\right) v^4+\frac{4\,q}{9}\,v^5+ \left(-\frac{49\,q^2}{108}-\frac{26\,\text{eulerlog}_1(v^2)}{63}+\frac{11\,706\,720\,301}{6\,129\,723\,600}\right)\,v^6
\nonumber\\
&&+\left(\frac{q^3}{9}-\frac{2\,579\,q}{5\,346}\right) v^7+ \left(\frac{169\,\text{eulerlog}_1(v^2)}{567}+\frac{2\,606\,097\,992\,581}{4\,854\,741\,091\,200}\right)\,v^8 \,,
\end{eqnarray}
\begin{eqnarray}
\rho_{44}&=&1-\frac{269}{220} v^2-\frac{2\,q}{3}\,v^3+\left(\frac{q^2}{2}-\frac{14\,210\,377}{8\,808\,800}\right) v^4-\frac{69\,q}{55}\,v^5
\nonumber\\
&&\qquad\qquad+ \left(\frac{217\,q^2}{3\,960}-\frac{12\,568\,\text{eulerlog}_4(v^2)}{3\,465}+\frac{16\,600\,939\,332\,793}{1\,098\,809\,712\,000}\right)\,v^6 \,,\\
\rho^L_{43}&=&1-\frac{111}{88} v^2+\left(\frac{3\,q^2}{8}-\frac{6\,894\,273}{7\,047\,040}\right) v^4-\frac{12\,113\,q}{6\,160}\,v^5+ \left(\frac{1\,664\,224\,207\,351}{195\,343\,948\,800}-\frac{1\,571\,\text{eulerlog}_3(v^2)}{770}\right)\,v^6 \,,\\
\rho_{42}&=&1-\frac{191}{220} v^2-\frac{2\,q}{3}\,v^3+\left(\frac{q^2}{2}-\frac{3\,190\,529}{8\,808\,800}\right) v^4-\frac{7\,q}{110}\,v^5
\nonumber\\
&&\qquad\qquad+ \left(\frac{2\,323\,q^2}{3\,960}-\frac{3\,142\,\text{eulerlog}_2(v^2)}{3\,465}+\frac{848\,238\,724\,511}{219\,761\,942\,400}\right)\,v^6 \,,\\
\rho^L_{41}&=&1-\frac{301}{264} v^2+\left(\frac{3\,q^2}{8}-\frac{7\,775\,491}{21\,141\,120}\right) v^4+\left(-\frac{5\,q^3}{6}-\frac{20\,033\,q}{55\,440}\right) v^5
\nonumber\\
&&\qquad\qquad+ \left(\frac{1\,227\,423\,222\,031}{1\,758\,095\,539\,200}-\frac{1571\,\text{eulerlog}_1(v^2)}{6\,930}\right)\,v^6 \,. 
\end{eqnarray}
\end{subequations}
\end{widetext}
Lastly, we may use $E(r)$ instead of $|\vL|$ as the source term in Eq.~(\ref{hlm}) 
for the odd-parity modes. The corresponding $f_{\ell m}^H$ and $\rho_{\ell m}^H$ expressions are 
given in Appendices~\ref{AppendixB} and ~\ref{AppendixC}, respectively.

In the next section we shall investigate the numerical (exact) $\rho_{\ell m}$'s, and compare 
them with the analytical ones. We shall find that the agreement between the numerical and 
analytical $\rho_{\ell m}$ is quite good, except for some modes. Guided by the comparison 
with numerical results, we shall apply the Pad\'e summation to the $\rho_{\ell m}$'s, and also 
work out an improved resummation which consists in factoring out the lower-order PN terms 
in the Taylor-expanded $\rho_{\ell m}$'s. We find that this factorization brings 
the zeros of the analytical $\rho_{\ell m}$ closer to the numerical (exact) ones.

\section{Comparison between analytical and numerical results for the 
test-particle limit case}
\label{sec:tmlcomparison}
\begin{figure*}
\begin{center}
\begin{tabular}{ccc}
  \includegraphics[width=0.316\linewidth]{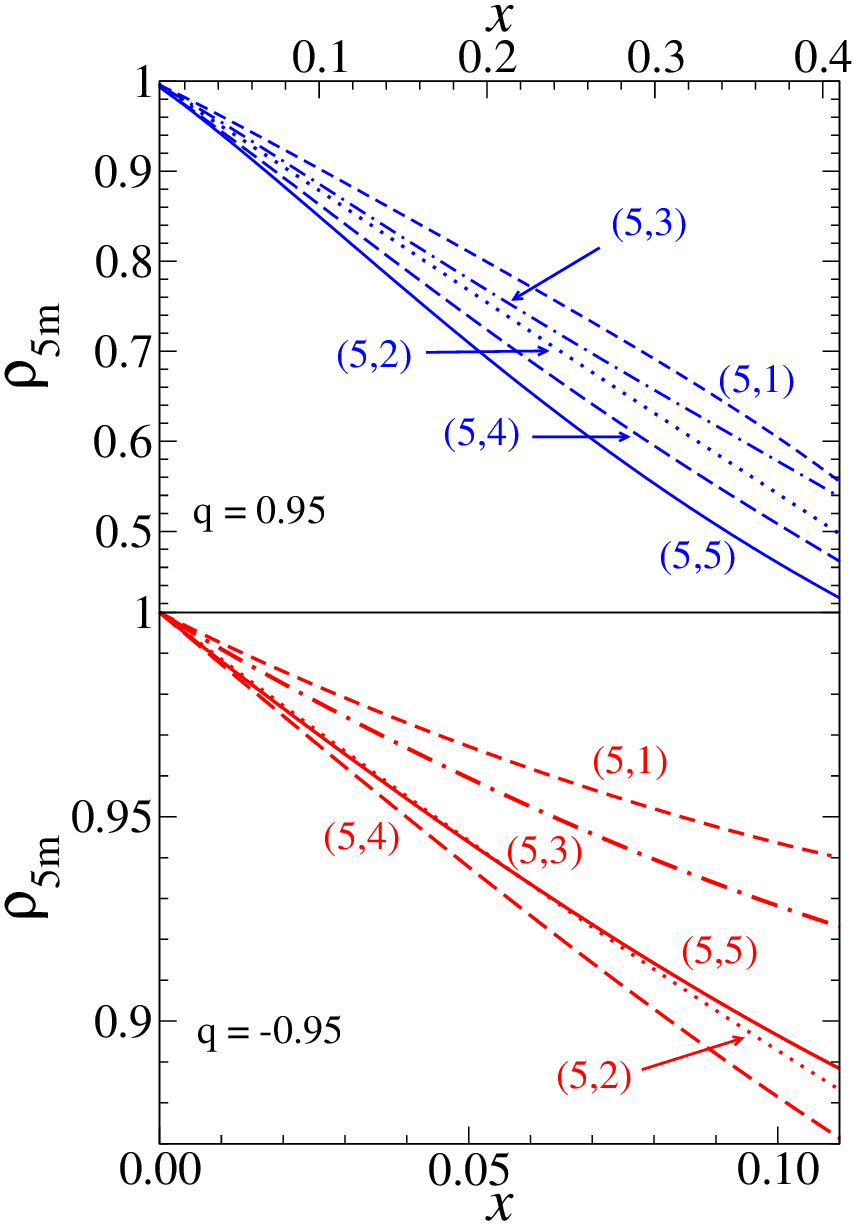} \quad \quad
  \includegraphics[width=0.314\linewidth]{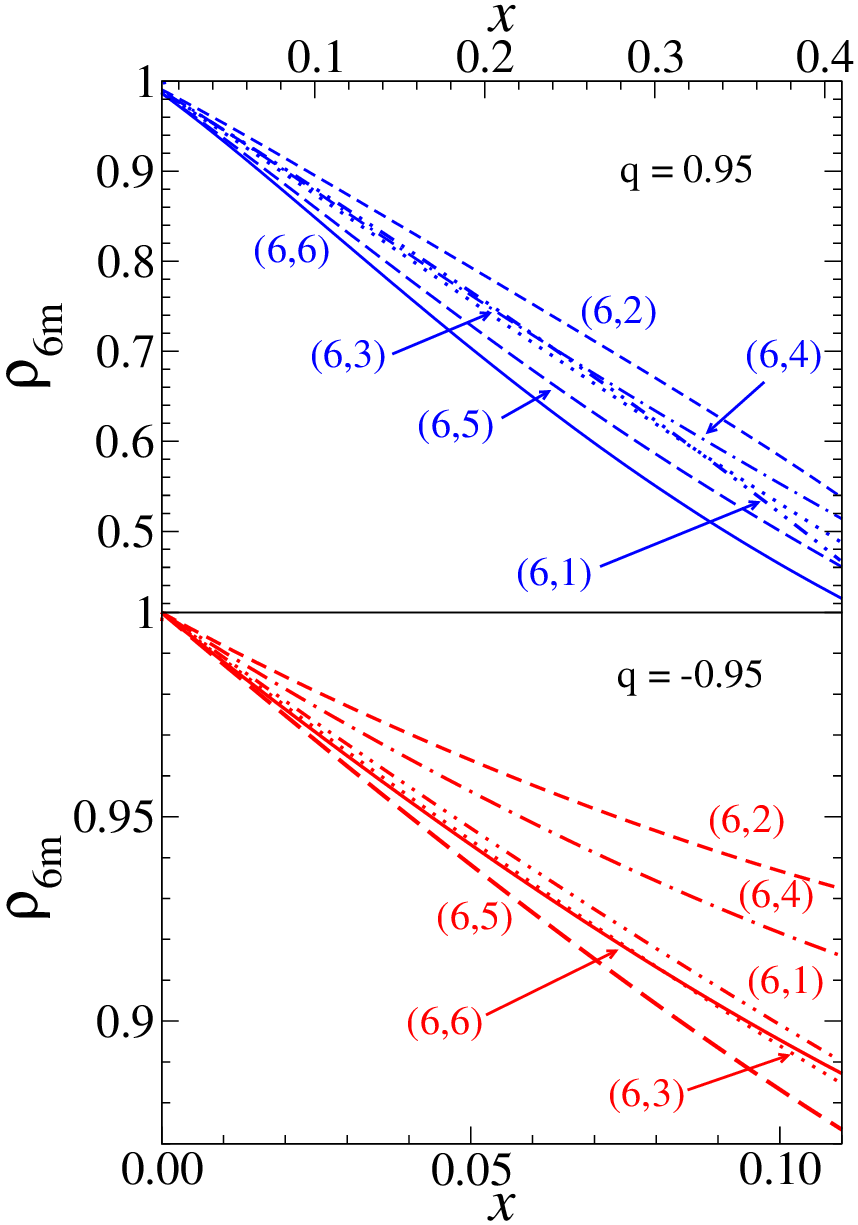} 
\end{tabular}
\end{center}
\caption{We plot the $\rho_{\ell m}$'s extracted from the numerical data as function of $x \equiv v^2$. 
The upper panels (blue colors) refer to $q = 0.95$, the lower panels (red colors) to $q =-0.95$. The variable $x$ ranges between 
$ 0 < x < x_{\rm LSO}(a)$.}
\label{fig:rholm56}
\end{figure*}
We have two goals to achieve in this section. The first is to accurately model 
the amplitude of the $(l,m)$ modes for several values of the spin 
parameter $q$ and velocity $v$. The second is to obtain the best agreement 
between the numerical (exact) and analytical energy fluxes  
{\it without} introducing adjustable parameters in the analytical model.

The numerical values of the energy flux used in this paper are
obtained with a high precision numerical code which solves the
Teukolsky equation~\cite{FujitaTagoshi04,FujitaTagoshi05,FujitaHikidaTagoshi09}.
The homogeneous solution of the radial Teukolsky equation is obtained
numerically by using a formalism developed by Mano, Suzuki and
Takasugi~\cite{ManoSuzukiTakasugi96}. In this method, the homogeneous
solutions are expressed in terms of series of two kinds of special
functions, hypergeometric functions and confluent hypergeometric
functions. In Refs.~\cite{FujitaTagoshi04,FujitaTagoshi05}, it was
shown that the series converges very fast and one can compute
numerically the homogeneous solutions very accurately.  The
homogeneous solution obtained with this method was applied to the
numerical calculation of gravitational waves emitted by a particle in
a quasi-circular and equatorial orbit around a Kerr black
hole~\cite{FujitaTagoshi04,FujitaTagoshi05}.  In this paper, for the
comparison with analytical formulas, we compute the $Z_{\ell
  m\omega_0}$ (and thus the $C_{\ell m}$) as well as $(dE/dt)_{\ell m}$
for various $q$ and $\Omega$. The computation is done with the double
precision accuracy, and the estimated accuracy of $Z_{\ell m\omega_0}$
(and thus the $C_{\ell m}$) as well as $(dE/dt)_{\ell m}$ is about 14
significant figures. As in Ref.~\cite{FujitaTagoshi04}, 
the accuracy is estimated by comparing 
the energy flux with that of Ref.~\cite{TagoshiNakamura1994}
in which the accuracy was estimated as about 20 significant figures.

\subsection{Hierarchy between the $(l,m)$'s modes}
\label{sec:hierarchy}

In Fig.~\ref{fig:modeamp} we study the hierarchy among the numerically-computed modes 
and plot $|h_{\ell m}|/|h_{22}|$ versus $v$ for the representative spin cases: 
$q=0.95$, $0$, and $-0.95$. The parameter $v$ varies between $0.1$ and $v_{\rm LSO}(q)$, 
where we denote with LSO the last stable orbit for a test-particle in the Kerr geometry. 

The strain waveforms $h_{\ell m}$'s are computed from the $C_{\ell m}$'s under the 
quasi-circular adiabatic assumption, i.e.,  
$h_{\ell m}=-C_{\ell m}/(m\Omega)^2$. As we shall discuss in Sec.~\ref{sec:energy}, 
the energy flux for quasi-circular adiabatic orbits 
can be computed through the well-known relation 
\begin{equation}
\label{energyflux}
F(v)= \frac{1}{16\pi}\,\sum_{\ell}\,\sum_{m=-\ell}^{\ell}(m\,\Omega)^2\,
\left|h_{\ell m}(v)\right|^2\,.
\end{equation}
Thus, when analyzing the contribution of the $\dot{h}_{\ell m}$'s to the energy flux, we need to 
remember that $\dot{h}_{\ell m}=i\,C_{\ell m}/(m\Omega)$. Thus, the 
dependence of $\dot{h}_{\ell m}$'s on $m$ is different than the one of  
${h}_{\ell m}$'s, and, as a consequence, the hierarchy of the modes 
in the energy flux is different.

Denoting by $|h_{\ell m}|/|h_{22}|$ the relative strain amplitude and
by $|\dot{h}_{\ell m}|^2/|h_{22}|^2$ the relative radiation power, we
find the following trends. In the anti-aligned case $q=-0.95$ and the
nonspinning case, the $(3,3)$, $(2,1)$ and $(4,4)$ modes are the
largest subdominant modes in terms of strain amplitude. In terms of
radiation power, they are also among the largest subdominant modes
although their hierarchy changes. The $(4,4)$ mode contributes more
power than the $(2,1)$ mode because of its larger $m$. For the same
reason, in the nonspinning case, the $(5,5)$ mode contributes more
power than the $(2,1)$ mode and becomes the third strongest
subdominant mode. In fact, in the anti-aligned and nonspinning cases,
relative to the $(2,2)$ mode, the $(3,3)$ mode contributes $>10\%$ of
radiation power at the LSO, only the $(3,3)$ and $(4,4)$ modes
contribute $>1\%$, and the $(5,5)$ mode contributes $1\%$ only in the
nonspinning case.  In the aligned case $q=0.95$, we plot in
Fig.~\ref{fig:modeamp} the relative strain amplitudes of 8 modes that
are larger than $5\%$ at the LSO. In terms of the relative radiation
power, the $(3,3)$, $(4,4)$, $(5,5)$, $(6,6)$, $(7,7)$ and $(8,8)$
modes are the largest subdominant modes. The $(3,3)$, $(4,4)$ and
$(5,5)$ modes each contributes $>10\%$ relative to the $(2,2)$ mode at
the LSO. In particular, the $(3,3)$ mode contributes $>30\%$ relative
to the $(2,2)$ mode to both the strain amplitude and the radiation
power. Accurate modeling of its amplitude is therefore crucial in
modeling the full gravitational-wave waveform and the energy flux.

\subsection{Comparison between the analytical and numerical modes}
\label{sec:comparison}

We now examine the amplitude agreement of the numerical and analytical 
waveforms, focusing mainly on the dominant modes: $(2,2)$, 
$(2,1)$, $(3,3)$, $(3,2)$, $(4,4)$ and $(5,5)$. 

In Figs.~\ref{fig:rholm234} and \ref{fig:rholm56} we show several numerical 
$\rho_{\ell m}$'s versus $x \equiv v^2$ for three representative spin cases: $q = -0.95,0,0.95$. 
Since the latter are real, the numerical $\rho_{\ell m}$'s are obtained using 
Eq.~(\ref{hlm}) with $f_{\ell m} = \rho_{\ell m}^\ell$, 
that is dividing the numerical $|h_{\ell m}|^{1/\ell}$ by $(|T_{\ell m}|\,
\hat{S}_{\rm eff}^{(\epsilon)})^{1/\ell}$. The numerical $h_{\ell m}$ are computed 
from the numerical $C_{\ell m}$ through the relation $h_{\ell m}=-C_{\ell m}/(m\Omega)^2$

Using the 0.5PN (1.5PN) order spin terms in the odd (even)-parity modes computed 
in Appendix~\ref{AppendixD} for generic $\ell$ and $m$, and the non-spinning 1PN terms 
derived in Refs.~\cite{Kidder2008,DINresum}, we have 
\begin{eqnarray}
f_{\ell m}^{\rm even}(x) &=& 1 
- \left (1 - \frac{1}{\ell} + \frac{m^2\,(\ell+9)}{2 \ell\,(\ell +1)\,(2\ell + 3)}
\right )\,\ell\,x\nonumber \\
&& - \frac{2}{3}\,\ell\,q\,x^{3/2}  + {\cal O}(x^2)\,,
\end{eqnarray}
and 
\begin{eqnarray}
f_{\ell m}^{L}(x) &=& 1 - \frac{3}{2}\,q\,x^{1/2}\delta_{\ell 2}\,\delta_{m 1} \nonumber \\
&& - \left (1 + \frac{1}{\ell} - \frac{2}{\ell^2} + \frac{m^2\,(l+4)}{2 \ell\,(\ell +2)\,(2\ell + 3)}
\right )\,\ell\,x\nonumber \\
&&  + {\cal O}(x^{3/2})\,.
\end{eqnarray}
Note that the 1.5PN spin terms in the odd-parity modes are not known 
for generic $\ell$ and $m$, but they are available through $\ell =6$ 
in this paper. 

Reference~\cite{DINresum} pointed out that because the 1PN order term in the $f_{\ell m}^{\rm even}$ 
and $f_{\ell m}^{L}$ scale as $\ell$ and is negative, for large $\ell$ it can cause the $f_{\ell m}$ to 
go to zero even before reaching the LSO. For example if we consider the LSO in Schwarzschild, 
$x_{\rm LSO}(0) = 1/6$ ($v_{\rm LSO}(0) = 1/\sqrt{6} \simeq 0.4082$), 
$f_{66}$ at 1PN order has a zero at $v = 0.3634$~\cite{DINresum}. In the even-parity 
case, the inclusion of the 1.5PN spin term with $q>0$ can cause the zero to occur even at smaller 
values of $v$. In particular, for $q=0.95$, $f_{66}$ has 
a zero at $v = 0.3362$ ($v_{\rm LSO}(0.95) = 0.6497$). 
By contrast, the cases with $q<0$ can push the zero to negative or imaginary values,  
or to values of $v$ above the LSO, thus making it harmless. For example, for $q=-0.95$, $f_{66}$ has 
a zero at $v = 0.4075$ ($v_{\rm LSO}(-0.95) = 0.3373$). 
Similarly, when considering the odd-parity modes for large $\ell$, e.g., 
the $f^L_{65}$ mode, we find that in the non-spinning case 
the 1PN term causes $f^L_{65}$ 
to have a zero at $v = 0.3602$, and the inclusion of 1.5PN spin term causes 
the zero to move to $v = 0.3502$ for $q = 0.95$, and to $v = 0.3717$ for $q = -0.95$.

In the spin case, the above problem can be even worst than in the non-spinning 
case for lower values of $\ell$. For example, the 1PN term causes a zero 
in the $f_{33}$ at $v = 0.5345$ which is above $v_{\rm LSO}(0)$, 
but the inclusion of the 1.5PN spin term moves the zero 
to $v = 0.4764$ for $q = 0.95$ which is quite below $v_{\rm LSO}(0.95)$. 

Motivated by the above discussion and the result in Appendix~\ref{AppendixD} that 
shows that the even-parity 1.5PN spin terms scale as $\ell$ ($f_{\ell m}^{\rm even} = - 2\ell\,q\,v^3/3$), 
we adopt the $\rho$-resummation also for the spin terms. The $\rho_{\ell m}$'s through 1.5PN order read:
\begin{eqnarray}
\rho_{\ell m}^{\rm even}(x) &=& 1 
- \left (1 - \frac{1}{\ell} + \frac{m^2\,(\ell+9)}{2 \ell\,(\ell +1)\,(2\ell + 3)}
\right )\,x\nonumber \\
&& - \frac{2}{3}\,q\,x^{3/2}  + {\cal O}(x^2)\,,
\end{eqnarray}
and 
\begin{eqnarray}
\rho_{\ell m}^{L}(x) &=& 1 - \frac{3}{2}\,\frac{1}{\ell}\,q\,x^{1/2}\delta_{\ell 2}\,\delta_{m 1} - \frac{9}{8}\frac{\ell-1}{\ell^2}\,q^2\,x\,\delta_{\ell 2}\,\delta_{m 1}\nonumber \\
&& - \left (1 + \frac{1}{\ell} - \frac{2}{\ell^2} + \frac{m^2\,(l+4)}{2 \ell\,(\ell +2)\,(2\ell + 3)}
\right )\,x\nonumber \\
&&  + {\cal O}(x^{3/2})\,.
\end{eqnarray}
We notice that the 1PN and 1.5PN terms in $\rho_{66}$ cause a zero at
$v = 0.8902$ for $q =0$, and at $v = 0.7577$ for $q =0.95$.  The zero
in $\rho_{33}$ occurs at $v = 0.9258$ for $q =0$, and at $v = 0.7765$
for $q =0.95$.  All these numbers are larger than $v_{\rm
  LSO}(q)$. Note however that the $\rho$-resummation may be less
effective for $q>0.95$, since at $q=1$, the zero in $\rho_{66}$ occurs
at $v=0.7530$ and the zero in $\rho_{33}$ occurs at $v=0.7713$, both
smaller than $v_{\rm LSO}(1)=0.7937$.  Of course all this discussion
does not take into account the higher-order PN terms which can also
move the zero to lower or higher values. However, as we shall see
below, the behavior of the numerical $\rho_{\ell m}$ is captured
by the 0.5PN, 1PN and 1.5PN terms.

In Figs.~\ref{fig:rholm234} and \ref{fig:rholm56} we plot the $\ell =
2,3,4,5,6$ ($ m = \ell, \ell -1, \dots ,1$) numerical modes versus
$x$. First, as observed in Ref.~\cite{DINresum} for the nonspinning
case, also for the spin case, the behavior of the $\rho_{\ell m}$ is
reasonably {\it simple}. In particular, except for the $(2,1)$ case
which shows a special shape due to the presence of the 0.5PN term
($\sqrt{x}$), all the other modes are well represented by (broken) 
straight lines with one or two changes in the slope at high frequency. As
  in the nonspinning case, but less pronounced here, for each
  value of $\ell$, the (negative) slopes of the dominant $m = \ell$
  (even-parity), and subdominant $m=\ell-1$ (odd-parity) modes are
  close to each other, and these slopes become somewhat closer as
  $\ell$ increases. This property is reproduced by the analytical
  $\rho_{\ell m}$'s truncated at 1.5PN order through $\ell=6$ modes,
  whose 1.5PN terms are known. 
 
In Figs.~\ref{fig:rho22} and \ref{fig:rho33} we compare the numerical and 
analytical $\rho_{22}$ and $\rho_{33}$, respectively. We use the following notation for the 
analytical models. We indicate with ${\rm T}_{\rm N}[\rho_{\ell m}]$ 
the $\rho_{\ell m}$ expanded in Taylor series of $v$ through $v^N$. 
We indicate with  ${\rm P}_{\rm N}^{\rm M}[\rho_{\ell m}]$ the Pad\'e-summed $\rho_{\ell m}$ with 
$M$ and $N$ denoting the order of the polynomial in $v$ in the numerator
and denominator, respectively. When applying the Pad\'e summation 
in presence of logarithms (i.e., $\log(v)$) we treat the latter as constants.
We indicate with $\rho^f_{\ell m}$ an improved 
resummation of the Taylor-expanded $\rho_{\ell m}$'s which 
consists in factoring out their 0.5PN, 1PN and 1.5PN order terms, that is we write 
\begin{equation}
\label{rhof}
\rho^f_{\ell m} = (1 + c_{1/2}^{\ell m}\,v + c_1^{\ell m}\,v^2 + c_{3/2}^{\ell m}\,v^3)\,
(1 + d_2^{\ell m}\,v^4 + \cdots)\,,
\end{equation}
where the coefficients $c_{1/2}^{\ell m}$, $c_1^{\ell m}$ and $c_{3/2}^{\ell m}$ 
are the 0.5PN, 1PN and 1.5PN order terms 
in the $\rho_{\ell m}$, and the coefficients 
$d_i^{\ell m}$ with $i \geq 2$ 
in Eq.~(\ref{rhof}) are obtained by imposing that the Taylor-expanded $\rho^f_{\ell m}$ 
coincides with $\rho_{\ell m}$. We shall motivate the introduction 
of the $\rho^f_{\ell m}$'s in the discussion below, but basically  
we find that the first factor on the right-hand side of Eq.~(\ref{rhof}) 
can capture reasonably well the zeros of the numerical (exact) $\rho_{\ell m}$'s. 

\ 

For the modes $\ell < 4$, we find the following $\rho^f_{\ell m}$'s: 
\begin{figure*}
\begin{center}
\begin{tabular}{ccc}
  \includegraphics[width=0.316\linewidth]{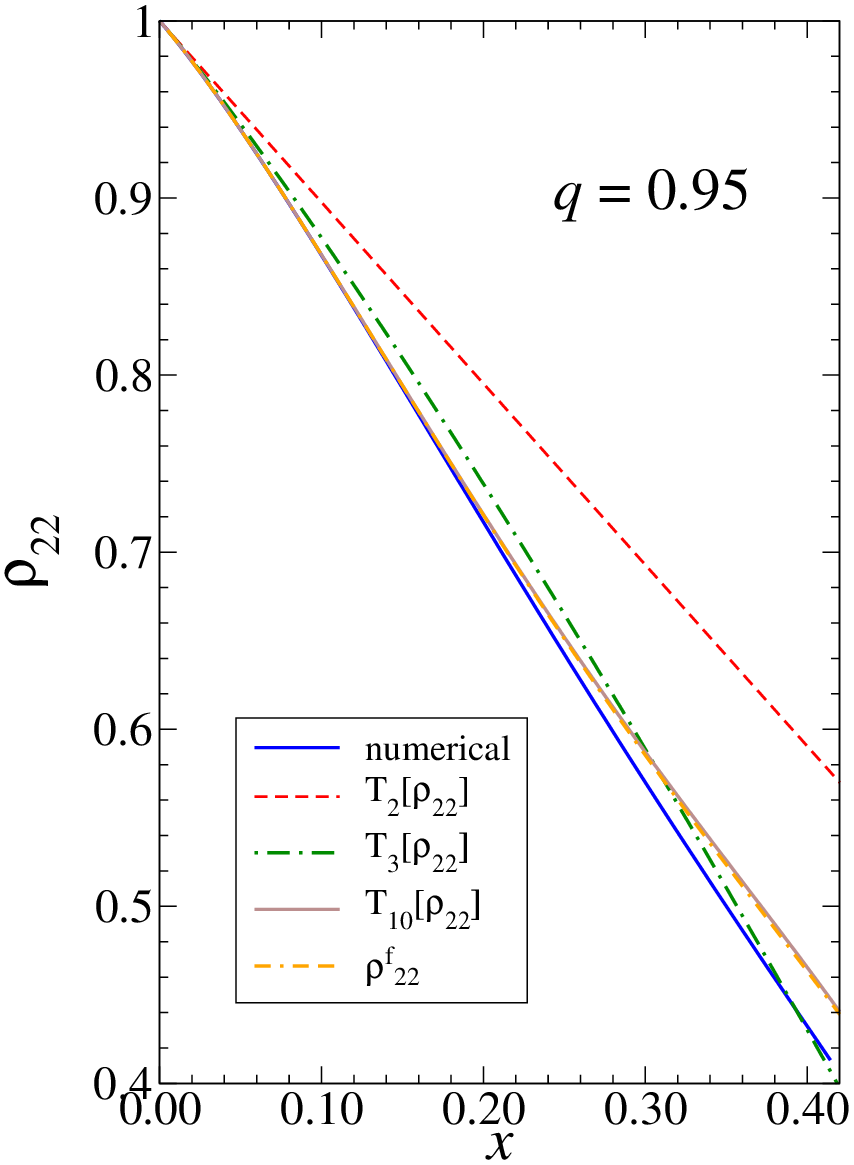} \quad
  \includegraphics[width=0.314\linewidth]{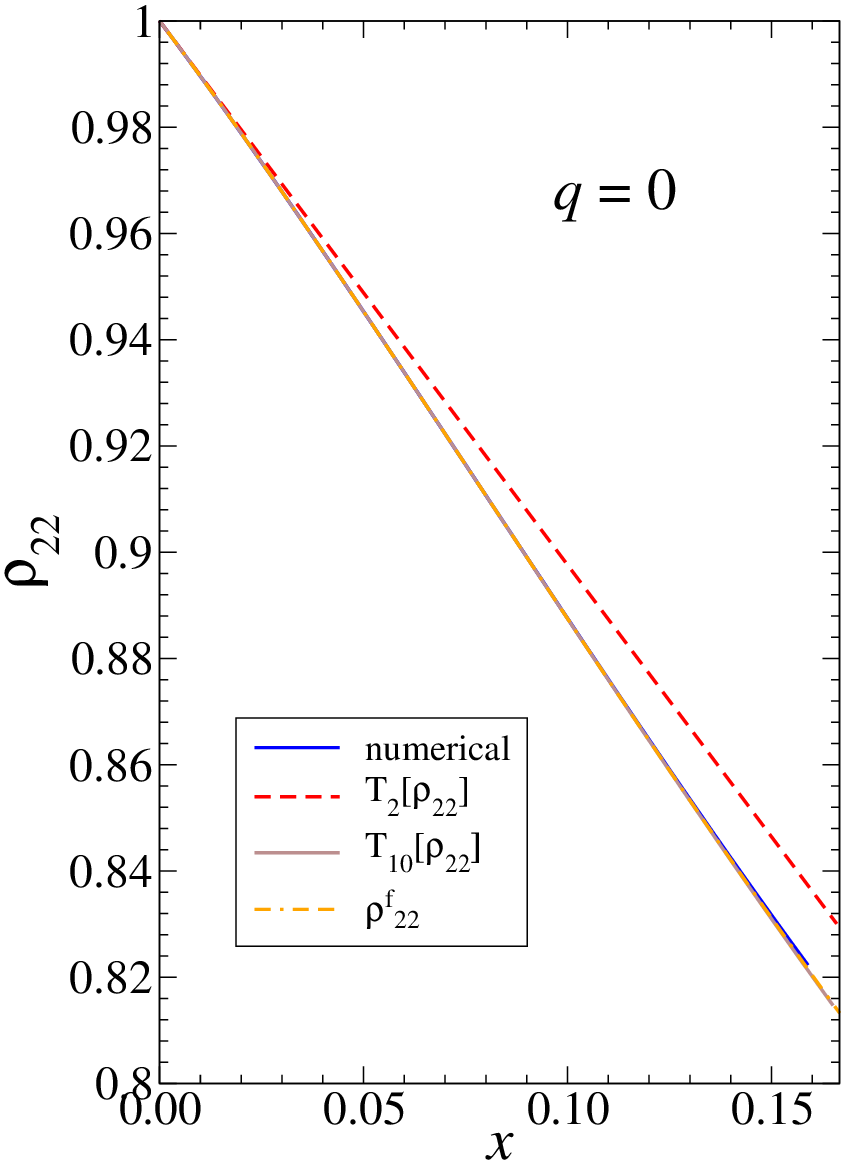} \quad
  \includegraphics[width=0.315\linewidth]{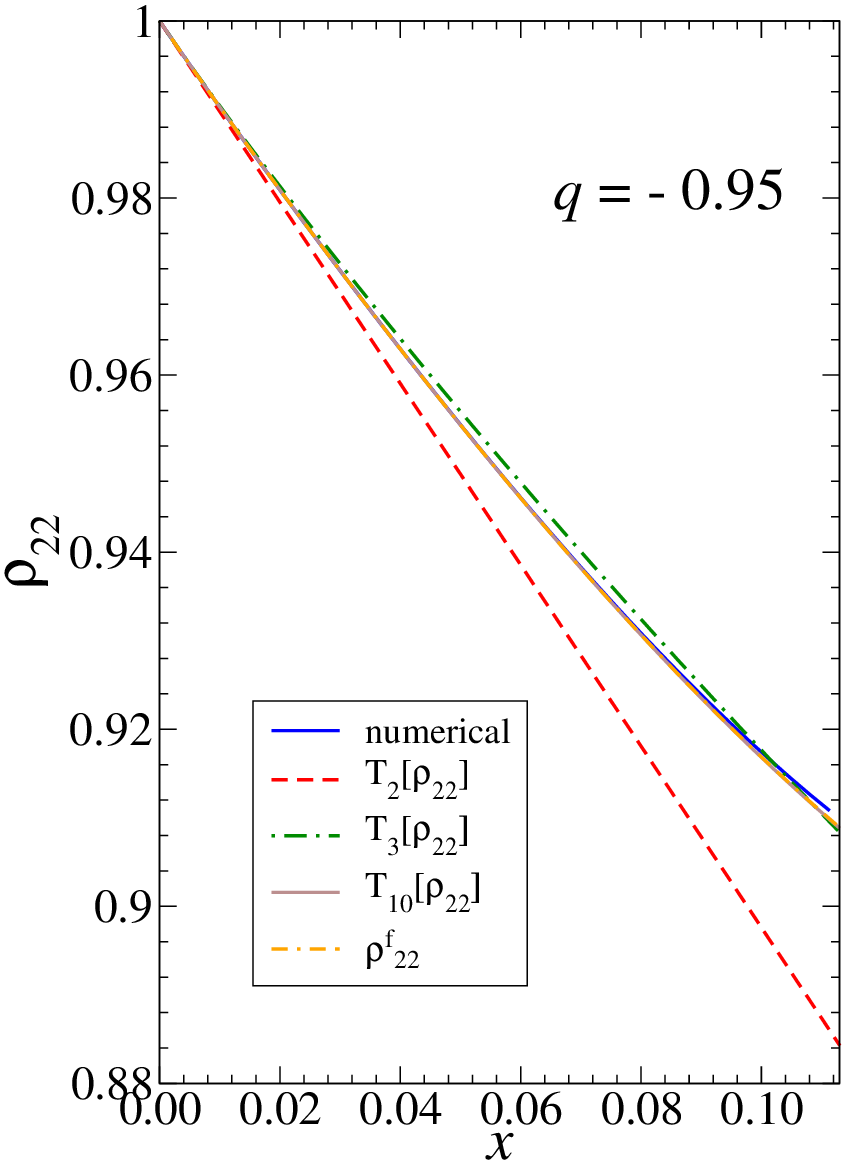} 
\end{tabular}
\end{center}
\caption{Numerical and analytical $\rho_{22}$'s as functions of
  $x=v^2$. The three panels are for spin values $q=0.95, 0$ and
  $-0.95$. The notation of the analytical $\rho_{22}$ models follows
  the definition in Sec.~\ref{sec:comparison}. The $T_{10}[\rho_{22}]$
  and $\rho^f_{22}$ lines overlap with each other and in the $q=0$
  case they also overlap with the numerical $\rho_{22}$.}
\label{fig:rho22}
\end{figure*}
\begin{figure*}
\begin{center}
\begin{tabular}{ccc}
  \includegraphics[width=0.316\linewidth]{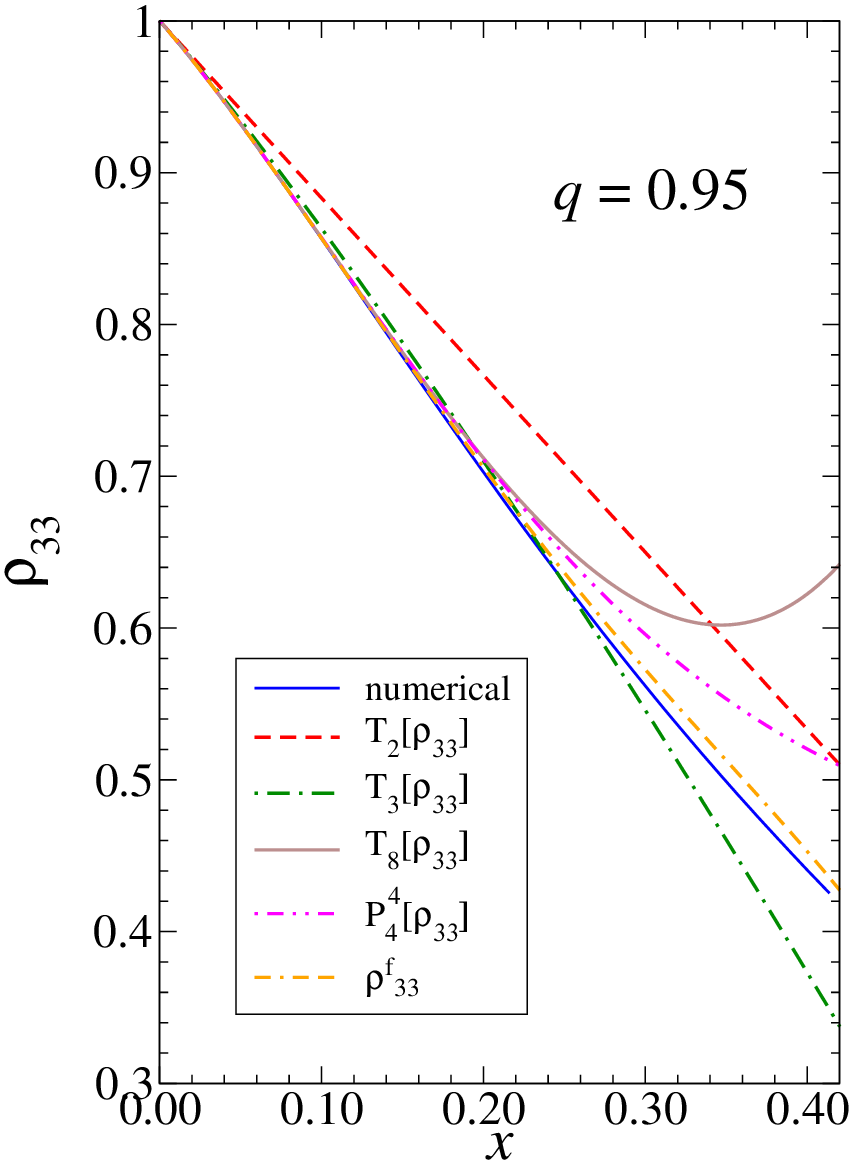} \quad
  \includegraphics[width=0.314\linewidth]{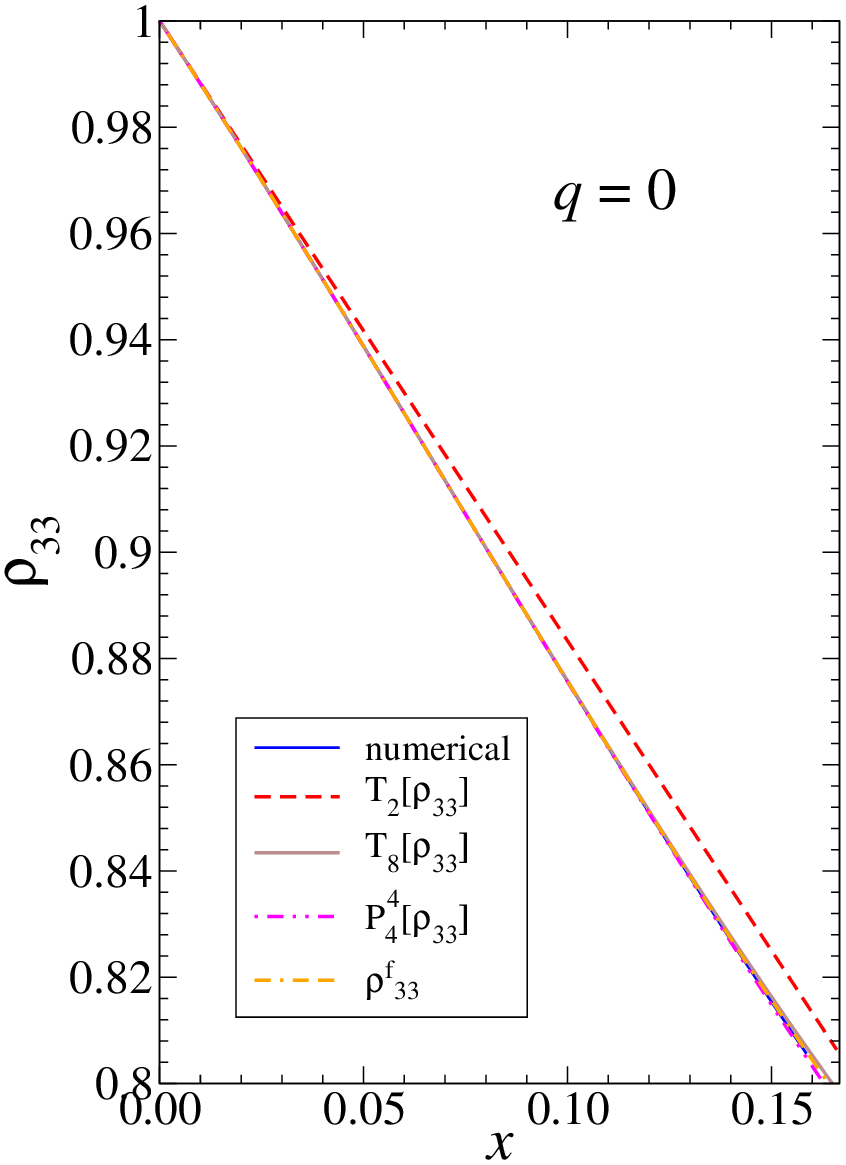} \quad
  \includegraphics[width=0.315\linewidth]{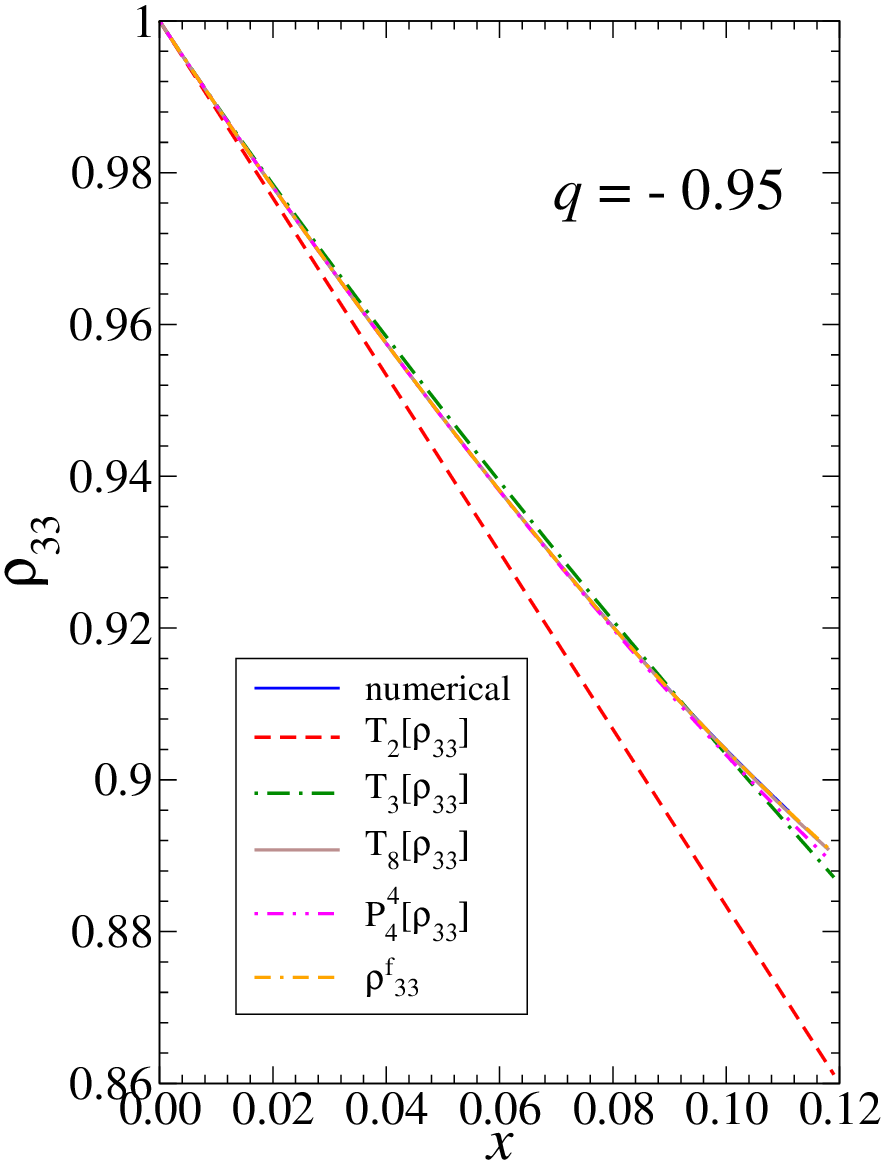} 
\end{tabular}
\end{center}
\caption{Numerical and analytical $\rho_{33}$'s as functions of
  $x=v^2$. The three panels are for spin values $q=0.95, 0$ and
  $-0.95$. The notation of the analytical $\rho_{33}$ models follows
  the definition in Sec.~\ref{sec:comparison}. In the antialigned
  $q=-0.95$ case, the numerical, $T_8[\rho_{33}]$ and $\rho^f_{33}$
  lines overlap, while in the nonspinning $q=0$ case, they also
  overlap with $P^4_4[\rho_{33}]$.}
\label{fig:rho33}
\end{figure*}
\begin{widetext}
\begin{subequations}\label{rholmtplfac}
\begin{eqnarray}
\rho^f_{22}&=& \left ( 1-\frac{43}{42}\,v^2 -\frac{2\,q}{3}\,v^3 \right)\,\left[ 1+ \left(\frac{q^2}{2}-\frac{20\,555}{10\,584}\right)\,v^4 -\frac{34\,q}{21}\,v^5 + \left(-\frac{428\,\text{eulerlog}_2(v^2)}{105}+\frac{109
  \,q^2}{126}+\frac{656\,928\,119}{61\,122\,600}\right)\,v^6 \right. \nonumber \\
&& + \left(\frac{2\,q^3}{3}-\frac{14\,069\,q}{7\,938}\right)\,v^7 + \left(-\frac{q^4}{8}+\frac{4\,751
  \,q^2}{7\,056}+\frac{6\,877\,264\,829\,389}{800\,950\,550\,400}\right)\,v^8 + \left(-\frac{856\,\text{eulerlog}_2(v^2)\,q}{315}+\frac{34\,q^3}{27}\right.\nonumber\\
&& \left.\left. +\frac{245\,281\,097
  \,q}{45\,841\,950}\right)\,v^9
 + \left(\frac{439\,877\,\text{eulerlog}_2(v^2)}{55\,566}+\frac{319\,q^4}{1\,008}-\frac{1\,312\,819
  \,q^2}{2\,667\,168}-\frac{179\,558\,258\,690\,231}{8\,409\,980\,779\,200}\right)\,v^{10} \right] \,,\\
\rho^{f\,L}_{21}&=& \left [ 1-\frac{3\,q}{4}\,v + \left(-\frac{9\,q^2}{32}-\frac{59}{56}\right)\,v^2 + \left(\frac{1\,177\,q}{672}-\frac{27\,q^3}{128}\right)\,v^3 \right ]\,\left[ 1+ \left(-\frac{405\,q^4}{2\,048}-\frac{865\,q^2}{1\,792}-\frac{47\,009}{56\,448}\right)\,v^4 \right. \nonumber\\
&&+ \left(-\frac{729\,q^5}{2048}-\frac{141\,q^3}{1\,792}-\frac{12\,137
  \,q}{6\,272}\right)\,v^5 + \left(-\frac{107\,\text{eulerlog}_1(v^2)}{105}-\frac{18\,225\,q^6}{32\,768}-\frac{9\,477\,q^4}{57\,344}+\frac{2\,534\,545
  \,q^2}{903\,168}\right. \nonumber\\
&&  \left.+\frac{2\,662\,510\,933}{1\,303\,948\,800}\right)\,v^6
\left. + \left(-\frac{54\,675\,q^7}{65\,536}+\frac{837\,q^5}{114\,688}-\frac{734\,519
  \,q^3}{602\,112}-\frac{1\,240\,566\,577\,q}{521\,579\,520}\right)\,v^7 + \left(-\frac{321\,\text{eulerlog}_1(v^2)\,q^2}{1\,120} \right.\right.\nonumber\\
&&\left. -\frac{898\,857\,q^8}{1\,048\,576}-\frac{4\,617
  \,q^6}{229\,376} -\frac{915\,459\,q^4}{1\,605\,632}+\frac{139\,532\,257\,q^2}{27\,165\,600}+\frac{1\,799\,642\,241\,599}{2\,071\,144\,857\,600}\right)\,v^8 \nonumber\\
&& + \left(-\frac{963
  \,\text{eulerlog}_1(v^2)\,q^3}{2\,240}+\frac{125\,939 \text{eulerlog}_1(v^2)\,q}{70\,560}-\frac{1\,043\,199\,q^9}{1\,048\,576} +\frac{12\,393
  \,q^7}{262\,144}+\frac{380\,169\,q^5}{3\,211\,264}\right.\nonumber\\
&& \left. -\frac{107\,920\,920\,827\,q^3}{41\,726\,361\,600}-\frac{494\,887\,939\,808\,057\,q}{91\,130\,373\,734\,400}\right)\,v^9\nonumber\\
&&\left. + \left(-\frac{2\,889
  \,\text{eulerlog}_1(v^2)\,q^4}{7\,168}+\frac{195\,061 \text{eulerlog}_1(v^2)\,q^2}{188\,160}+\frac{5\,029\,963
  \,\text{eulerlog}_1(v^2)}{5\,927\,040}-\frac{3\,9031\,389\,q^{10}}{33\,554\,432}\right.\right. \nonumber\\
&& \left.\left.+\frac{34\,610\,733\,q^8}{58\,720\,256} -\frac{18\,644\,823
  \,q^6}{51\,380\,224}+\frac{7\,997\,241\,271\,q^4}{14\,836\,039\,680}+\frac{83\,8234\,689\,365\,819
  \,q^2}{145\,808\,597\,975\,040}-\frac{1\,133\,240\,747\,153}{386\,613\,706\,752}\right)\,v^{10}  \right] \,,
\end{eqnarray}
\begin{eqnarray}
\rho^f_{33}&=& \left ( 1 -\frac{7}{6}\,v^2 -\frac{2\,q}{3}\,v^3 \right )\,\left[ 1+ \left(\frac{q^2}{2}-\frac{6\,719}{3\,960}\right)\,v^4 -\frac{4\,q}{3}\,v^5 + \left(-\frac{26\,\text{eulerlog}_3(v^2)}{7}+\frac{13
  \,q^2}{18}+\frac{688\,425\,313}{56\,756\,700}\right)\,v^6 \right. \nonumber\\
&&\left. + \left(\frac{2\,q^3}{3}-\frac{1\,073\,q}{1\,188}\right)\,v^7 + \left(\frac{7\,066\,253\,659}{951\,350\,400}-\frac{5\,q^2}{108}\right)\,v^8 \right] \,,\\
\rho^{f\,L}_{32}&=& \left( 1-\frac{164}{135}\,v^2 + \frac{2\,q}{9}\,v^3 + \left(\frac{q^2}{3}-\frac{180\,566}{200\,475}\right)\,v^4 \right)\,\left[ 1 -\frac{2\,788\,q}{1\,215}\,v^5 + \left(-\frac{104\,\text{eulerlog}_2(v^2)}{63}+\frac{488
  \,q^2}{405}\right.\right.\nonumber\\
&&\left.\left.  +\frac{5\,849\,948\,554}{940\,355\,325}\right)\,v^6 
 -\frac{457\,232\,q}{164\,025}\,v^7 + \left(\frac{107\,912
  \,q^2}{54\,675}+\frac{3\,002\,382\,469\,466}{731\,462\,106\,375}\right)\,v^8 \right] \,,\\
\rho^f_{31}&=& \left( 1-\frac{13}{18}\,v^2 -\frac{2\,q}{3}\,v^3 \right)\,\left[ 1+ \left(\frac{101}{7\,128}-\frac{5\,q^2}{6}\right)\,v^4 +\frac{4\,q}{9}\,v^5 + \left(\frac{2\,942\,362\,219}{1\,532\,430\,900}-\frac{26\,\text{eulerlog}_1(v^2)}{63}-\frac{19
  \,q^2}{18}\right)\,v^6 \right. \nonumber\\
&&\left. - \left(\frac{4\,q^3}{9}+\frac{1\,625\,q}{10\,692}\right)\,v^7 + \left(\frac{16\,469\,528\,659}{8\,562\,153\,600}-\frac{151\,q^2}{324}\right)\,v^8 \right] \,,
\end{eqnarray}
\begin{eqnarray}
\rho^f_{44}&=& \left( 1-\frac{269}{220}\,v^2 \right)\,\left[ 1-\frac{2\,q}{3}\,v^3+ \left(\frac{q^2}{2}-\frac{14\,210\,377}{8\,808\,800}\right)\,v^4 -\frac{683\,q}{330}\,v^5 - \left(\frac{12\,568
  \,\text{eulerlog}_4(v^2)}{3\,465}-\frac{1\,319\,q^2}{1\,980}\right.\right.\nonumber\\
&&\left.\left.-\frac{7\,216\,765\,000\,811}{549\,404\,856\,000}\right)\,v^6 \right] \,,\\ 
\rho^{f \,L}_{43}&=& 1-\frac{111}{88} v^2+\left(\frac{3\,q^2}{8}-\frac{6\,894\,273}{7\,047\,040}\right) v^4-\frac{12\,113\,q}{6\,160}\,v^5+ \left(\frac{1\,664\,224\,207\,351}{195\,343\,948\,800}-\frac{1\,571\,\text{eulerlog}_3(v^2)}{770}\right)\,v^6 \,, \nonumber \\ \\
\rho^{f\,L}_{42}&=& \left ( 1-\frac{191}{220}\,v^2 -\frac{2\,q}{3}\,v^3 \right )\,\left[ 1+ \left(\frac{q^2}{2}-\frac{3\,190\,529}{8\,808\,800}\right)\,v^4 -\frac{7\,q}{110}\,v^5 + \left(-\frac{3\,142\,\text{eulerlog}_2(v^2)}{3\,465}+\frac{2\,021
  \,q^2}{1\,980}\right.\right.\nonumber\\
&&\left.\left.  +\frac{1\,947\,834\,451\,721}{549\,404\,856\,000}\right)\,v^6 \right] \,,\\
\rho^{f \,L}_{41}&=& \left ( 1-\frac{301}{264}\,v^2 + \left(\frac{3\,q^2}{8}-\frac{7\,775\,491}{21\,141\,120}\right)\,v^4 \right)\,\left[ 1+ \left(-\frac{5\,q^3}{6}-\frac{20\,033\,q}{55\,440}\right)\,v^5 + \left(\frac{1\,227\,423\,222\,031}{1\,758\,095\,539\,200}\right.\right.\nonumber\\
&&\left.\left.-\frac{1\,571
  \,\text{eulerlog}_1(v^2)}{6\,930}\right)\,v^6 \right] \,.
\end{eqnarray}
\end{subequations}
\end{widetext}
\begin{figure*}
\begin{center}
\begin{tabular}{cc}
  \includegraphics[width=0.45\linewidth]{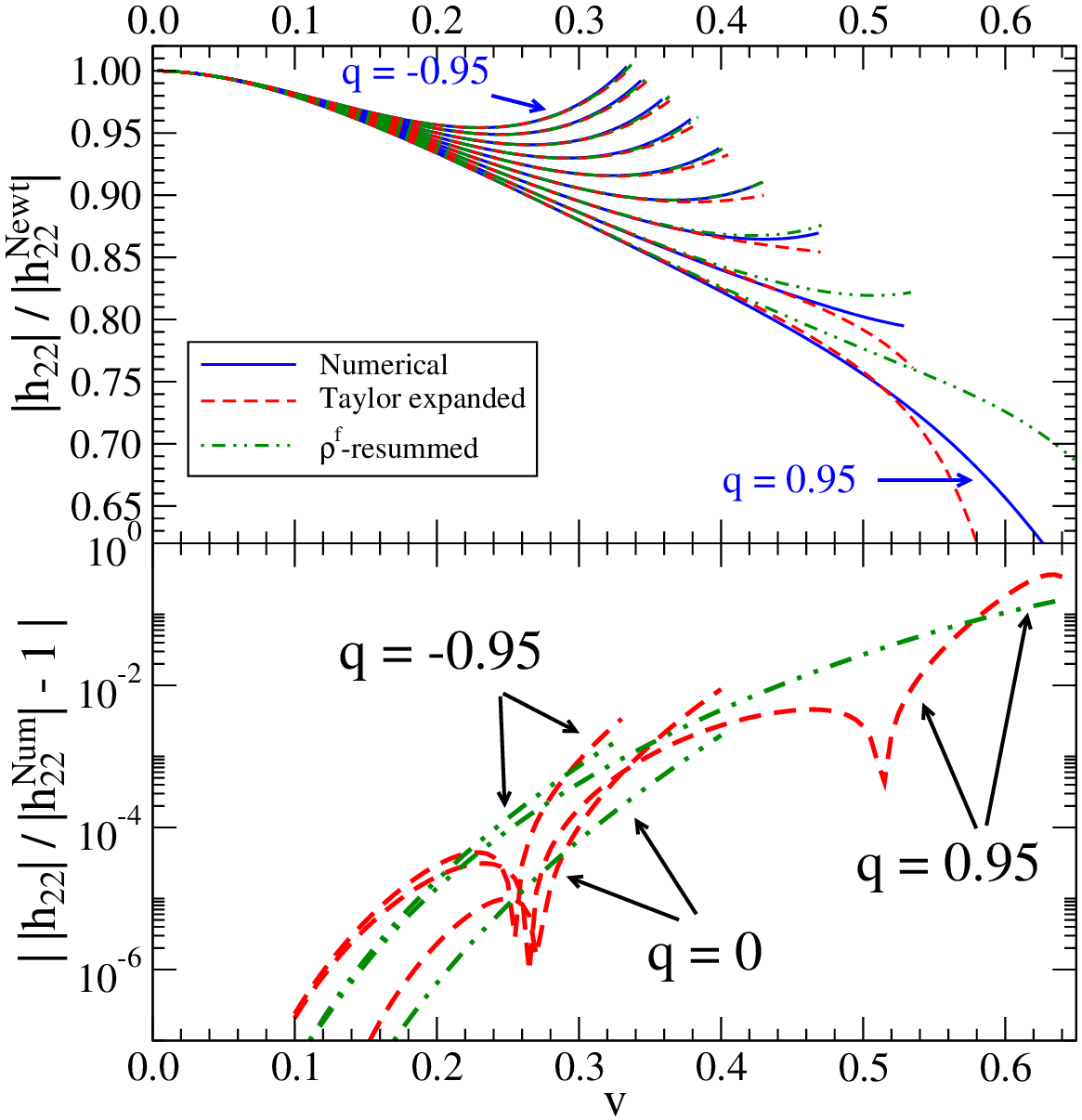} \qquad \qquad
  \includegraphics[width=0.45\linewidth]{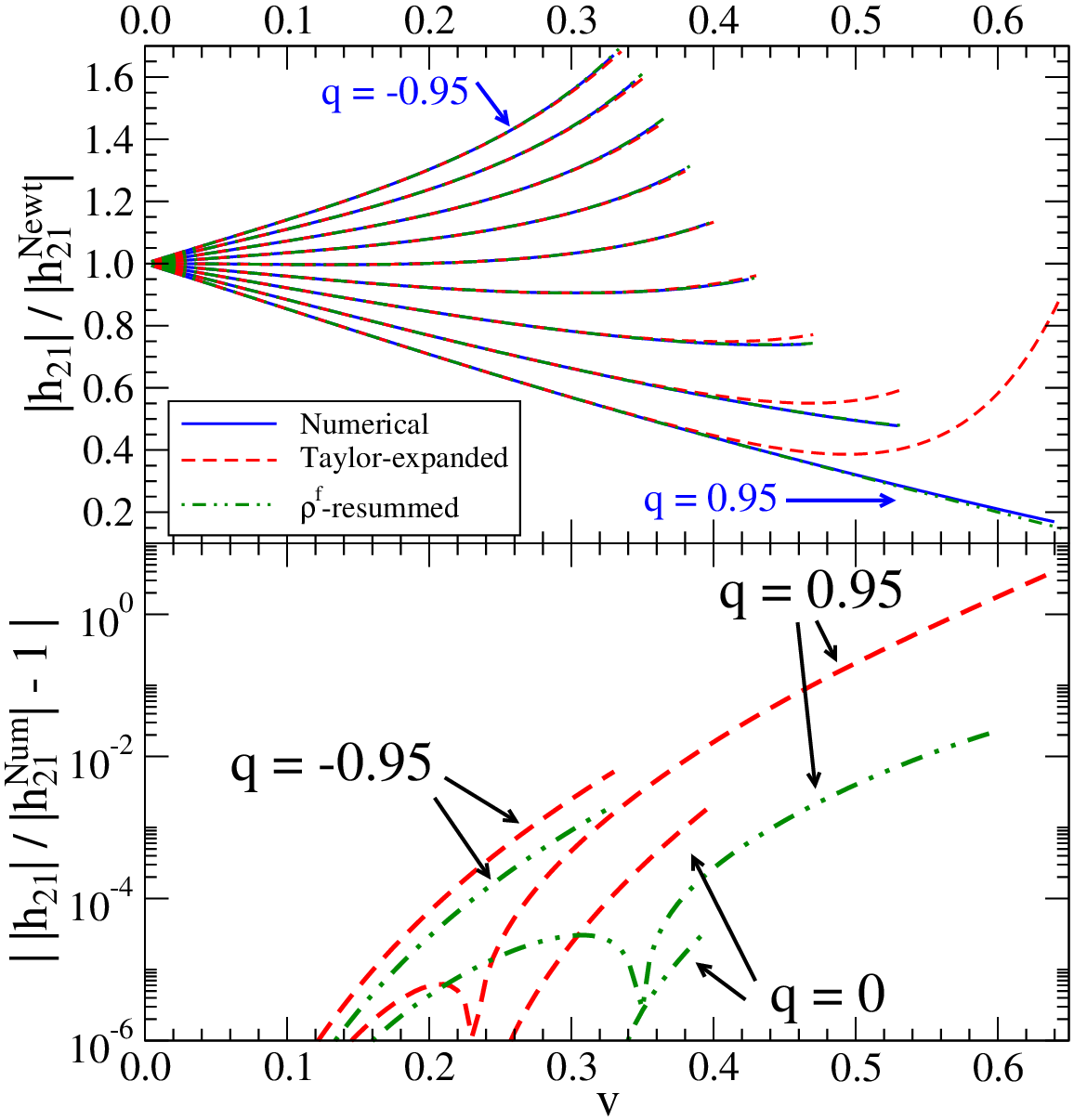} 
\end{tabular}
\end{center}
    \caption{Upper panel: Comparison between the numerical and analytical Newtonian normalized $|h_{22}|$ and $|h_{21}|$ modes for a
    test-particle orbiting around a Kerr black hole in the equatorial
    plane. For the numerical data and analytical models (Taylor-expanded and $\rho^f$-resummed), we have nine curves 
    corresponding to different spin values of the Kerr black hole. From top to bottom, 
    the spins are $q =
    -0.95, -0.75, -0.5, -0.25, 0, 0.25, 0.5, 0.75$ and
    $0.95$. Lower panel: relative fractional difference between analytical and numerical $|h_{\ell m}|$ for the 
representative spin values $q = -0.95, 0, 0.95$. }
\label{fig:waveform2}
\end{figure*}
\begin{figure*}
\begin{center}
\begin{tabular}{cc}
  \includegraphics[width=0.45\linewidth]{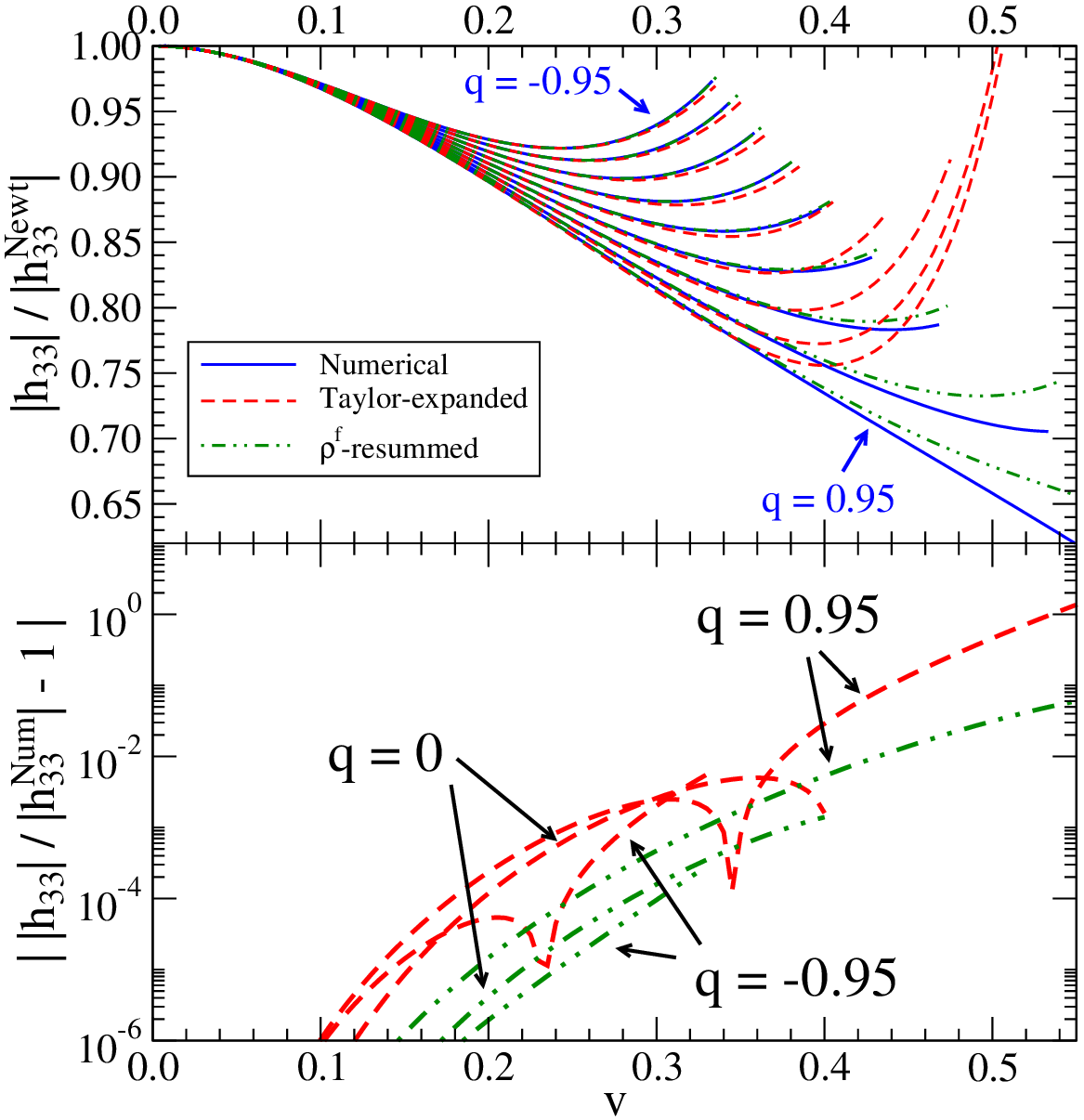} \qquad \qquad
  \includegraphics[width=0.45\linewidth]{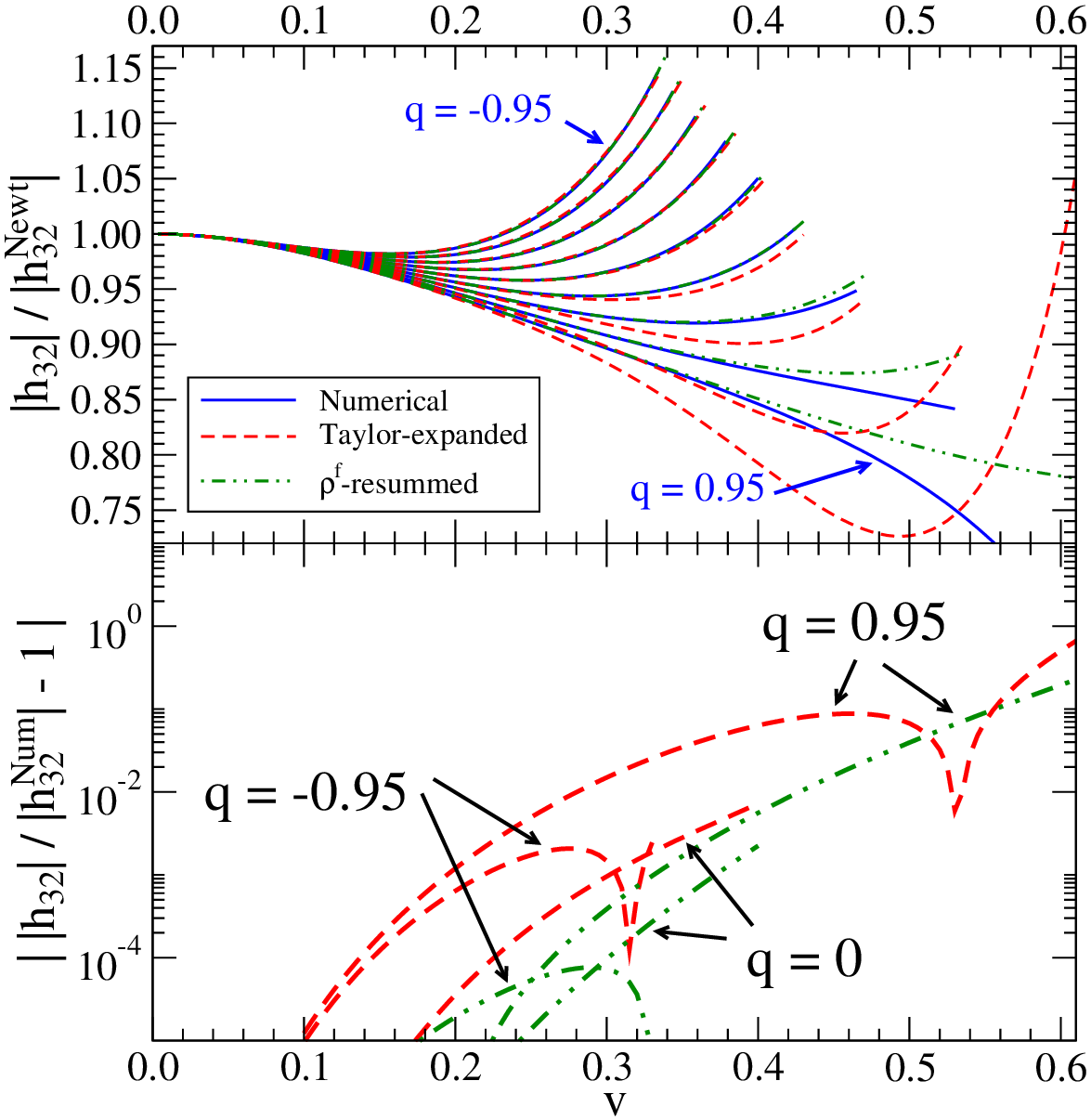} 
\end{tabular}
\end{center}
  \caption{Upper panel: Comparison between the numerical and analytical Newtonian normalized $|h_{33}|$ and $|h_{32}|$ modes for a
    test-particle orbiting around a Kerr black hole in the equatorial
    plane. For the numerical data and analytical models (Taylor-expanded and $\rho^f$-resummed), we have nine curves 
    corresponding to different spin values of the Kerr black hole. From top to bottom, 
    the spins are $q =
    -0.95, -0.75, -0.5, -0.25, 0, 0.25, 0.5, 0.75$ and
    $0.95$. Lower panel: relative fractional difference between analytical and numerical $|h_{\ell m}|$ for the 
representative spin values $q = -0.95, 0, 0.95$.}
\label{fig:waveform3}
\end{figure*}
We notice that for a few modes, it is convenient to factor out even the
2PN order term. The procedure of factoring out zeros of $\rho_{\ell
  m}$ can be improved in the future by introducing appropriate
adjustable parameters and calibrate them to the numerical result.

In Figs.~\ref{fig:rho22} and \ref{fig:rho33} we also show results when adopting 
the Pad\'e summation. We find that the diagonal and nearest-diagonal Pad\'e-summation   
improve the agreement with the numerical results not only for the $(3,3)$ mode, 
but also for the $(3,1)$ and $(4,2)$ modes. An even better agreement for several modes  
is obtained when using the farthest-diagonal Pad\'e-summation. However, this 
quite interesting result suffers by the presence of spurious poles appearing 
for several $q$ values, and for this reason we will no longer discuss the Pad\'e-summation 
in this paper. 

Finally, we observe that close to the LSO the even-parity modes $\rho^L$ agree slightly 
better to the numerical results than $\rho^H$'s. Thus, we adopt 
in this paper the multipolar waveforms built with the $\rho^L$.
\begin{figure}
  \includegraphics[width=0.9\linewidth]{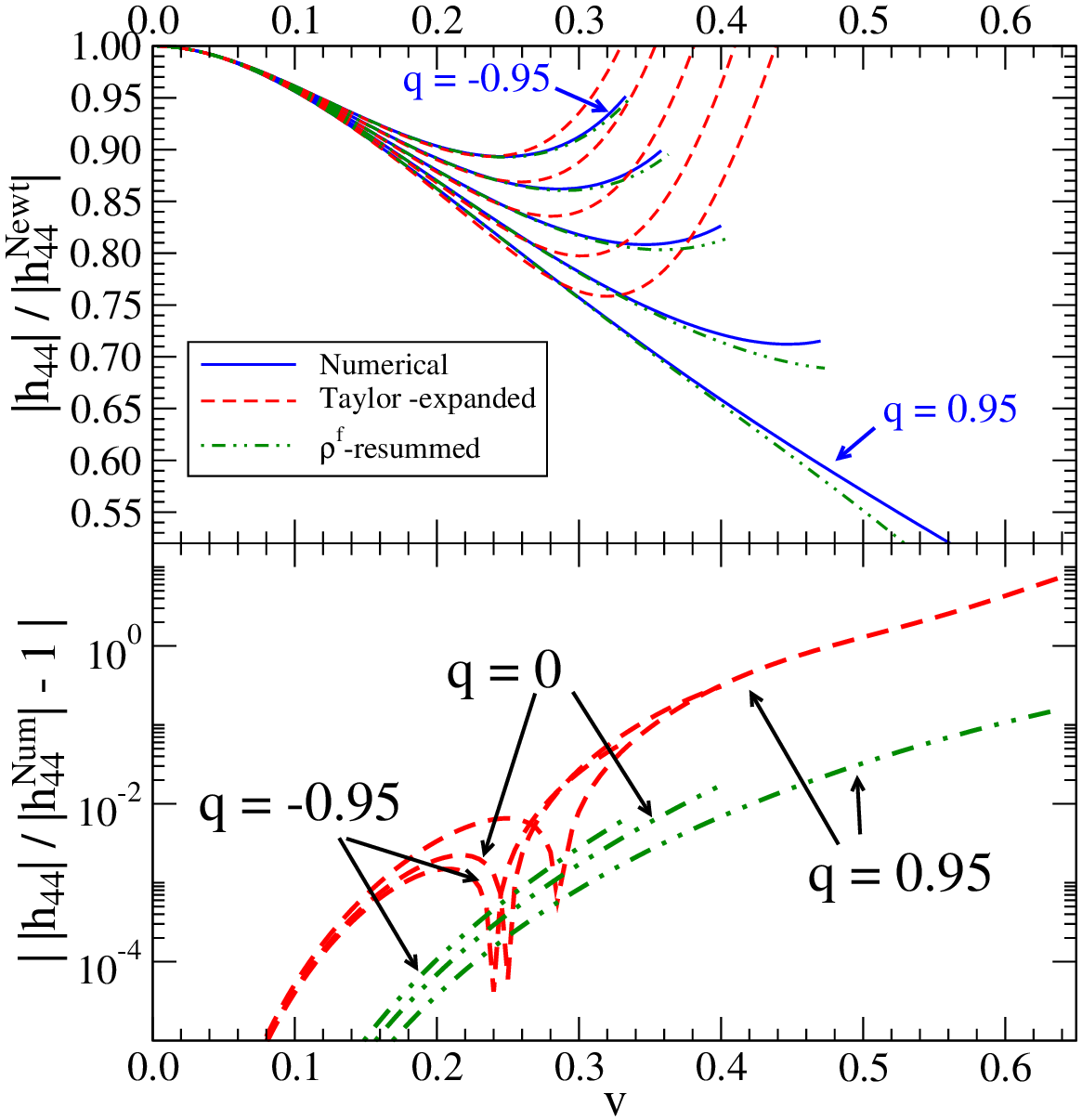} 
\caption{Upper panel: Comparison between the numerical and analytical 
Newtonian normalized $|h_{44}|$ mode for a
    test-particle orbiting around a Kerr black hole in the equatorial
    plane. For the numerical data and analytical models (Taylor-expanded and $\rho^f$-resummed), 
    we have four curves corresponding to different spin values of the Kerr black hole. From top to
    bottom, the spins are $q = -0.95, -0.5, 0, 0.5$ and
    $0.95$. Lower panel: relative fractional difference between analytical and numerical $|h_{\ell m}|$ for the representative spin values $q = -0.95, 0$ and $0.95$.} 
\label{fig:waveform4}
\end{figure}
In Figs.~\ref{fig:waveform2}, \ref{fig:waveform3} and
\ref{fig:waveform4} we compare the Taylor-expanded, 
$\rho^f$-resummed and numerical Newtonian-normalized multipolar
amplitudes for the dominant modes. In general, the $\rho^f$ and
$\rho$-resummed amplitudes agree better with the numerical amplitudes
than Taylor-expanded amplitudes do, especially for higher-order
modes. More specifically, we find that $\rho$-resummed amplitudes
  (not shown in Figs.~\ref{fig:waveform2}, \ref{fig:waveform3} and
  \ref{fig:waveform4}) differ from the numerical ones by
  $\laq 0.6\%$ up to $v \leq 0.4$ for the (2,2), (2,1) and (3,2) modes
  and by $\laq 1.8\%$ for the (3,3) and (4,4) modes.  Their fractional
  difference grows up to $\sim 1$--$10$ at the LSO when $q = 0.95$. 

When applying the $\rho^f$-resummation, we find that the fractional amplitude 
difference between the numerical and analytical (2,2) amplitude at the LSO 
is $16\%\,(33\%)$, $0.18\%\,(0.32\%)$ and $0.20\%\,(0.85\%)$ 
for $q = 0.95, 0, -0.95$, respectively. We indicated in parenthesis 
the numbers when Taylor-expanded amplitudes are employed. 
For the (2,1), (3,3) and (4,4) modes, for which fewer spin PN terms are known (see 
Table~\ref{tab:PNorder}), the improvement due to the $\rho^f$-resummation 
is more striking. In fact, for the (2,1), (3,3) and (4,4) modes we obtain a fractional 
amplitude difference of $2.4\%\,(4.2)$, $0.2\%\,(0.58\%)$ and 
$0.0036\%\,(0.15\%)$, $7.5\%\,(2)$, $0.027\%\,(0.55\%)$ and $0.13\%\,(0.2\%)$, 
$16\%\,(7.5)$, $1.7\%\,(28\%)$ and $0.6\%\,(5.8\%)$, for $q = 0.95, 0, -0.95$, 
respectively.

We summarize the results of Figs.~\ref{fig:waveform2},
  \ref{fig:waveform3} and \ref{fig:waveform4} as follows. 
  First, we remark that the Taylor-expanded amplitudes agree with
  the numerical ones quite well for the (2,2) mode where the PN expansion
  is known through the highest order today (5.5 PN for nonspinning terms and
  4PN for spin terms). Thus, for the (2,2) mode the improvement due  
  to the resummation technique is marginal. We expect that a similar 
  result holds for higher modes when sufficient PN terms are known. 
  Second, the factorized resummed waveforms 
  consistently improve the amplitude agreement with 
  numerical waveforms for several values of $q$ and large spanning of $v$. 
  In the lower panels of Figs.~\ref{fig:waveform2}, \ref{fig:waveform3} and
  \ref{fig:waveform4}, we observe that the fractional amplitude
  difference between the numerical and $\rho^f$-resummed
  waveforms is always smaller than the difference between the
  numerical and Taylor-expanded waveforms, except around the 
  $v$ values where the numerical and Taylor-expanded
  amplitudes coincide. For all modes [except the (2,2) mode] and all spin values shown in
  the figures, we find that $\rho^f$-resummed amplitudes are typically closer to the numerical amplitudes than 
  Taylor-expanded are by an order of magnitude or more.

Finally, for $\ell\ge 5$ modes, the $\rho^f$-resummation is not very successful in modeling 
the numerical amplitudes, but it is better than Taylor-expanded amplitudes. 
We know nonspinning and spin corrections only through 
2.5PN order in the $(5,5)$ mode (see Table~\ref{tab:PNorder}), thus it is not surprising that we cannot model those modes very well. Since the contribution of the $\ell\ge 5$ modes to the radiation power and strain amplitude is not negligible, it would be very useful to calculate higher order corrections in those modes in the future.

\begin{figure*}
\begin{center}
\begin{tabular}{cc}
  \includegraphics[width=0.45\linewidth]{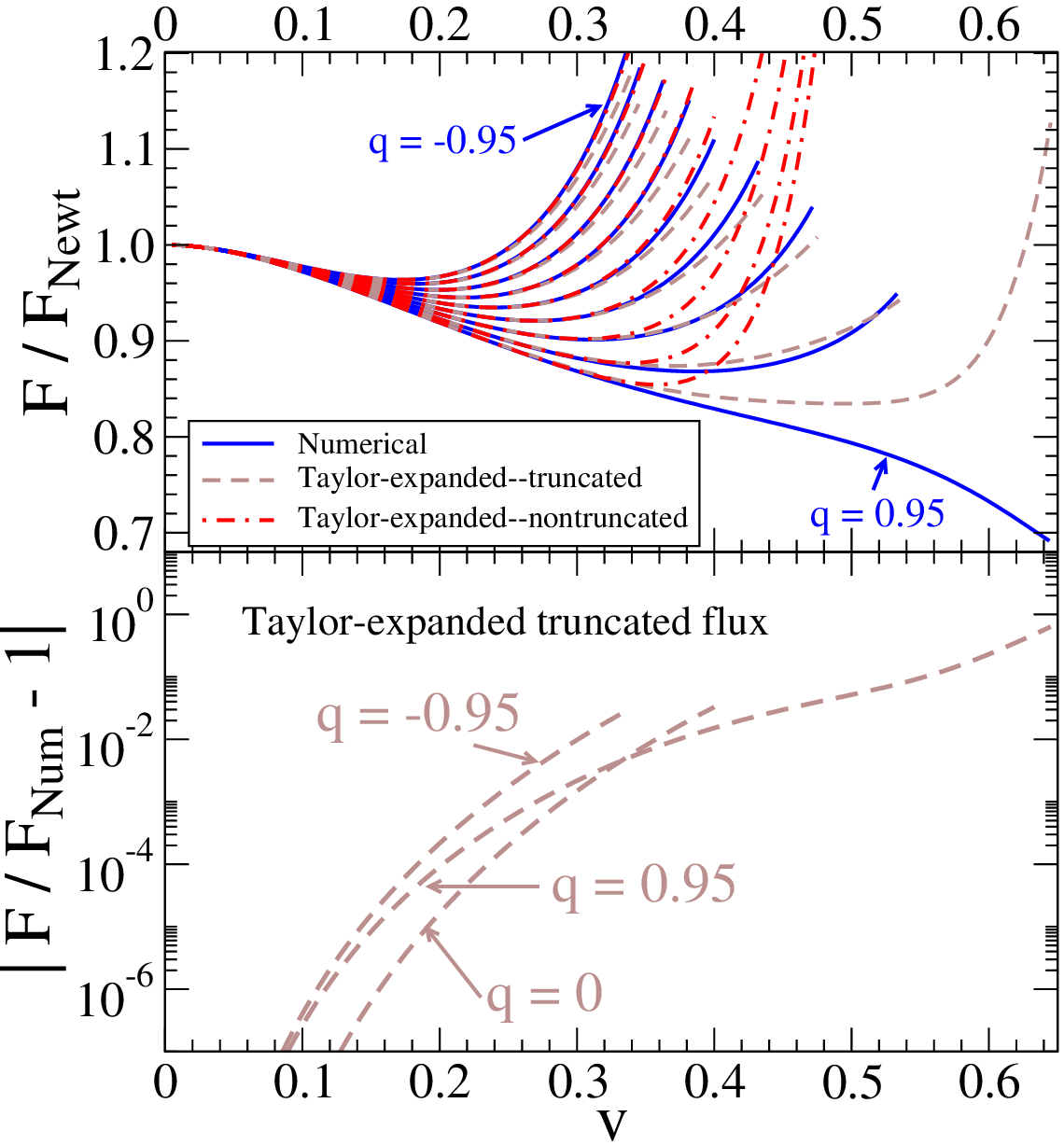} \qquad
  \includegraphics[width=0.45\linewidth]{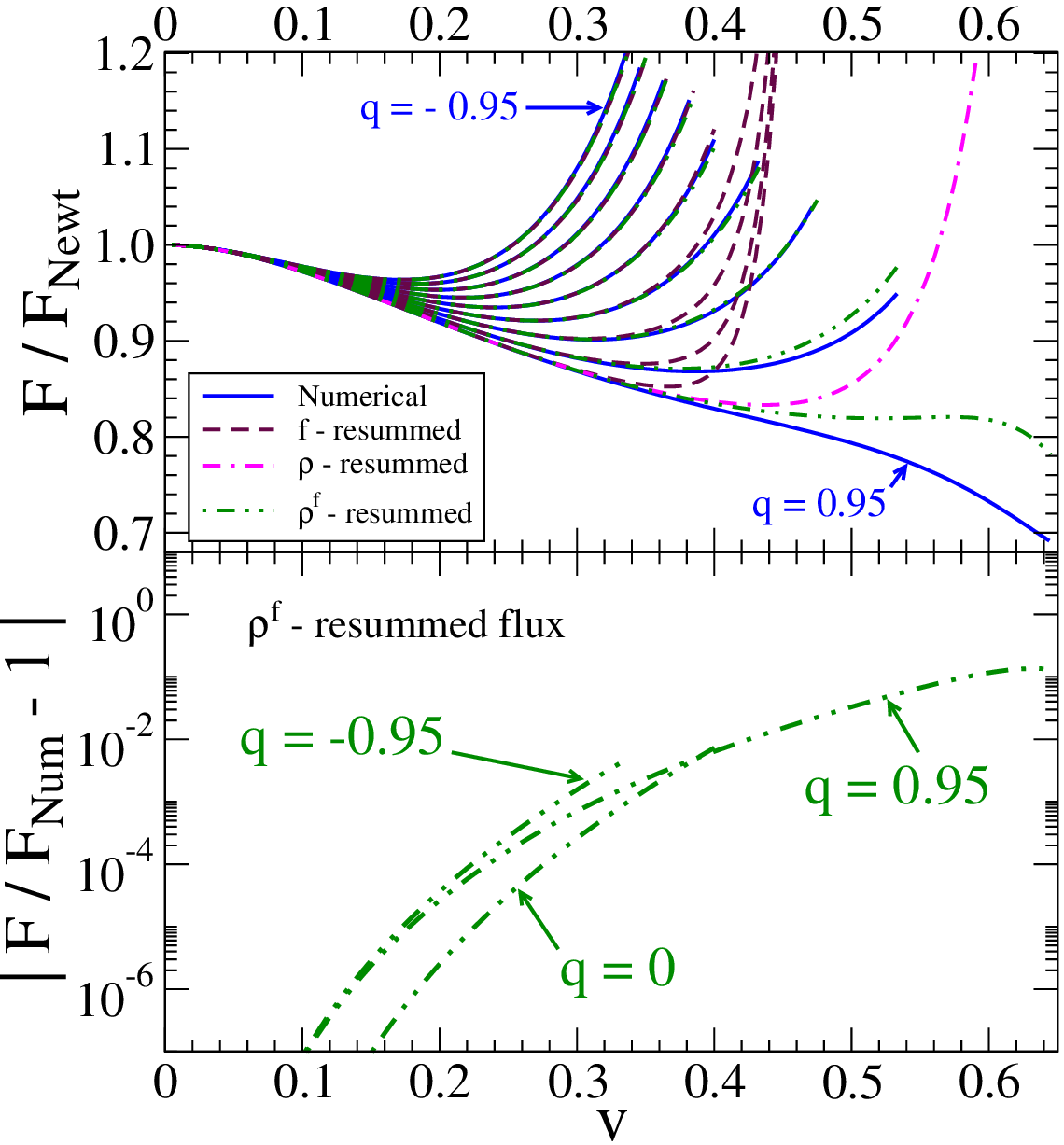}
\end{tabular}
\end{center}
\caption{Upper panels: Comparison between the numerical and analytical
  Newtonian normalized energy flux for a test-particle orbiting a Kerr
  black hole in the equatorial plane. There are nine curves for each
  waveform's model and numerical data. They correspond to different
  spins of the Kerr black hole. From top to bottom, the spins are $q =
  -0.95, -0.75, -0.5, -0.25, 0, 0.25, 0.5, 0.75,$ and $0.95$. The
  Taylor-expanded--truncated flux includes test-particle spin terms
  through 4PN order \cite{TSTS1996}, and nonspinning terms through
  5.5PN order \cite{TTS55PN}. The Taylor-expanded--nontruncated flux
  includes higher-order PN terms originated by the new PN terms in the
  $h_{\ell m}$'s computed by Tagoshi and
  Fujita~\cite{TagoshiFujita2010} (see
  Table~\ref{tab:PNorder}). The $\rho$-resummed flux is plotted
    for one spin value $q=0.95$ in the upper right panel, to show the
    large improvement when using $\rho^f$-resummed instead of $\rho$-resummed
    flux. Lower panels: fractional difference
  between the numerical and analytical energy fluxes for the
  representative spin cases: $q = -0.95, 0, 0.95$.}
\label{fig:flux}
\end{figure*}

\subsection{Comparison between analytical and numerical energy fluxes}
\label{sec:energy}

Here we compare numerical and analytical Newtonian-normalized energy fluxes for a test-particle orbiting 
a Kerr black hole in the equatorial plane. The fluxes are computed by 
summing the power radiated using Eq.~(\ref{energyflux}) and setting 
$\ell = 8$. For a test-particle moving along a quasi-circular equatorial orbit, 
the Newtonian-normalized flux is $F(v)/F_{\rm Newt}(v)$ 
where $F_{\rm Newt}(v) = 32\nu^2\,v^{10}/5$.

We note that the dominant error source of the numerical calculation
of the total flux is the truncation at $\ell=8$ of the mode summation. 
Let $F_{\ell=8}(v)$ be the contribution from $\ell=8$ mode for $F(v)$. 
The fraction, $F_{\ell=8}(v)/F(v)$, is about $10^{-10}$ around 
$v=0.1$ and $10^{-5}$ to $10^{-2}$ around the LSO. 

In Fig.~\ref{fig:flux}, we compare numerical and analytical 
Newtonian-normalized energy fluxes for different spin values 
of the Kerr black hole. In the left panel of Fig.~\ref{fig:flux} 
we consider two Taylor-expanded fluxes computed from 
the Taylor-expanded $h_{\ell m}$'s: one that truncates all terms
beyond 5.5PN order and spin terms beyond 4PN order (Taylor-expanded--truncated) 
and one that keeps all higher order terms (Taylor-expanded--nontruncated). [The former 
is the Taylor-expanded flux that consistently includes nonspinning 
effects through 5.5PN order and spin effects through 4PN order~\cite{TTS55PN,TSTS1996}; the latter 
includes new higher-order PN terms computed by Tagoshi and Fujita~\cite{TagoshiFujita2010}.] 

In the left panel of Fig.~\ref{fig:flux}, we do not show the
Taylor-expanded flux truncated at 4PN order~\cite{TSTS1996} since its
agreement with the numerical flux is rather poor. Figures~2 and 3 of
Ref.~\cite{Porter:2004eb} show that in this case the Taylor-expanded
flux starts to differ from the numerical one at a relatively low
velocity of $v=0.2$ for all spin values. By contrast, the agreement is
substantially improved when we include the 5.5PN order nonspinning
terms in the Taylor-expanded--truncated flux. The
Taylor-expanded--nontruncated flux agrees better with the numerical
flux than the Taylor-expanded--truncated flux for retrograde orbits
with $q<0$, while its agreement is worse for prograde orbits with
$q>0$. For spin values $q>0.5$, the agreement is especially bad, as
the Taylor-nontruncated flux grows too fast when $v>0.4$. We find that
this difference is mainly due to the large new spin
term~\cite{TagoshiFujita2010} in the $(3,3)$ mode, i.e. $\left(-q^2+ 
{9\,\pi\,q^2}/{2} + q\,{89}/{5}\right)\,v^7$ in
$\hat{Z}_{33\omega_0}$ (real part only).  Without any resummation, 
the Taylor-expanded--truncated flux agrees well with the numerical flux 
for all spin values except for $q=0.95$. 
The lower left panel shows that the fractional differences between the 
numerical and the Taylor-expanded--truncated fluxes are below $1\%$ 
until $v=0.3$, and are below $10\%$ for $q=0.95$ until $v=0.55$, 
and below $10\%$ for all other spin values until the LSO.

  In the right panel of Fig.~\ref{fig:flux} we consider three
  analytical flux models which use the $f_{\ell m}$, $\rho_{\ell m}$
  (for $q=0.95$ only) and $\rho_{\ell m}^f$, respectively.  The
  fractional difference between the numerical flux and $f$, $\rho$ or
  $\rho^f$-resummed fluxes is $<0.3\%$ for all spin values when
  $v<0.3$. Larger differences appear only when $v>0.3$ for large and
  aligned spins, and the $f$-resummed flux performs especially bad when
  $v>0.4$. In the case of $q=0.95$, we show the significant
  improvements achieved from the $f$-resummed to the $\rho$-resummed and
  eventually to the $\rho^f$-resummed flux. The fractional difference
  with numerical flux at the LSO is reduced from $\sim 3.5\times 10^4$ to $\sim
  3$ to $13\%$. The main reason for the bad performance of the
  $f$-resummed flux is caused by the new spin
  term~\cite{TagoshiFujita2010} in the (3,3) mode, i.e. $\left(-q^2+ {9\,\pi\,q^2}/{2} +
    q\,{89}/{5}\right)\,v^7$ in $\hat{Z}_{33\omega_0}$ (real part
  only), as is in the case of the Taylor-expanded--nontruncated flux.
As a matter of fact, we notice that if we did not include this new term
  computed in Ref.~\cite{TagoshiFujita2010}, and applied the
  $f$-resummation, or the $\rho$-resummation only to the
  nonspinning terms~\cite{EOBNRS}~\footnote{The new higher-order PN
    terms computed in Ref.~\cite{TagoshiFujita2010} were not available
    at the time Ref.~\cite{EOBNRS} appeared.}, we would find a flux
  not very different from the $\rho$-resummed flux in the right panel
  of Fig.~\ref{fig:flux}.  In the $\rho$ or $\rho^f$-resummation, this
  new term is suppressed by an order of magnitude, which leads to the
  improvements in their performance in modeling the numerical
  flux. Specifically, this term becomes
  $\left(q\,{6187}/{330}-q^3\right)\,v^7$ in $f_{33}$,
  $\left(q\,{5297}/{2970} + {q^3}/{3} \right)\,v^7$ in $\rho_{33}$,
  and $\left(-q\,{1073}/{1188} + q^3\,{2}/{3} \right)\,v^7$ in
  $\rho^f_{33}$.  

  Finally, for large aligned spin $q=0.95$ at the LSO, the 
  $\rho^f$-resummed flux is closer to the numerical flux 
  than the Taylor-expanded--truncated flux. Furthermore, 
  we want to emphasize that the $\rho^f$-resummation improves 
  the Taylor-expanded flux substantially over a large range of 
  $v$ and spin values. 
  The differences between numerical and
  $\rho^f$-resummed fluxes are smaller than those between the
  numerical and Taylor-expanded--truncated fluxes, by a factor of 3 --
  5 at low velocities. Considering the large number of orbits an
  extreme mass-ratio binary spends in this range of velocities or
  frequencies, such an improvement is indeed significant in correcting
  the orbital dynamics (see Ref.~\cite{Yunes:2009ef} for a
  quantitative analysis in the nonspinning case).

\section{Factorized multipolar waveforms for generic mass-ratio 
spinning, non-precessing black holes}
\label{sec:gmrresum}

In this section we extend the calculation of Sec.~\ref{sec:tmlresum} to generic mass-ratio 
spinning, non-precessing black-hole binaries. 

In Ref.~\cite{Kidder2008,BFIS}, the non-spinning Taylor-expanded multipolar
waveforms were computed through 3PN order. In Ref.~\cite{ABFO}, spinning Taylor-expanded
multipolar waveforms were computed through 1.5PN order. Using the definitions:
\begin{subequations}
\begin{eqnarray}
M&\equiv&m_1+m_2 \\
\delta m &\equiv& m_1-m_2\,, \\
\chi_S &\equiv& \frac{1}{2}\,\left(\frac{S_1}{m_1^2}+\frac{S_2}{m_2^2}\right)\,, \\
\chi_A &\equiv& \frac{1}{2}\,\left(\frac{S_1}{m_1^2}-\frac{S_2}{m_2^2}\right)\,, 
\end{eqnarray}
\end{subequations}
and restricting ourselves to circular, equatorial orbits, we obtain the following 
modes decomposed with respect to -2 spin-weighted spherical harmonics 
\begin{widetext}
\begin{subequations}
\begin{eqnarray}
\label{h22}
h_{22}&=&-8\,\sqrt{\frac{\pi}{5}}\,\frac{\nu\, M}{R}\,e^{-2i\phi}\,v^2\,\left\{1-\left(\frac{107}{42}-\frac{55}{42}\nu\right)\,v^2+\left[2\pi+12\,i\,\log\left(\frac{v}{v_0}\right) \right. \right. \nonumber \\
&& \left. \left. -\frac{4}{3}(1-\nu)\,\chi_S-\frac{4}{3}\frac{\delta m}{M}\,\chi_A\right]\,v^3\right\}\,, \\
h_{21}&=&-\frac{8i}{3}\,\sqrt{\frac{\pi}{5}}\,\frac{\nu\,\delta m}{R}\,e^{-i\phi}\,v^3\,\left [1-\left(\frac{3}{2}\chi_S+\frac{3}{2}\frac{M}{\delta m}\,\chi_A\right)\,v\right ]\,, \\
h_{33}&=&3i\sqrt{\frac{6\pi}{7}}\,\frac{\nu\,\delta m}{R}\,e^{-3i\phi}\,v^3\,\left \{1-(4-2\nu)\,v^2+ 
\left[3\pi-\frac{21i}{5}+6i\log\frac{3}{2}+18\,i\,\log\left(\frac{v}{v_0}\right) \right. \right. \nonumber \\
&& \left. \left. -\left(2-\frac{5}{2}\nu\right)\chi_S-\left(2-\frac{19}{2}\nu\right)\frac{M}{\delta m}\,\chi_A\right]\,v^3\right\}\,, \\
h_{32}&=&-\frac{8}{3}\,\sqrt{\frac{\pi}{7}}\,\frac{\nu\, M}{R}\,e^{-2i\phi}\,v^4\,(1-3\nu+4\nu\chi_S\,v)\,, \\
h_{31}&=&-\frac{i}{3}\,\sqrt{\frac{2\pi}{35}}\,\frac{\nu\,\delta m}{R}\,e^{-i\phi}\,v^3\,\left\{1-\left(\frac{8}{3}+\frac{2}{3}\nu\right)\,v^2 +\left[\pi-\frac{7i}{5}-2i\log2+6\,i\,\log\left(\frac{v}{v_0}\right) \right. \right. \nonumber \\
&& \left. \left. -\left(2-\frac{13}{2}\nu\right)\chi_S-\left(2-\frac{11}{2}\nu\right)\,\frac{M}{\delta m}\,\chi_A\right]\,v^3 \right\}\,, \\
h_{44}&=&\frac{64}{9}\,\sqrt{\frac{\pi}{7}}\,\frac{\nu\, M}{R}\,e^{-4i\phi}\,v^4\,\left \{1-3\nu-
\left(\frac{593}{110}-\frac{1273}{66}\nu+\frac{175}{22}\nu^2\right)\,v^2+
\left [-\frac{42}{5}i+\frac{1193}{40}i\,\nu+\left(4\,\pi+8i\log2\right)(1-3\,\nu) \right.\right. \nonumber\\
&&\left.\left.+24i(1-3\,\nu)\log\left(\frac{v}{v_0}\right)+
\left (- \frac{8}{3} +\frac{164}{15}\nu - \frac{56}{5}\nu^2 \right )\,\chi_S +\left (-\frac{8}{3} + \frac{52}{5}\nu \right )\,\frac{\delta m}{M}\,\chi_A\right ]\,v^3\right \}\,, \\
h_{43}&=&\frac{9i}{5}\,\sqrt{\frac{2\pi}{7}}\,\frac{\nu\,\delta m}{R}\,e^{-3i\phi}\,v^5\,\left [1-2\nu+\left(\frac{5}{2}\nu\,\chi_S-\frac{5}{2}\nu\,\frac{M}{\delta m}\,\chi_A\right)\,v\right ]\,, \\
h_{42}&=&-\frac{8}{63}\,\sqrt{\pi}\,\frac{\nu\, M}{R}\,e^{-2i\phi}\,v^4\,\left \{1-3\nu-\left(\frac{437}{110}-\frac{805}{66}\,\nu+\frac{19}{22}\,\nu^2\right)\,v^2+\left [-\frac{21}{5}i(1-4\,\nu)+2\,\pi(1-3\,\nu) \right.\right.\nonumber\\
&&\left.\left.+12i(1-3\,\nu)\log\left(\frac{v}{v_0}\right)+
\left (- \frac{8}{3} +\frac{236}{15}\nu - \frac{104}{5}\nu^2 \right )\,\chi_S + 
\left (-\frac{8}{3} + \frac{28}{5}\nu \right )\,\frac{\delta m}{M}\,\chi_A\right ]\,v^3\right \}\,, \\
h_{41}&=&-\frac{i}{105}\,\sqrt{2\pi}\,\frac{\nu\,\delta m}{R}\,e^{-i\phi}\,v^5\,\left [1-2\nu+\left(\frac{5}{2}\nu\,\chi_S-\frac{5}{2}\nu\,\frac{M}{\delta m}\,\chi_A\right)\,v\right ]\,.
\label{h41}
\end{eqnarray}
\end{subequations}
\end{widetext}
The 1.5PN, 0.5PN and 1PN order spin terms in the modes $h_{22}$, $h_{21}$, $h_{33}$, respectively, 
were obtained in Ref.~\cite{ABFO}. The 1.5PN-order (0.5PN-order) spin terms in the even (odd) parity 
modes are computed in Appendix \ref{AppendixD}. The higher-order non-spinning PN terms can 
be found in Refs.~\cite{Kidder2008,BFIS,DINresum}.

To compute the factorized multipolar waveforms for generic mass ratios we use Eq.~(\ref{hlm}). 
For the source terms $\hat{S}_{\rm eff}^{(\epsilon_p)}$ we employ the energy and angular momentum for circular, 
equatorial orbits computed from the effective-one-body 
Hamiltonian of Ref.~\cite{DJS08} (at the PN order at which we derive the factorized modes, the Taylor-expanded 
Hamiltonian of Ref.~\cite{DJS08} coincides with the Hamiltonian of Ref.~\cite{BB}). More explicitly, 
when expanding the effective-one-body energy and angular momentum for circular, equatorial orbits through 
1.5PN order, we find
\begin{eqnarray}
\label{Ham}
\frac{E(v)}{\mu}&=&1-\frac{\nu}{2}\,v^2\,\left\{1-\frac{(9+4\nu)}{12}\,v^2 \right. \nonumber \\
&& \left. +\frac{8}{3}\left[\left(1-\frac{1}{2}\nu\right)\,\chi_S+\frac{\delta m}{M}\,\chi_A\right]\,v^3 \right \}\,,
\end{eqnarray}
and 
\begin{eqnarray}
\label{Lorb}
\frac{L(v)}{\mu}&=&\nu\,v^{-1}\,\left\{1+\frac{9+\nu}{6}\,v^2 \right.
\nonumber \\
&& \left.-\frac{10}{3}\left[\left(1-\frac{1}{2}\nu\right)\,\chi_S+\frac{\delta m}{M}\,\chi_A\right]\,v^3\right\} \,.\nonumber \\
\end{eqnarray}
Equations~(\ref{Ham}), (\ref{Lorb}) are sufficient for computing
  the quantity $f_{\ell m}$ in Eq.~(\ref{hlm}). In fact, similarly to
  the test-particle case analyzed in Sec.~\ref{sec:tmlresum}, the
  factor $T_{\ell m}$ in the generic mass-ratio case is not modified
  by spin effects.  The factor $\delta_{\ell m}$ is not modified by
  spin effects either since there is no imaginary spin terms in
  Eqs.~(\ref{h22})--(\ref{h41}). The nonspinning $\delta_{\ell m}$
  expressions for generic mass ratios are given in Eqs.~(20)--(29) of
  Ref.~\cite{DINresum}.  Thus, inserting Eqs.~(\ref{Ham}) and
  (\ref{Lorb}) in Eq.~(\ref{hlm}), and using
  Eqs.~(\ref{h22})--(\ref{h41}), we derive the even-parity $f_{\ell
    m}$ and $\rho_{\ell m}$ and odd-parity $f_{\ell m}^L$ and
  $\rho_{\ell m}^L$ up to the highest PN accuracy known today. We obtain
\begin{subequations}
\begin{eqnarray}
f_{22}&=&1+\frac{1}{42}(55\nu-86)\,v^2 \nonumber\\
&& -\frac{4}{3}\left[(1-\nu)\,\chi_S+\frac{\delta m}{M}\,\chi_A\right]\,v^3 \,, \\
f_{21}^L&=&1-\frac{3}{2}\left(\chi_S+\frac{M}{\delta m}\,{\chi_A}\right)\,v \,, \\
f_{33}&=&1+\left(2\nu-\frac{7}{2}\right)\,v^2-
\left[\left(2-\frac{5}{2}\nu\right)\chi_S \right. \nonumber \\ 
&& \left. +\left(2-\frac{19}{2}\nu\right)\,\frac{M}{\delta m}\,{\chi_A}\right]\,v^3\,, \\
f_{32}^L&=&1-\frac{4\nu}{3\nu-1}\chi_S\,v\,, \\
f_{31}&=&1+\left(-\frac{2}{3}\nu-\frac{13}{6}\right)\,v^2
-\left[\left(2-\frac{13}{2}\nu\right)\chi_S \right. \nonumber \\
&& \left. +\left(2-\frac{11}{2}\nu\right)\,\frac{M}{\delta m}\,{\chi_A}\right]\,v^3\,,
\end{eqnarray}
\begin{eqnarray}
f_{44}&=& 1 - \frac{2\,625 \nu^2-5\,870 \nu +1\,614}{330 (1-3 \nu)}\,v^2 \nonumber\\
&& - \frac{4}{15}\left[\frac{42 \nu^2-41 \nu +10}{1-3\nu}\chi_S+\frac{10-39 \nu}{1-3 \nu}\frac{\delta m}{M}\chi_A\right]\,v^3\,,\nonumber\\ \\
f_{43}^L&=&1-\frac{5\nu}{2(2\nu-1)}\left(\chi_S-\frac{M}{\delta m}\,{\chi_A}\right)\,v\,, \\
f_{42}&=& 1 - \frac{285 \nu ^2-3\,530 \nu +1\,146}{330 (1-3 \nu)}\,v^2 \nonumber\\
&& - \frac{4}{15}\left[\frac{78 \nu^2-59 \nu +10}{1-3\nu}\chi_S+\frac{10-21 \nu}{1-3\nu}\frac{\delta m}{M}\chi_A\right]\,v^3\,,\nonumber\\ \\
f_{41}^L&=&1-\frac{5\nu}{2(2\nu-1)}\left(\chi_S-\frac{M}{\delta m}\,{\chi_A}\right)\,v\,,
\end{eqnarray}
\end{subequations}
and \footnote{We have realized that the odd-$m$ factorized $h_{\ell m}$ built from 
$(\rho_{\ell m})^\ell$ and $(\rho^L_{\ell m})^\ell$ are singular in the equal-mass limit $\delta m\rightarrow 0$. This happens because the leading factor $h_{\ell m}^{(N,\epsilon_p)}$ in the odd-$m$ case is proportional to $\delta m$,
$h_{\ell m}^{(N,\epsilon_p)}\propto c_{\ell+\epsilon_p}(\nu) = -
\frac{\delta m}{M}\left(\frac{m_1}{M}\right)^{m-2}\,\sum_{i=0}^{m-1}\left(\frac{m_2}{m_1}\right)^i \,,$
but $(\rho_{\ell m})^\ell$ and $(\rho^L_{\ell m})^\ell$ are at least quadratic in
$\chi_A/\delta m$. This problem was resolved in Ref.~\cite{SEOBNRv1} 
by replacing the $(\rho_{\ell m})^\ell$ and $(\rho^L_{\ell m})^\ell$ 
with the nonspinning (NS) $(\rho_{\ell m})^\ell$ and $(\rho^L_{\ell m})^\ell$ 
plus the spinning (S) $f_{\ell m}$ and $f^L_{\ell m}$, defined in Eqs.~(41b), (41c), (41e), (41g), and (41i). 
In Ref.~\cite{SEOBNRv1}, $\rho^{L\,\rm NS}_{21}$, $\rho^{\rm NS}_{33}$, 
$\rho^{\rm NS}_{31}$, $\rho^{L\,\rm NS}_{43}$, and
$\rho^{L\,\rm NS}_{41}$ are given in Eqs.~(A8b), (A9a), (A9c), (A10b),
and (A10d), respectively, and $f^{L\,\rm S}_{21}$, $f^{\rm S}_{33}$,
$f^{\rm S}_{31}$, $f^{L\,\rm S}_{43}$, and $f^{L\,\rm S}_{41}$ are
given in Eqs.~(A15a)--(A15d).}
\begin{subequations}\label{rholm}
\begin{eqnarray}
\rho_{22}&=&1+\frac{1}{84}(55\nu-86)\,v^2 \nonumber\\
&& -\frac{2}{3}\left[(1-\nu)\,\chi_S+\frac{\delta m}{M}\,\chi_A\right]\,v^3 \,, \\
\rho_{21}^L&=&1-\frac{3}{4}\left(\chi_S+\frac{M}{\delta m}\,\chi_A\right)\,v \,, \\
\rho_{33}&=&1+\left(\frac{2}{3}\nu-\frac{7}{6}\right)\,v^2
-\left[\left(\frac{2}{3}-\frac{5}{6}\nu\right)\chi_S \right. \nonumber \\
&& \left. +\left(\frac{2}{3}-\frac{19}{6}\nu\right)\,\frac{M}{\delta m}\,\chi_A\right]\,v^3\,, \\
\rho_{32}^L&=&1-\frac{4\nu}{3(3\nu-1)}\chi_S\,v\,, \\
\rho_{31}&=&1+\left(-\frac{2}{9}\nu-\frac{13}{18}\right)\,v^2-\left[\left(\frac{2}{3}-\frac{13}{6}\nu\right)\chi_S \right. \nonumber \\
&& \left. +\left(\frac{2}{3}-\frac{11}{6}\nu\right)\,\frac{M}{\delta m}\,\chi_A\right]\,v^3\,,
\end{eqnarray}
\begin{eqnarray}
\rho_{44}&=& 1 - \frac{2\,625 \nu^2-5\,870 \nu +1\,614}{1\,320 (1-3 \nu)}\,v^2 \nonumber\\
&& - \left[\frac{42 \nu ^2-41 \nu +10}{15 (1-3\nu)}\chi_S+\frac{10-39 \nu}{15 (1-3 \nu)}\frac{\delta m}{M}\chi_A\right]\,v^3\,,\nonumber\\ \\
\rho_{43}^L&=& 1-\frac{5\nu}{8(2\nu-1)}\left(\chi_S-\frac{M}{\delta m}\,\chi_A\right)\,v\,,\\
\rho_{42}&=& 1 - \frac{285 \nu ^2-3\,530 \nu +1\,146}{1320 (1-3 \nu)}\,v^2 \nonumber\\
&& - \left[\frac{78 \nu ^2-59 \nu +10}{15 (1-3\nu)}\chi_S+\frac{10-21 \nu}{15 (1-3\nu)}\frac{\delta m}{M}\chi_A\right]\,v^3\,,\nonumber\\ \\
\rho_{41}^L&=& 1-\frac{5\nu}{8(2\nu-1)}\left(\chi_S-\frac{M}{\delta m}\,\chi_A\right)\,v\,.
\end{eqnarray}
\end{subequations}
We may use $E(r)$ instead of $|\vL|$ as the source term in the
odd-parity modes. However, there is no difference between $f_{lm}^L$
and $f_{lm}^H$, and correspondingly between $\rho_{lm}^L$ and
$\rho_{lm}^H$, through PN orders where spin effects of binaries with
generic mass ratio are known.

\begin{figure}
  \includegraphics[width=1.\linewidth]{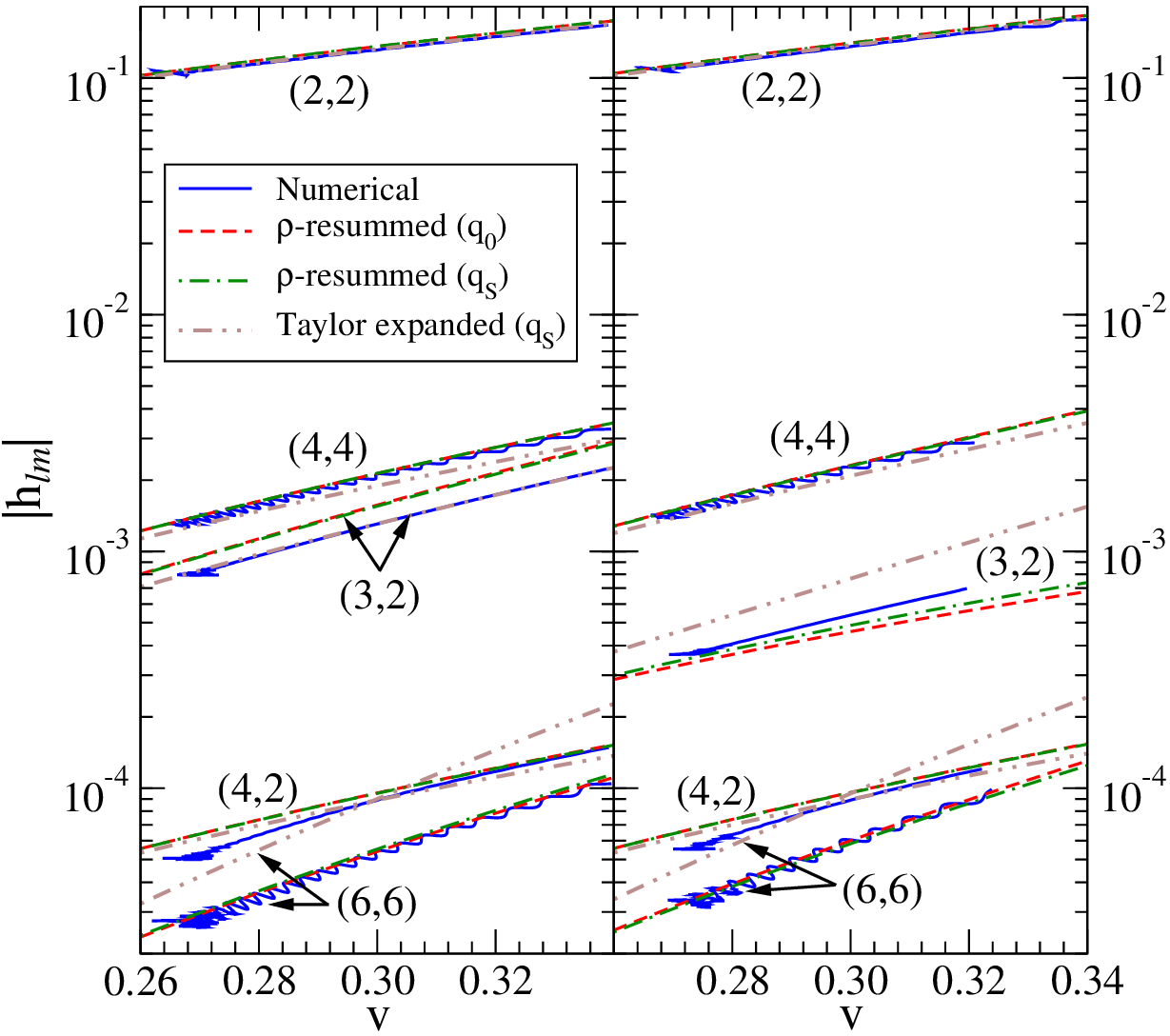}
  \caption{Comparisons between the numerical, Taylor-expanded and $\rho$-resummed 
    amplitudes of the dominant modes for an equal-mass
    equal-spin black hole binary as functions of the orbital velocity
    $v$. In the left panel, the component spins are $\chi_1=\chi_2=0.43655$;
    in the right panel, the component spins are
    $\chi_1=\chi_2=-0.43757$. The numerical amplitudes were produced by the Caltech/Cornell/CITA 
collaboration.}
\label{fig:hampUUDD}
\end{figure}
In the nonspinning case, using 1PN, 2PN and 3PN corrections,  
it was shown \cite{DINresum} that the
dependence of $\rho_{\ell m}$ on the mass-ratio $\nu$ is mild. As a 
consequence, it was considered meaningful to use test-particle results at PN orders where
generic mass-ratio results are unknown. Since for each mode 
only the leading order generic mass-ratio spin terms are known, it 
is not possible to carry out an exhaustive study and understand how the 
spin terms in $\rho_{\ell m}$ depend on $\nu$. As obtained 
in Appendix~\ref{AppendixD}, at leading order, the 0.5PN spin terms 
in the odd-parity modes are proportional to $\nu$. Thus, they are zeros in 
the test-particle limit, but finite in the comparable mass case. 
Moreover, we find that the dependence on $\nu$ of the 1.5PN spin terms 
in the even-parity modes is not that simple. 
Depending on the values of $\chi_S$ and $\chi_A$, the relative 
difference between $h^{(0),\rm 1.5PN}_{\ell m}(\nu=0.25)$ and 
$h^{(0),\rm 1.5PN}_{\ell m}(\nu=0)$ varies from zero to order of unity. 
Therefore, also the dependence of $h^{(0),\rm 1.5PN}_{\ell m}$ on $\nu$ is 
not mild. 

Nevertheless, it is still reasonable to include the test-particle limit 
spin terms in $f_{\ell m}$ and $\rho_{\ell m}$ such that at least part of the higher
order spin effects are included, and to check the results against 
available numerical (exact) data. Specifically, we
  combine the test-particle and generic mass ratio results by
  replacing all the test-particle terms in $f_{\ell m}$ and $\rho_{\ell
    m}$ whose generic mass-ratio counterparts are known with their
  generic expressions.

Thus, in the generic mass-ratio, spinning case, we propose to add to the 
$f_{\ell m}$'s and $\rho_{\ell m}$'s derived in this section 
the test-particle limit terms derived in Sec.~\ref{sec:tmlresum}.  
In applying this procedure we need to make a choice for the 
dimensionless spin variable $q$ appearing in the test-particle limit 
$f_{\ell m}$'s and $\rho_{\ell m}$'s.  For a black-hole binary with
component masses $m_1$ and $m_2$ and spins $\chi_1$ and $\chi_2$, 
we consider here two possibilities motivated by the choice of the 
deformed-Kerr--spin in the effective-one-body 
formalism. References~\cite{DJS08,EOBNRS} used for the deformed-Kerr--spin 
\begin{eqnarray}
\label{S0q}
q_0&=& \frac{\left|\vS_0\right|}{M^2}=
\frac{1}{M^2}\,\left|\left(1+\frac{m_2}{m_1}\right)\vS_1+\left(1+\frac{m_1}{m_2}\right)\vS_2\right|\,,
\nonumber \\ 
&=& \sqrt{1 - 4\nu}\,\chi_A + \chi_S\,,
\end{eqnarray}
while Ref.~\cite{BB} used the following deformed-Kerr spin 
\begin{eqnarray}
\label{Sq}
q_S &=& \frac{\left|\vS\right|}{M^2}=
\frac{1}{M^2}\,\left|\vS_1+\vS_2\right|\,,\nonumber \\
&=& \sqrt{1 - 4\nu}\,\chi_A + (1 - 2 \nu)\,\chi_S\,.
\end{eqnarray}
Moreover, in the generic mass-ratio, spinning case, we 
also propose to use as effective sources in Eq.~(\ref{hlm}) the Hamiltonian and 
angular momentum for quasi-circular orbits computed using the effective-one-body 
Hamiltonians~\cite{DJS08,BB}. 

In Fig.~\ref{fig:hampUUDD}, we compare the amplitudes of the
numerical, the Taylor-expanded and the $\rho$-resummed modes for the
five most dominant modes and for the two configurations of
equal-mass, equal-spin black hole binaries of the Caltech-Cornell-CITA collaboration of 
Ref.~\cite{Chu:2009md,EOBNRS}. We employ the effective sources built using the 
Hamiltonian and angular momentum for quasi-circular orbits of Ref.~\cite{DJS08,EOBNRS}.  
The dimensionless spins in the two configurations are $\chi_1=\chi_2=0.43655$ and
$\chi_1=\chi_2=-0.43757$, respectively.  The numerical amplitudes are
derived from the numerical simulations published in
Ref.~\cite{EOBNRS}. Oscillations in the numerical amplitudes are
due to numerical artifacts in the simulations. For the (2,2) mode, the 
Taylor-expanded amplitudes agree quite well with the 
numerical amplitude, at least up to the frequency considered. 
Thus, the improvement due to the $\rho$-resummation 
is  marginal. For higher-order modes, there are large differences
  between numerical and Taylor-expanded amplitudes, and we find a substantial
  improvement when we adopt the $\rho$-resummation, except for the (3,3)
  mode in the spin aligned case ($\chi_1=\chi_2=0.43655$) whose
  numerical and Taylor-expanded amplitudes overlap, likely by
  coincidence.  For the (2,2), (4,4) and (6,6) modes, the relative
difference between numerical and $\rho$-resummed amplitudes is within
$5\%$ \footnote{Note that Ref.~\cite{EOBNRS} found a relative difference on the
    order $1\%$ in the (2,2) mode comparison. However,  Ref.~\cite{EOBNRS}
    compared effective-one-body waveforms generated using the full {\it non-adiabatic} effective-one-body
     evolution. In Fig.~\ref{fig:hampUUDD}, the analytical amplitudes are
    generated using the {\it adiabatic} quasi-circular
    effective-one-body evolutions.}. For the (3,2) and (4,2) modes, the relative
difference is between $10\% \mbox{--}20\%$. 
We find that the results in Fig.~\ref{fig:hampUUDD} depend weakly on 
the choice of $q$. In fact, using $q=q_0$ defined in Eq.~(\ref{S0q}) and
$q=q_S = q_0/2$ (when $\nu=0.25$) defined in Eq.~\eqref{Sq}, the
relative amplitude difference is $<2\%$ for the (2,2), (4,4), (4,2)
and (6,6) modes, and $\sim 5\%$ for the (3,2) mode.  Therefore, the
uncertainty in the $\rho$-resummed amplitude due to the choice of $q$
is less than half their systematic difference from the numerical
results.

\begin{figure}
  \includegraphics[width=1.\linewidth]{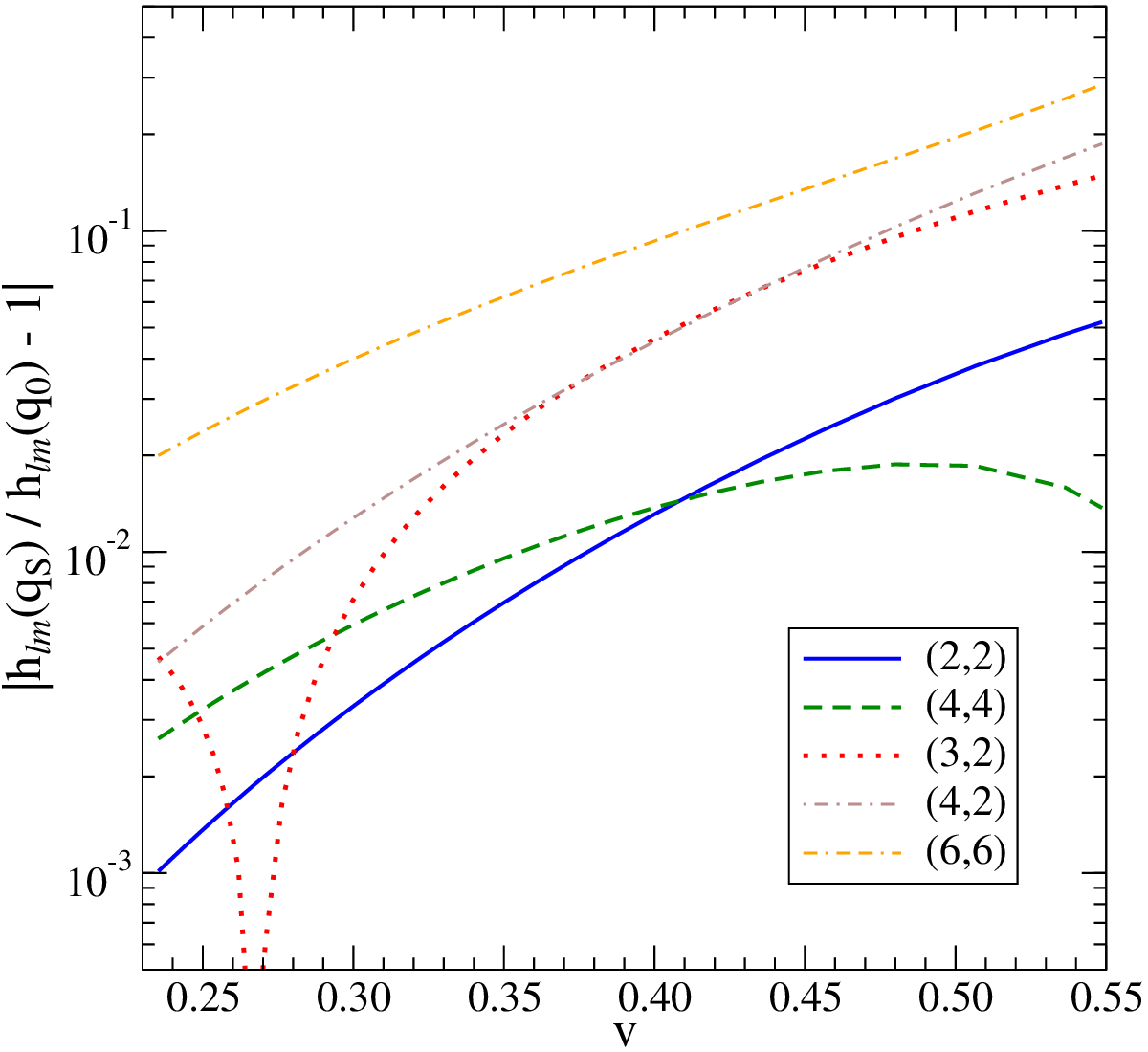}
  \caption{Relative difference between $\rho$-resummed amplitudes of 
    the dominant modes for an equal-mass, equal-spin black-hole binary 
    when the test-particle spin is set to $q=q_0$ and $q=q_S$.
    The component spins of the binary are $\chi_1=\chi_2=0.95$.
    The relative amplitude difference is plotted as a function of the orbital velocity $v$. 
    }
\label{fig:hampS0St}
\end{figure}

Since we expect a stronger amplitude dependence on $q$ in the case of
larger spin magnitudes, we study the $\rho$-resummed amplitude
dependence on the choice of $q$ in Fig.~\ref{fig:hampS0St}, where we
show the difference between the amplitudes $|h_{\ell m}(q=q_0)|$ and
$|h_{\ell m}(q=q_S)|$ for an equal-mass, equal-spin black-hole binary
with component spins $\chi_1=\chi_2=0.95$. The relative amplitude
differences are $<5\%$ for the dominating (2,2) and (4,4) modes, and
$<10\%$ for the weaker (3,2) and (4,2) modes at $v<0.45$. For the
(6,6) mode, since the test-particle spin terms in $\rho_{66}$ are
known only through 2PN order, i.e. only one more term is known beyond
the generic mass ratio results, the amplitude dependence on $q$ is
entirely determined by this term and is somewhat stronger --- reaching
$30\%$ at the LSO.  

Finally, we check the effect of the
  test-particle spin terms by comparing $|h_{\ell m}(q=q_0)|$ and
  $|h_{\ell m}(q=0)|$ (i.e., removing the test-particle spin terms
  from the generic mass-ratio amplitudes in the latter) for this
  binary configuration. The difference, compared to
  Fig..~\ref{fig:hampS0St}, becomes larger by a factor of a few and
  reaches $10$--$25\%$ for the (2,2), (4,4), (3,2) and (4,2) modes in
  the range of frequencies investigated in this paper. These terms may
  provide non-negligible corrections to the waveform and flux modeling.

\section{Conclusions}
\label{sec:conclusions}

In our study we employed the spin PN multipolar waveforms derived and decomposed with respect to the 
-2 spin-weighted spheroidal harmonics in Refs.~\cite{TSTS1996}, and transformed them 
in -2 spin-weighted spherical harmonics.  We also took advantage of the new, 
recently computed~\cite{TagoshiFujita2010}, higher-order nonspinning and 
spin PN contributions in several subdominant modes. We also augmented our knowledge of 
the higher-order spin terms for generic mass-ratios, computing the generic 
expressions for the half and, one and half post-Newtonian contributions to the 
odd-parity (current) and even-parity (odd) multipoles, respectively (see Appendix~\ref{AppendixD}). 

Using the above results we extended the resummation method of factorized multipolar 
waveforms introduced in Ref.~\cite{DINresum} to spinning, non-precessing 
black-hole binaries. This factorized multipolar decomposition consists 
in a multiplicative decomposition of the $h_{\ell m}$ waveform 
into the product of several factors corresponding to various physical 
effects and the replacement of the factor $f_{\ell m}$ by its $\ell$-th 
root $\rho_{\ell m} = (f_{\ell m})^{1/\ell}$. 

In the case of a nonspinning test-particle orbiting a Kerr black hole in the 
equatorial plane, we found that the $\rho$-resummation is quite effective 
in reproducing the numerical multipolar amplitudes and energy flux 
up to $q \geq 0.75$ and $v \geq 0.4$. However, for larger values of $q$, 
we observed that the analytical $\rho_{\ell m}(v)$'s either have a slope larger 
than the numerical one, or they tend to grow as function of $v$ instead of decreasing. 
This behavior can be cured by factoring out the 
lower-order PN terms in the $\rho_{\ell m}$, notably the 
0.5PN, 1PN and 1.5PN order terms. Being the lower-order PN terms 
negative (for $q>0$), this procedure corresponds to factoring out the zeros 
of $\rho_{\ell m}$ which turns out to capture the numerical (exact) zeros.. 

When applying the $\rho^f$-resummation, we found that the fractional amplitude 
difference between the numerical and analytical (2,2) mode at the LSO 
is $16\%\,(33\%)$, $0.18\%\,(0.32\%)$ and $0.20\%\,(0.85\%)$ 
for $q = 0.95, 0, -0.95$, respectively. We indicated in parenthesis 
the numbers when Taylor-expanded amplitudes are employed. Thus, we found that 
for the (2,2) mode the improvement of the resummation is marginal. This might be  
due to the fact that the (2,2) mode is known at rather high PN order 
(5.5 PN for nonspinning terms and 4PN for spin terms). 
For the (2,1), (3,3) and (4,4) modes, for which less spin PN terms are known (see 
Table~\ref{tab:PNorder}) the improvement due to the $\rho^f$-resummation 
is even more striking. In fact, for those modes we obtained a fractional 
amplitude differences $2.4\%\,(4.2)$, $0.2\%\,(0.58\%)$ and 
$0.0036\%\,(0.15\%)$, $7.5\%\,(2)$, $0.027\%\,(0.55\%)$ and $0.13\%\,(0.2\%)$, 
$16\%\,(7.5)$, $1.7\%\,(28\%)$ and $0.6\%\,(5.8\%)$, for $q = 0.95, 0, -0.95$, 
respectively. For $\ell \geq 5$, the $\rho^f$-resummed amplitudes 
are certainly better than the Taylor-expanded amplitudes, but they 
differ from the numerical results quite substantially at high frequency. 
This is due to the fact that for those modes the spin effects are known only 
up to 2.5PN order or lower. In summary, we found that the multipolar amplitudes 
computed with the $\rho^f$-resummation are systematically closer to the numerical (exact) results 
than Taylor-expanded ones over a large range of $v$ and spin values. The agreement 
can be further improved by including suitable adjustable parameters and calibrating them 
to the numerical results, as done in the non-spinning case in Ref.~\cite{Yunes:2009ef}.

Moreover, the numerical energy flux can also be successfully modeled by the 
$\rho^f$-resummation --- for example we found that the fractional difference
between the numerical and $\rho^f$-resummed flux is $13\%\,(63\%)$, $0.70\%\,(3.3\%)$ 
and $0.48\%\,(2.9\%)$ for $q = 0.95, 0, -0.95$, respectively, where the numbers 
in parenthesis refer to the Taylor-expanded--truncated PN flux. 
For large aligned spins, the  $\rho^f$-resummed flux 
is much closer to the numerical flux at the LSO than the Taylor-expanded--truncated flux. 
Furthermore, we emphasize again that the $\rho^f$-resummation improves the 
Taylor-expanded flux substantially over a large range of $v$ and spin values, and 
especially at low frequency where the majority of the signal-to-noise ratio of a binary accumulates.

We have also extended the factorized resummation to generic mass-ratio, non-precessing,
 spinning black-hole binaries, and proposed, as in Ref.~\cite{DINresum}, to augment 
the generic mass-ratio $\rho_{\ell m}$ with higher-order test-particle spin contributions. 
Unlike in the nonspinning case~\cite{DINresum}, in the spinning case only 
the leading-order generic mass-ratio spin terms are known. Using this limited information 
we found that the dependence on $\nu$ of the spin terms is not necessarily mild. It depends on the mass 
ratio and the spin values. Nevertheless, we explored the possibility of adding 
the spin contributions from the test-particle limit case to the generic mass-ratio amplitudes.

When adding the test-particle limit contributions, we proposed to identify $q$ with 
the Kerr-deformed spin in the effective-one-body description. Using the two choices currently available 
in the literature, that is $q = |\mathbf{S}_0|/M^2$~\cite{DJS08,EOBNRS} 
or $q = |\mathbf{S}|/M^2$~\cite{BB}, we found that 
the resummed amplitudes of the (2,2), (4,4), (4,2) and (6,6) modes agree with numerical 
simulation results \cite{EOBNRS} to within $2\%$, for equal-mass, equal-spin binaries 
with spins $|\chi_1|=|\chi_2|\simeq 0.44$. The (3,2) mode amplitude agrees with numerical 
results at $5\%$ level. The relative difference between the two choices of resummed amplitudes 
is less than half their difference from numerical results. 
When the spins are near extremal, e.g., $\chi_1=\chi_2=0.95$, we found a mild, but non-negligible 
$q$ dependence of the resummed amplitudes. Finally, when setting $q =0$, that is removing the test-particle 
spin terms from the generic mass-ratio amplitudes, we obtain that the results vary by $10$--$20\%$ 
for the (2,2), (4,4), (3,2) and (4,2) modes in the range of frequencies investigated in this paper.

The study carried out in this paper should be considered as a first step in the 
modeling of extreme--mass-ratio inspirals and comparable mass black-hole binaries 
in presence of spins. We expect that in the extreme--mass-ratio inspiral case, the amplitude and flux 
agreement can be further improved by including in our 
$\rho^f_{\ell m}$ a few adjustable parameters and calibrate them to the numerical data, as already 
done in Ref.~\cite{Yunes:2009ef} for nonspinning extreme--mass-ratio inspirals. 
In the comparable-mass case, more detailed comparisons with accurate numerical-relativity 
simulations will allow us to nail down the choice of the spin parameter $q$, and allow 
us to carry out direct comparisons between the numerical and analytical $\rho_{\ell m}$, 
thus helping in modeling the latter.

\begin{acknowledgments}
We thank Larry Kidder and Andreas Ross for useful discussions, and Bala Iyer for useful comments.

A.B. and Y.P. acknowledge support from NSF Grants PHY-0603762 and PHY-0903631. 
A.B. acknowledges support also from NASA grant NNX09AI81G. 
H.T. acknowledges support from JSPS Grand-in-Aid for Scientific Research No.20540271.
\end{acknowledgments}

\appendix

\section{Taylor-expanded multipolar waveforms $\hat{Z}_{\ell m \omega_0}$}
\label{AppendixA}

In order to compute the multipolar waveforms for a test particle around a Kerr black hole,
we transform the Teukolsky equation into the frequency domain, 
and expand it into the -2 spin-weighted spheroidal harmonics. 
The resulting equation is an ordinary differential equation about the radial coordinate.
This radial Teukolsky equation can be solved formally by using the Green function.
Since the Green function is represented by homogeneous solutions of the radial Teukolsky equation,
the central issue of this problem is to obtain the homogeneous solutions. 
There are two methods for obtaining them.

In the first method, we transform the radial Teukolsky equation into the Sasaki-Nakamura
equation. In the Schwarzschild case, the homogeneous Sasaki-Nakamura equation 
becomes the homogeneous Regge-Wheeler equation.
We expand the homogeneous Sasaki-Nakamura or Regge-Wheeler equation in terms of 
$\epsilon\equiv G M\omega$ where $\omega$ is the angular frequency of the wave.
In the case of circular orbit, $\omega$ becomes $\omega_0=m\Omega$.
(we revive the gravity constant $G$ here). We look for the solution in power series in $\epsilon$. 
This is thus a kind of post-Minkowskian expansion. 
One difference between the ordinary post-Minkowskian approximation and this approximation 
is that we must impose correct boundary conditions at the horizon.
Closed analytic representation of the solution at each order is necessary in order to 
obtain the asymptotic amplitudes which constitute the Green function.
The lowest order solutions are represented by 
spherical Bessel functions. The higher-order solutions can, in principle, be derived iteratively. 
However, it becomes more difficult to perform this iteration and to 
derive the solution in closed analytic form at higher orders. 
The highest order computation so far was done in the Schwarzschild case
by Tanaka et al. \cite{TTS55PN} in which the closed analytic formulas 
for a homogeneous solution is obtained up to ${\cal O}(\epsilon)$ for arbitrary $\ell$,
and up to ${\cal O}(\epsilon^3)$ for $\ell=2$ and $3$,
and up to ${\cal O}(\epsilon^2)$ for $\ell=4$.
The formulas are explicitly given in a review paper \cite{MSSTT1997}.
Those computations are sufficient for obtaining the 
energy flux through 5.5PN order. 
Since the formulas for the $\hat{Z}_{\ell m\omega_0}$'s are not given in the literature, 
we write them below. For each mode, we write the terms up to ${\cal O}(v^{11-2(\ell-2)})$
relative to the lowest-order term. 

Furthermore, in the Kerr case, so far the highest order 
computation was done by Tagoshi et al. \cite{TSTS1996} 
in which the closed analytic formulas for a homogeneous solution is obtained 
at ${\cal O}(\epsilon)$ for arbitrary $\ell$ modes, and at ${\cal O}(\epsilon^2)$ for $\ell=2$ and $3$ modes. 
These computations are sufficient for obtaining 
the energy flux through 4PN order.

Two of the authors have recently obtained the ${\cal O}(\epsilon^2)$ closed analytic formulas 
for $\ell=4$ mode \cite{TagoshiFujita2010}. This order is necessary to derive the multipolar waveforms  
through 3PN order beyond $C_{4m}^{(N,0)}$, i.e. 4PN order beyond $C_{22}^{(N,0)}$ 
(see Table~\ref{tab:PNorder}). 
More details of the computation and complete results are given elsewhere \cite{TagoshiFujita2010}.
Here we show only the explicit formulas for $\hat{Z}_{\ell m\omega_0}$ defined in 
Eq.~(\ref{hatZlmw0}). We write the spin-dependent 4PN-order $\hat{Z}_{\ell m\omega_0}$ 
in which each mode contains terms up to ${\cal O}(v^{8-(\ell-2)-\epsilon_p})$ 
relative to the lowest-order term.
($\epsilon_p$ is the parity of each mode).

The second method to obtain the homogeneous Teukolsky function is 
based on the Mano-Suzuki-Takasugi formalism \cite{ManoSuzukiTakasugi96}.
In this formalism, the homogeneous solutions of the Teukolsky equation is
represented with the series of hypergeometric functions and confluent hypergeometric
functions. The expansion coefficients of the two series solutions are the same, and 
they are closely related to the series expansion in power of $\epsilon$. 
Thus, if we compute this series up to higher order, we automatically obtain the
higher order PN expansion formulas.
Such computation was applied to the evaluation of the PN expansion of the 
black hole absorption effect in the Kerr case \cite{TagoshiManoTakasugi97}.
This method was also applied to the energy flux  
through 5.5PN order in the Schwarzschild case, confirming 
the results obtained with the above iteration method~\cite{TTS55PN}.
We apply this method to the Kerr case and obtain the 4PN-order multipolar waveforms
which agree with the results obtained with the above iteration method.
This method has also been recently applied to the computation of the 5.5PN-order multipolar waveforms
in the Schwarzschild case by Fujita and Iyer \cite{FujitaIyer2010}.
The non-spinning terms of expressions below agree with their results up to ${\cal O}(v^{11-2(\ell-2)})$.

We have
\begin{widetext}
\begin{eqnarray}\label{hatZ22w}
\hat{Z}_{22\omega_0}&=&
1 - {{107\,{v^2}}\over {42}} +
      \left[2\,\pi  - {{4\,q}\over 3} +12i\log\left(\frac{v}{v_0}\right) \right] \,{v^3} +
      \left(-{{2\,173}\over {1\,512}} + {q^2} \right) \,{v^4} +
      \left[-{{107\,\pi }\over {21}} - {{460\,q}\over {189}}-\frac{214}{7}\,
i\log\left(\frac{v}{v_0}\right)\right]\,{v^5}
\nonumber\\
&&+ \left[ {{27\,027\,409}\over {646\,800}} -
         {{856\,{\rm eulerlog}_2(v^2)}\over {105}} + {{2\,{{\pi }^2}}\over
	 3} + \frac{428}{105}i\pi -
         {{8\,\pi \,q}\over 3} - \frac{4}{3}\,i\,q + {{32\,{q^2}}\over
	 {567}} -8\,i\,(2\,q-3\,\pi)\log\left(\frac{v}{v_0}\right) \right.
\nonumber\\
&&	\left.-72\log^2\left(\frac{v}{v_0}\right) \right]\,{v^6}
-  \left[ {{2\,173\,\pi }\over {756}} - {{24\,509\,q}\over {3\,402}} -
2\,\pi \,{q^2}+\frac{2\,173}{126}\,i\,\log\left(\frac{v}{v_0}\right)-12\,i\,q^2\log\left(\frac{v}{v_0}\right) \right] \,{v^7}
\nonumber\\
&& + \left[ -{{846\,557\,506\,853}\over {12\,713\,500\,800}} +
         {{45\,796\,{\rm eulerlog}_2(v^2)}\over {2\,205}} - {{107\,{{\pi }^2}}\over {63}} +
         {{2\,671\,117\,{q^2}}\over {666\,792}}-\frac{22\,898}{2\,205}\,i\,\pi+\left(\frac{214}{63}\,i -
         {{920\,\pi}\over {189}}\right)\,q \right.
\nonumber\\
&&- \left.\frac{4}{63}\,i\,(460\,q+963\,\pi)\log\left(\frac{v}{v_0}\right) +\frac{1\,284}{7}\log^2\left(\frac{v}{v_0}\right) \right]\,{v^8} + \left[ -\frac{1\,712}{105}\,\pi\,\text{eulerlog}_2(v^2)-\frac{64\,i\,\zeta (3)}{3}-\frac{4\,\pi\,^3}{3}\right.
\nonumber\\
&&+\left.\frac{856\,i\,\pi ^2}{63}+\frac{27\,027\,409\,\pi\,}{323\,400}-\frac{259\,i}{81} -\left(\frac{3\,424}{35}\,i\,\text{eulerlog}_2(v^2)- 8\,i\,\pi ^2+\frac{1712\,\pi}{35}-\frac{27\,027\,409\,i}{53\,900}\right) \log \left(\frac{v}{v_0}\right)\right.
\nonumber\\
&&-\left.144\,\pi\,\log ^2\left(\frac{v}{v_0}\right)-288\,i\,\log ^3\left(\frac{v}{v_0}\right) \right]\,{v^9}
+ \left[ \frac{232\,511}{39\,690}\left(\,2\,\text{eulerlog}_2(v^2)-\,i\,\pi\right)\right.
\nonumber\\
&&-\left.\frac{2\,173\,\pi\,^2}{2\,268}-\frac{866\,305\,477\,369}{9\,153\,720\,576}-\frac{2\,173}{63}\,i\,\pi  \log\left(\frac{v}{v_0}\right)+\frac{2\,173}{21} \log ^2\left(\frac{v}{v_0}\right) \right]\,{v^{10}}
\nonumber\\
&&+ \left[ \frac{91\,592\,\pi\,\text{eulerlog}_2(v^2)}{2\,205}+\frac{3\,424\,i\,\zeta (3)}{63}+\frac{214\,\pi\,^3}{63}-\frac{45796\,i\,\pi ^2}{1\,323}-\frac{846\,557\,506\,853\,\pi\,}{6\,356\,750\,400}+\frac{3\,959\,i}{486} \right.
\nonumber\\
&&+ \left(\frac{183\,184}{735}\,i\,\text{eulerlog}_2(v^2)-\frac{428\,i\,\pi^2}{21}+\frac{91\,592\,\pi\,}{735}-\frac{846\,557\,506\,853\,i}{1\,059\,458\,400}\right) \log \left(\frac{v}{v_0}\right)
\nonumber\\
&&+ \left.\frac{2\,568}{7}\,\pi\, \log ^2\left(\frac{v}{v_0}\right)+\frac{5\,136}{7}\,i\,\log ^3\left(\frac{v}{v_0}\right) \right]\,{v^{11}}
\,,
\end{eqnarray}
\begin{eqnarray}\label{hatZ21w}
\hat{Z}_{21\omega_0}&=& 1 - {{3q}\over{2}}v- {{17}\over {28}}v^2 + \left[\pi - {61\,q \over 126} - \frac{i}{2}(1+4\log 2)+6\,i\,\log\left(\frac{v}{v_0}\right)\right]v^3
\nonumber\\
&&+ \left[ -{{43}\over {126}} - {3\,q\pi \over 2} + 3\,q^2
       +\frac{3}{4}\,i\,q(1+4\log 2)-9\,i\,q\log\left(\frac{v}{v_0}\right) \right] \,
       {v^4}
\nonumber\\
&&+ \left[ -{{17\pi }\over {28}} -
         {{68\,q}\over {27}} - {3\,q^3 \over 4} + \frac{17}{56}\,i\,(1+4\log 2)-\frac{51}{14}\,i\,\log\left(\frac{v}{v_0}\right) \right] \,{v^5}
\nonumber\\
&&+ \left[ {{15\,223\,771}\over {1\,455\,300}} - {{61\,\pi \,q}\over 126} +
         {{28\,565\,{q^2}}\over {4\,536}} + {\pi^2 \over 6} -
	 \frac{214\,{\rm eulerlog}_1(v^2)}{105} + \frac{109}{210}\,i\,\pi
	 + \left(-\frac{65}{252}\,i+\frac{61}{63}\,i\,\log 2\right)\,q \right.
\nonumber\\
&& \left. - \left(1+2\log 2+2\,i\,\pi\right)\log 2 + 3\left(1+2\,i\,\pi-\frac{61}{63}\,i\,q+4\log
    2\right)\log\left(\frac{v}{v_0}\right)-18\log^2\left(\frac{v}{v_0}\right)
    \right] \,{v^6}
\nonumber\\
&&+ \left[ \frac{107}{35}\,q\,\text{eulerlog}_1(v^2)-\frac{2\,689\,q^3}{756}+3\,\pi\,q^2-\frac{3\,i\,q^2}{4}-6\,i\,q^2 \log  2 -\frac{\pi ^2\,q}{4}-\frac{109\,i\,\pi \,q}{140}-\frac{374\,592\,223\,q}{17\,463\,600} \right.
\nonumber\\
&&+ 3\,i\,\pi \,q\,\log  2 +3\,q\,\log ^2 2 +\frac{3}{2}\,q\,\log  2 -\frac{43\,\pi\,}{126}+\frac{43\,i}{252}+\frac{43}{63}\,i\,\log  2 
\nonumber\\
&&+ \left.\left(18\,i\,q^2-9\,i\,\pi \,q-\frac{9\,q}{2}-18\,q\,\log  2 -\frac{43\,i}{21}\right) \log \left(\frac{v}{v_0}\right)+27\,q\,\log^2\left(\frac{v}{v_0}\right) \right] \,{v^7}
\nonumber\\
&&+ \left[ \frac{187\,310\,657}{42\,378\,336}+\frac{1\,819\,\text{eulerlog}_1(v^2)}{1\,470}-\frac{17\,\pi\,^2}{168}-\frac{1\,853\,i\,\pi}{5\,880}+\frac{17 \log ^2 2 }{14}+\frac{17}{14}\,i\,\pi  \log  2 +\frac{17 \log  2 }{28} \right.
\nonumber\\
&&+ \left.\left(-\frac{51}{28}-\frac{51\,i\,\pi}{14}-\frac{51 \log  2 }{7}\right) \log \left(\frac{v}{v_0}\right)+\frac{153}{14} \log ^2\left(\frac{v}{v_0}\right) \right] \,{v^8}
\nonumber\\
&&+ \left[ \frac{107}{105}\left( 2\,i\,\pi + 1 + 4\log  2 \right)\,i\,\text{eulerlog}_1(v^2)-\frac{8\,i\,\zeta(3)}{3}-\frac{\pi ^3}{6}+\frac{407\,i\,\pi ^2}{252}+\frac{15\,965\,281\,\pi\,}{1\,455\,300}- \frac{24\,580\,669\,i}{4\,365\,900} \right.
\nonumber\\
&&+\frac{4}{3}\,i\,\log ^3 2 -2\,\pi\, \log ^2 2 +i \log ^2 2 -\frac{1}{3}\,i\,\pi ^2 \log  2 +\frac{109}{105}\,\pi\, \log  2 -\frac{15\,223\,771}{727\,650}\,i\,\log  2 
\nonumber\\
&&+ \left(-\frac{428}{35}\,i\,\text{eulerlog}_1(v^2)+i\,\pi\,^2-\frac{109\,\pi\,}{35}+\frac{15\,223\,771\,i}{242\,550}-12\,i\,\log ^2 2 +12\,\pi\,\log  2 -6\,i\,\log  2 \right)\log \left(\frac{v}{v_0}\right)
\nonumber\\
&&+ \left.(9\,i-18\,\pi\,+9\,i\,\log (16)) \log ^2\left(\frac{v}{v_0}\right)-36\,i\,\log^3\left(\frac{v}{v_0}\right) \right] \,{v^9}
\nonumber\\
&&+ \left[ \frac{54\,222\,281\,699}{4\,767\,562\,800}+\frac{4\,601\,\text{eulerlog}_1(v^2)}{6\,615}-\frac{43\,\pi\,^2}{756}-\frac{4\,687\,i\,\pi}{26\,460}+\frac{43 \log ^2 2 }{63}+\frac{43}{63}\,i\,\pi  \log  2 +\frac{43 \log  2 }{126} \right.
\nonumber\\
&&- \left.\frac{43}{42}(2\,i\,\pi + 1+4\log 2 ) \log \left(\frac{v}{v_0}\right)+\frac{43}{7} \log ^2\left(\frac{v}{v_0}\right) \right] \,{v^{10}}
\nonumber\\
&&+ \left[ -\frac{1\,819}{2\,940}\left(2\,i\,\pi+1+4\log 2 \right)\,i\,\text{eulerlog}_1(v^2)+\frac{34\,i\,\zeta (3)}{21}+\frac{17\,\pi^3}{168}-\frac{6\,919\,i\,\pi ^2}{7\,056}+\frac{871\,003\,801\,\pi\,}{211\,891\,680}\right.\nonumber\\
&&-\frac{500\,228\,329\,i}{254\,270\,016} 
 -\frac{17}{21}\,i\,\log^3 2 +\frac{17}{14}\,\pi\, \log ^2 2 -\frac{17}{28}\,i\,\log ^2 2 +\frac{17}{84}\,i\,\pi ^2 \log  2 -\frac{1\,853\,\pi\, \log 2 }{2\,940}\nonumber\\
&& -\frac{187\,310\,657\,i\,\log  2 }{21\,189\,168}+\left(\frac{1\,819}{245}\,i\,\text{eulerlog}_1(v^2)-\frac{17\,i\,\pi ^2}{28}+\frac{1\,853\,\pi\,}{980}+\frac{187\,310\,657\,i}{7\,063\,056}+\frac{51}{7}\,i\,\log ^2 2 \right.\nonumber\\
&&\left.\left.-\frac{51}{7}\,\pi\, \log  2 +\frac{51}{14}\,i\,\log  2 \right)\log \left(\frac{v}{v_0}\right)- \frac{153}{28}\,i\,\left(2\,i\,\pi+ 1+ 4\log  2 \right) \log ^2\left(\frac{v}{v_0}\right)+\frac{153}{7}\,i\,\log ^3\left(\frac{v}{v_0}\right)\right] \,{v^{11}}
\,,\nonumber\\
\end{eqnarray}
\begin{eqnarray}\label{hatZ33w}
\hat{Z}_{33\omega_0}&=& 1 - 4\,v^2 +
     \left[3\,\pi-2\,q-\frac{21}{5}\,i-6\,i\,\log\left(\frac{2}{3}\right)+18\,i\,\log\left(\frac{v}{v_0}\right)\right]\,v^3
     + \left(\frac{123}{110}+\frac{3\,q^2}{2}\right)\,v^4
\nonumber\\
&&+ \left[\frac{29\,q}{60} - 12\,\pi
      +\frac{84}{5}\,i+24\,i\,\log\left(\frac{2}{3}\right)-72\,i\,\log\left(\frac{v}{v_0}\right)\right]\,v^5
\nonumber\\
&&+ \left[ \frac{19\,388\,147}{280\,280} - \frac{67\,q^2}{32} - \frac{78\,{\rm
     eulerlog}_3(v^2)}{7} - 6\,\pi\,q +
     \frac{3\,\pi^2}{2} -\frac{246}{35}\,i\,\pi+\frac{87}{20}\,i\,q\right.
\nonumber\\
&&
 \left.+12\,i\,q\log\left(\frac{2}{3}\right)
     -18\log^2\left(\frac{2}{3}\right) -\frac{126}{5}\log\left(\frac{2}{3}\right)-18\,i\,\pi\log\left(\frac{2}{3}\right)\right.
\nonumber\\
&& \left. +\frac{18}{5}\left(21-10\,i\,q+15\,i\,\pi+30\log\left(\frac{2}{3}\right)-45\log\left(\frac{v}{v_0}\right)\right)\log\left(\frac{v}{v_0}\right)\right]\,{v^6} + \left[ -q^3+\frac{9\,\pi\,q^2}{2}-\frac{63\,i\,q^2}{10}\right.
\nonumber\\
&& \left. +9\,i\,q^2 \log \left(\frac{3}{2}\right)+\frac{89\,q}{5}+\frac{369\,\pi\,}{110}-\frac{2\,583\,i}{550}+\frac{369}{55}\,i\,\log\left(\frac{3}{2}\right)+\left(27\,i\,q^2+\frac{1\,107\,i}{55}\right) \log \left(\frac{v}{v_0}\right) \right]\,{v^7}
\nonumber\\
&& + \left[ \frac{312\,\text{eulerlog}_3(v^2)}{7}-6\,\pi\,^2+\frac{984\,i\,\pi}{35}-\frac{72614419}{350350}+72 \log ^2\left(\frac{3}{2}\right)-72\,i\,\pi  \log \left(\frac{3}{2}\right)-\frac{504}{5} \log \left(\frac{3}{2}\right) \right.
\nonumber\\
&& + \left. \left(-\frac{1\,512}{5}-216\,i\,\pi +432 \log\left(\frac{3}{2}\right)\right) \log \left(\frac{v}{v_0}\right)+648 \log ^2\left(\frac{v}{v_0}\right) \right]\,{v^8} + \left[ -\frac{234}{7}\,\pi\,\text{eulerlog}_3(v^2)\right.
\nonumber\\
&& +\frac{234}{5}\,i\,\text{eulerlog}_3(v^2)-\frac{468}{7}\,i\,\log \left(\frac{3}{2}\right)\,\text{eulerlog}_3(v^2)-72\,i\,\zeta (3)-\frac{9\,\pi\,^3}{2}+\frac{1\,509\,i\,\pi ^2}{70}+\frac{64\,722\,993\,\pi}{280\,280}
\nonumber\\
&& - \frac{92\,343\,253\,i}{300\,300}-36\,i\,\log ^3\left(\frac{3}{2}\right)-54\,\pi\, \log^2\left(\frac{3}{2}\right)+\frac{378}{5}\,i\,\log ^2\left(\frac{3}{2}\right)+\left(9\,i\,\pi ^2+\frac{1\,476}{35}\,\pi\right.\nonumber\\
&& \left.+ \frac{58\,164\,441}{140\,140}\,i\right) \log\left(\frac{3}{2}\right)
 + \left(27\,i\,\pi ^2-\frac{1\,404}{7}\,i\,\text{eulerlog}_3(v^2)+\frac{4\,428\,\pi\,}{35}
+\frac{174\,493\,323\,i}{140\,140}-324\,i\,\log^2\left(\frac{3}{2}\right)\right.\nonumber\\
&&\left.-\left(324\,\pi-\frac{2\,268}{5}\,i\right)\log\left(\frac{3}{2}\right)\right)\log \left(\frac{v}{v_0}\right)\nonumber\\
&& + \left. \left(\frac{3\,402\,i}{5}-486\,\pi\,-972\,i\,\log \left(\frac{3}{2}\right)\right) \log^2\left(\frac{v}{v_0}\right) -972\,i\,\log^3\left(\frac{v}{v_0}\right) \right]\,{v^9}
\,,
\end{eqnarray}
\begin{eqnarray}\label{hatZ32w}
  \hat{Z}_{32\omega_0}&=& 1 - \frac{4\,q}{3}\,v - \frac{193}{90}\,v^2 +
 \left[\frac{7\,q}{9} + 2\,\pi -3\,i
  +12\,i\,\log\left(\frac{v}{v_0}\right) \right]\,v^3 + \left[- \frac{1\,451}{3\,960} - \frac{8\,\pi\,q}{3} + \frac{100\,q^2}{27} +4\,i\,q\right.\nonumber\\
  &&\left. -16\,i\,q\log\left(\frac{v}{v_0}\right) \right] \,{v^4}
  + \left[ -\frac{4\,q^3}{3}-\frac{560\,q}{891}-\frac{193\,\pi\,}{45}+\frac{193\,i}{30}-\frac{386}{15}\,i\,\log \left(\frac{v}{v_0}\right) \right]\,{v^5} + \left[\frac{2\,160\,500\,827}{75\,675\,600} \right.
\nonumber\\
&&\left.-\frac{104\,\text{eulerlog}_2(v^2)}{21}+\frac{1\,283\,q^2}{405}  +\frac{14\,\pi\,q}{9} - \frac{19\,i\,q}{5}+\frac{2\,\pi^2}{3}-\frac{74\,i\,\pi }{21} + \left(\frac{28\,i\,q}{3}+24\,i\,\pi +36\right) \log\left(\frac{v}{v_0}\right)\right.\nonumber\\
&&\left.-\,72 \log ^2\left(\frac{v}{v_0}\right) \right]\,{v^6}
-\frac{1\,451}{1\,980} \left[\pi-\frac{3\,i}{2}+6\,i\,\log \left(\frac{v}{v_0}\right)\right]\,{v^7}
+ \left[ -\frac{19\,111\,297\,859}{495\,331\,200}+\frac{10\,036}{945}\text{eulerlog}_2(v^2) \right.
\nonumber\\
&&-\frac{193\,\pi\,^2}{135}+\frac{7141\,i\,\pi}{945} + \left.\left(-\frac{386}{5}-\frac{772\,i\,\pi}{15}\right) \log \left(\frac{v}{v_0}\right) +\frac{772}{5} \log ^2\left(\frac{v}{v_0}\right) \right]\,{v^8} + \left[ -\frac{208}{21}\,\pi\,\text{eulerlog}_2(v^2)\right.
\nonumber\\
&&+\frac{104}{7}\,i\,\text{eulerlog}_2(v^2)-\frac{64\,i\,\zeta (3)}{3} - \frac{4\,\pi\,^3}{3}+\frac{394\,i\,\pi ^2}{63}+\frac{2\,441\,581\,627\,\pi}{37\,837\,800}-\frac{20\,967\,548\,963\,i}{227\,026\,800} +\left(\frac{2\,160\,500\,827\,i}{6\,306\,300}\right.\nonumber\\
&&\left.\left.-\,\frac{416}{7}\,i\,\text{eulerlog}_2(v^2)+\,8\,i\,\pi ^2+\frac{296\,\pi\,}{7}\right) \log\left(\frac{v}{v_0}\right)
 +(-144\,\pi\,+216\,i) \log^2\left(\frac{v}{v_0}\right) -288\,i\,\log ^3\left(\frac{v}{v_0}\right)\right]\,{v^9}
\,,\nonumber\\
\end{eqnarray}
\begin{eqnarray}\label{hatZ31w}
\hat{Z}_{31\omega_0}&=& 1 - \frac{8}{3}\,v^2 +
\left[\pi-\frac{50\,q}{9}-\frac{7}{5}\,i-2\,i\,\log 2+6\,i\,\log\left(\frac{v}{v_0}\right)\right]\,v^3 +
\left(\frac{607}{198} + \frac{17\,q^2}{6}\right)\,v^4
\nonumber\\
&&+ \left(\frac{73\,q}{12} - \frac{8\,\pi}{3} +\frac{56}{15}\,i
 +\frac{16}{3}\,i\log 2 -16\,i\,\log\left(\frac{v}{v_0}\right) \right) \,{v^5}
\nonumber\\
&&+ \left[ \frac{10\,753\,397}{1\,513\,512} + \frac{18\,173\,q^2}{2\,592} -
     \frac{26\,{\rm eulerlog}_1(v^2)}{21} - \frac{50\,\pi\,q}{9} +
     \frac{\pi^2}{6} +\left(\frac{1\,373}{180} + \frac{100}{9}\log 2\right)\,i\,q
     - \frac{82}{105}\,i\,\pi  \right.
\nonumber\\
&&- 2\,i\,\pi\log 2 - \left.\frac{14}{5}\log
     2-2\log^2 2 + \left(\frac{42}{5}-\frac{100}{3}\,i\,q + 6\,i\,\pi +
		    12\log 2\right)\log\left(\frac{v}{v_0}\right)-18\log^2\left(\frac{v}{v_0}\right) \right]\,v^6
\nonumber\\
&&+ \left[ -\frac{187\,q^3}{27}+\frac{17\,\pi\,q^2}{6}-\frac{119\,i\,q^2}{30}-\frac{17}{3}\,i\,q^2 \log  2 -\frac{6\,541\,q}{891}+\frac{607\,\pi\,}{198}-\frac{4\,249\,i}{990}-\frac{607}{99}\,i\,\log 2  \right.
\nonumber\\
&&+ \left.\left(17\,i\,q^2+\frac{607\,i}{33}\right) \log \left(\frac{v}{v_0}\right) \right]\,{v^7}
+ \left[ -\frac{12\,785\,953}{1\,091\,475}+\frac{208}{63}\,\text{eulerlog}_1(v^2)-\frac{4\,\pi\,^2}{9}+\frac{656\,i\,\pi }{315}+\frac{16 \log^2 2 }{3}\right.
\nonumber\\
&&+ \frac{16}{3}\,i\,\pi  \log  2   + \left. \frac{112 \log  2 }{15}+\left(-\frac{112}{5}-16\,i\,\pi -32 \log  2 \right)\log \left(\frac{v}{v_0}\right)+48 \log ^2\left(\frac{v}{v_0}\right) \right]\,{v^8}
\nonumber\\
&&+ \left[ \frac{26}{21} \left(-\pi+\frac{7}{5}\,i+2\,i\,\log  2 \right)\,\text{eulerlog}_1(v^2)-\frac{8\,i\,\zeta (3)}{3}-\frac{\pi ^3}{6}+\frac{503\,i\,\pi ^2}{630}+\frac{60\,325\,537\,\pi}{7\,567\,560}-\frac{85\,747\,069\,i}{8\,108\,100}\right.
\nonumber\\
&&+ \frac{4}{3}\,i\,\log ^3 2 -2\,\pi\, \log ^2 2 +\frac{14}{5}\,i\,\log ^2 2 -\frac{1}{3}\,i\,\pi ^2 \log  2 -\frac{164}{105}\,\pi\, \log  2 -\frac{10\,753\,397}{756\,756}\,i\,\log  2 
\nonumber\\
&&+ \left(-\frac{52}{7}\,i\,\text{eulerlog}_1(v^2)+i\,\pi\,^2+\frac{164\,\pi\,}{35}+\frac{10\,753\,397\,i}{252\,252}-12\,i\,\log ^2 2 +12\pi\log  2 -\frac{84}{5}\,i\,\log  2 \right)\log \left(\frac{v}{v_0}\right)
\nonumber\\
&&+ \left. \left(\frac{126\,i}{5}-18\,\pi\,+36\,i\,\log  2 \right) \log^2\left(\frac{v}{v_0}\right) -36\,i\,\log^3\left(\frac{v}{v_0}\right) \right]\,{v^9}
\,,
\end{eqnarray}
\begin{eqnarray}\label{hatZ44w}
\hat{Z}_{44\omega_0}&=& 1 - \frac{593}{110}\,v^2 +
\left(4\,\pi - \frac{8\,q}{3}-\frac{42}{5}\,i+8\,i\,\log 2 +24\,i\,\log\left(\frac{v}{v_0}\right)\right)\,v^3 + \left(\frac{1\,068\,671}{200\,200} + 2\,q^2\right)\,v^4
\nonumber\\
&&+ \left[ \frac{6\,774\,q}{1\,375}-\frac{1186\,\pi\,}{55}+\frac{12\,453\,i}{275}-\frac{2\,372}{55}\,i\,\log  2 -\frac{7\,116}{55}\,i\,\log \left(\frac{v}{v_0}\right) \right]\,{v^5}
\nonumber\\
&&+ \left[ \frac{42\,783\,901\,441}{499\,458\,960}-\frac{50\,272\,\text{eulerlog}_4(v^2)}{3\,465}-\frac{624\,721\,q^2}{123\,750}-\frac{32\,\pi\,q}{3}+\frac{1\,216\,i\,q}{75}-\frac{64}{3}\,i\,q \log  2 +\frac{8\,\pi\,^2}{3} \right.
\nonumber\\
&& -\, \frac{91\,288\,i\,\pi }{3\,465} -  32 \log ^2 2 +32\,i\,\pi  \log  2 +\frac{336 \log  2 }{5}+\left(-64\,i\,q+96\,i\,\pi +\frac{1\,008}{5}-192 \log 2 \right) \log \left(\frac{v}{v_0}\right)\nonumber\\
&&\left.-288 \log^2\left(\frac{v}{v_0}\right) \right]\,{v^6}+ \left[\frac{1\,068\,671\,\pi\,}{50\,050}-\frac{3\,206\,013\,i}{71\,500}+\frac{1\,068\,671}{25\,025}\,i\,\log  2 +\frac{3\,206\,013}{25\,025}\,i\,\log \left(\frac{v}{v_0}\right) \right]\,{v^7}
\,,
\end{eqnarray}
\begin{eqnarray}\label{hatZ43w}
\hat{Z}_{43\omega_0}&=& 1 - \frac{5\,q}{4}\,v - \frac{39}{11}\,v^2 + \left[ \frac{1\,438\,q}{825}+3\,\pi\,-\frac{32\,i}{5}+6\,i\,\log  3 -6\,i\,\log  2 +18\,i\,\log \left(\frac{v}{v_0}\right) \right]\,{v^3}
\nonumber\\
&&+ \left[ \frac{7\,206}{5\,005}+\frac{37\,q^2}{8}-\frac{15\,\pi\,q}{4}+8\,i\,q-\frac{15}{2}\,i\,q \log 3 +\frac{15}{2}\,i\,q \log  2 -\frac{45}{2}\,i\,q \log \left(\frac{v}{v_0}\right) \right]\,{v^4}
\nonumber\\
&&+ \left[ -\frac{15\,q^3}{8}+\frac{29\,941\,q}{200\,200}-\frac{117\,\pi\,}{11}+\frac{1\,248\,i}{55}-\frac{234}{11}\,i\,\log  3 +\frac{234}{11}\,i\,\log  2 -\frac{702}{11}\,i\,\log \left(\frac{v}{v_0}\right) \right]\,{v^5}
\nonumber\\
&&+ \left[ \frac{1\,272\,567\,389}{27\,747\,720}-\frac{3\,142}{385}\,\text{eulerlog}_3(v^2)+\frac{3\,\pi\,^2}{2}-\frac{5\,821\,i\,\pi}{385}-18 \log ^2\left(\frac{3}{2}\right)+18\,i\,\pi  \log\left(\frac{3}{2}\right)+\frac{192}{5} \log \left(\frac{3}{2}\right) \right.
\nonumber\\
&&+ \left. \left(\frac{576}{5}+54\,i\,\pi -108 \log\left(\frac{3}{2}\right)\right) \log \left(\frac{v}{v_0}\right) -162 \log ^2\left(\frac{v}{v_0}\right)\right]\,{v^6}
\nonumber\\
&&+ \left[ \frac{21\,618\,\pi\,}{5\,005}-\frac{230\,592\,i}{25\,025}+\frac{43\,236}{5\,005}\,i\,\log\left(\frac{3}{2}\right) +\frac{129\,708}{5\,005}\,i\,\log \left(\frac{v}{v_0}\right)\right]\,{v^7}
\,,
\end{eqnarray}
\begin{eqnarray}\label{hatZ42w}
\hat{Z}_{42\omega_0}&=& 1 - \frac{437}{110}\,v^2 + \left[2\,\pi - \frac{17\,q}{3}-\frac{21}{5}\,i+12\,i\log\left(\frac{v}{v_0}\right)\right]\,v^3 + \left(\frac{1\,038\,039}{200\,200} + \frac{20}{7}\,q^2\right)\,v^4
\nonumber\\
&&+ \left[ \frac{36\,353\,q}{2\,750}-\frac{437\,\pi\,}{55}+\frac{9\,177\,i}{550}-\frac{2\,622}{55}\,i\,\log \left(\frac{v}{v_0}\right) \right]\,{v^5}
\nonumber\\
&&+ \left[ \frac{44\,982\,355\,673}{2\,497\,294\,800}-\frac{12\,568\,\text{eulerlog}_2(v^2)}{3\,465}+\frac{1\,265\,648\,q^2}{275\,625}-\frac{34\,\pi\,q}{3}+\frac{1727\,i\,q}{75}+\frac{2\,\pi^2}{3}-\frac{22\,822\,i\,\pi }{3\,465} \right.
\nonumber\\
&&+ \left. \left(-68\,i\,q+24\,i\,\pi +\frac{252}{5}\right) \log\left(\frac{v}{v_0}\right)-72 \log ^2\left(\frac{v}{v_0}\right) \right]\,{v^6}  \nonumber\\ &&+\left[\frac{1\,038\,039\,\pi\,}{100\,100}-\frac{3\,114\,117\,i}{143\,000}+\frac{3\,114\,117}{50\,050}\,i\,\log \left(\frac{v}{v_0}\right)\right]\,{v^7}\,,
\end{eqnarray}
\begin{eqnarray}\label{hatZ41w}
\hat{Z}_{41\omega_0}&=& 1 - \frac{5\,q}{4}\,v - \frac{101}{33}\,v^2 + \left[ \frac{7\,q}{825}+\pi -\frac{32\,i}{15}-2\,i\,\log  2 +6\,i\,\log \left(\frac{v}{v_0}\right) \right]\,{v^3}
\nonumber\\
&&+ \left[ \frac{42\,982}{15\,015}+\frac{331\,q^2}{56}-\frac{5\,\pi\,q}{4}+\frac{8\,i\,q}{3}+\frac{5}{2}\,i\,q \log 2 -\frac{15}{2}\,i\,q \log \left(\frac{v}{v_0}\right) \right]\,{v^4}
\nonumber\\
&&+ \left[ -\frac{275\,q^3}{168}+\frac{5\,199\,317\,q}{1\,801\,800}-\frac{101\,\pi}{33}+\frac{3\,232\,i}{495}+\frac{202}{33}\,i\,\log  2 -\frac{202}{11}\,i\,\log \left(\frac{v}{v_0}\right) \right]\,{v^5}
\nonumber\\
&&+ \left[ \frac{1\,250\,147\,453}{249\,729\,480}-\frac{3\,142}{3\,465}\,\text{eulerlog}_1(v^2)+\frac{\pi ^2}{6}-\frac{5\,821\,i\,\pi }{3\,465}-2 \log ^2 2 -2\,i\,\pi  \log  2 -\frac{64 \log  2 }{15} \right.
\nonumber\\
&&+ \left. \left(\frac{64}{5}+6\,i\,\pi +12\log  2 \right)\log \left(\frac{v}{v_0}\right) -18 \log ^2\left(\frac{v}{v_0}\right)\right]\,{v^6}
\nonumber\\
&&+ \left[\frac{42\,982\,\pi\,}{15\,015}-\frac{1\,375\,424\,i}{225\,225}-\frac{85\,964}{15\,015}\,i\,\log 2  + \frac{85\,964}{5\,005}\,i\,\log \left(\frac{v}{v_0}\right)\right]\,{v^7}
\,,
\end{eqnarray}
\begin{eqnarray}\label{hatZ55w}
\hat{Z}_{55\omega_0}&=& 1-\frac{263}{39}\,v^2+\left[-\frac{10\,q}{3}+5\,\pi\,-\frac{181\,i}{14}+10\,i\,\log \left(\frac{5}{2}\right)+30\,i\,\log \left(\frac{v}{v_0}\right)\right]\,v^3 +\left(\frac{5\,q^2}{2}+\frac{9\,185}{819}\right)\,v^4
\nonumber\\
&&+ \left[\frac{26\,944\,q}{2\,457}-\frac{1\,315\,\pi\,}{39}+\frac{47\,603\,i}{546}-\frac{2\,630}{39}\,i\,\log\left(\frac{5}{2}\right)-\frac{2\,630}{13}\,i\,\log\left(\frac{v}{v_0}\right)\right]\,v^5
\,,
\end{eqnarray}
\begin{eqnarray}\label{hatZ54w}
\hat{Z}_{54\omega_0}&=& 1-\frac{6\,q}{5}\,v-\frac{4\,451}{910}\,v^2+ \left[\frac{7\,513\,q}{2\,925}+4\,\pi\,-\frac{52\,i}{5}+8\,i\,\log  2 +24\,i\,\log\left(\frac{v}{v_0}\right)\right]\,v^3
\nonumber\\
&&+ \left[\frac{10\,715}{2\,184}+\frac{707\,q^2}{125}-\frac{24\,\pi\,q}{5}+\frac{312\,i\,q}{25}-\frac{48}{5}\,i\,q \log  2 -\frac{144}{5}\,i\,q \log \left(\frac{v}{v_0}\right)\right]\,v^4
\nonumber\\
&&+ \left[-\frac{8\,902\,\pi\,}{455}+\frac{8\,902\,i}{175}-\frac{17\,804}{455}\,i\,\log  2  -\frac{53\,412}{455}\,i\,\log \left(\frac{v}{v_0}\right) \right]\,v^5
\,,
\end{eqnarray}
\begin{eqnarray}\label{hatZ53w}
\hat{Z}_{53\omega_0}&=& 1 -\frac{69}{13}\,v^2+\left[-\frac{442\,q}{75}+3\,\pi\,-\frac{543\,i}{70}+6\,i\,\log \left(\frac{3}{2}\right)+18\,i\,\log \left(\frac{v}{v_0}\right)\right]\,v^3+ \left[\frac{91\,q^2}{30}+\frac{12\,463}{1\,365}\right]\,v^4
\nonumber\\
&&+ \left[\frac{47\,296\,q}{2\,275}-\frac{207\,\pi\,}{13}+\frac{37\,467\,i}{910}-\frac{414}{13}\,i\,\log\left(\frac{3}{2}\right)-\frac{1\,242}{13}\,i\,\log\left(\frac{v}{v_0}\right)\right]\,v^5
\,,
\end{eqnarray}
\begin{eqnarray}\label{hatZ52w}
\hat{Z}_{52\omega_0}&=& 1-\frac{6\,q}{5}\,v-\frac{3\,911}{910}\,v^2+ \left[\frac{2\,317\,q}{2\,925}+2\,\pi\,-\frac{26\,i}{5}+12\,i\,\log \left(\frac{v}{v_0}\right)\right]\,v^3
\nonumber\\
&&+ \left[\frac{63\,439}{10\,920}+\frac{833\,q^2}{125}-\frac{12\,\pi\,q}{5}+\frac{156\,i\,q}{25}-\frac{72}{5}\,i\,q \log \left(\frac{v}{v_0}\right)\right]\,v^4
\nonumber\\
&&+ \left[-\frac{3\,911\,\pi\,}{455}+\frac{3\,911\,i}{175} -\frac{23\,466}{455}\,i\,\log \left(\frac{v}{v_0}\right) \right]\,v^5
\,,
\end{eqnarray}
\begin{eqnarray}\label{hatZ51w}
\hat{Z}_{51\omega_0}&=& 1-\frac{179}{39}\,v^2+\left[-\frac{538\,q}{75}+\pi -\frac{181\,i}{70}-2\,i\,\log  2 +6\,i\,\log \left(\frac{v}{v_0}\right)\right]\,v^3+ \left[\frac{33\,q^2}{10}+\frac{5\,023}{585}\right]\,v^4
\nonumber\\
&&+ \left[\frac{208\,192\,q}{8\,775}-\frac{179\,\pi\,}{39}+\frac{32\,399\,i}{2\,730}+\frac{358}{39}\,i\,\log  2 -\frac{358}{13}\,i\,\log\left(\frac{v}{v_0}\right)\right]\,v^5
\,,
\end{eqnarray}
\begin{eqnarray}\label{hatZ66w}
\hat{Z}_{66\omega_0}&=& 1-\frac{113}{14}\,v^2+ \left[-4\,q+6\,\pi\,-\frac{249\,i}{14}+12\,i\,\log  3 +36\,i\,\log \left(\frac{v}{v_0}\right)\right]\,v^3+\left(\frac{1\,372\,317}{73\,304}+3\,q^2\right)\,v^4
\,,
\end{eqnarray}
\begin{eqnarray}\label{hatZ65w}
\hat{Z}_{65\omega_0}&=& 1-\frac{7\,q}{6}\,v-\frac{149}{24}\,v^2+ \left[\frac{2\,927\,q}{882}+5\,\pi\,-\frac{104\,i}{7}+10\,i\,\log\left(\frac{5}{2}\right)+30\,i\,\log \left(\frac{v}{v_0}\right)\right]\,v^3
\,,
\end{eqnarray}
\begin{eqnarray}\label{hatZ64w}
\hat{Z}_{64\omega_0}&=& 1-\frac{93}{14}\,v^2+ \left[-\frac{56\,q}{9}+4\,\pi\,-\frac{83\,i}{7}+8\,i\,\log  2 +24\,i\,\log \left(\frac{v}{v_0}\right)\right]\,v^3+\left[\frac{109\,q^2}{33}+\frac{3\,261\,767}{219\,912}\right]\,v^4
\,,
\end{eqnarray}
\begin{eqnarray}\label{hatZ63w}
\hat{Z}_{63\omega_0}&=& 1-\frac{7\,q}{6}\,v-\frac{133}{24}\,v^2+ \left[\frac{461\,q}{294}+3\,\pi\,-\frac{312\,i}{35}+6\,i\,\log\left(\frac{3}{2}\right)+18\,i\,\log \left(\frac{v}{v_0}\right)\right]\,v^3
\,,
\end{eqnarray}
\begin{eqnarray}\label{hatZ62w}
\hat{Z}_{62\omega_0}&=& 1-\frac{81}{14}\,v^2+ \left[-\frac{68\,q}{9}+2\,\pi\,-\frac{83\,i}{14}+12\,i\,\log\left(\frac{v}{v_0}\right)\right]\,v^3+\left[\frac{115\,q^2}{33}+\frac{14\,482\,483}{1\,099\,560}\right]\,v^4
\,,
\end{eqnarray}
\begin{eqnarray}\label{hatZ61w}
\hat{Z}_{61\omega_0}&=& 1-\frac{7\,q}{6}\,v-\frac{125}{24}\,v^2+ \left[\frac{611\,q}{882}+\pi -\frac{104\,i}{35}-2\,i\,\log  2 +6\,i\,\log \left(\frac{v}{v_0}\right)\right]\,v^3
\,,
\end{eqnarray}
\begin{eqnarray}\label{hatZ77w}
\hat{Z}_{77\omega_0}&=& 1-\frac{319}{34}\,v^2+ \left[-\frac{14\,q}{3}+7\,\pi\,-\frac{4\,129\,i}{180}+14\,i\,\log\left(\frac{7}{2}\right)+42\,i\,\log \left(\frac{v}{v_0}\right)\right]\,v^3
\,,
\end{eqnarray}
\begin{eqnarray}\label{hatZ76w}
\hat{Z}_{76\omega_0}&=& 1 -\frac{8\,q}{7}\,v-\frac{1\,787}{238}\,v^2
\,,
\end{eqnarray}
\begin{eqnarray}\label{hatZ75w}
\hat{Z}_{75\omega_0}&=& 1-\frac{271}{34}\,v^2+ \left[-\frac{974\,q}{147}+5\,\pi\,-\frac{4\,129\,i}{252}+10\,i\,\log\left(\frac{5}{2}\right)+30\,i\,\log \left(\frac{v}{v_0}\right)\right]\,v^3
\,,
\end{eqnarray}
\begin{eqnarray}\label{hatZ74w}
\hat{Z}_{74\omega_0}&=& 1 -\frac{8\,q}{7}\,v-\frac{14\,543}{2\,142}\,v^2
\,,
\end{eqnarray}
\begin{eqnarray}\label{hatZ73w}
\hat{Z}_{73\omega_0}&=& 1-\frac{239}{34}\,v^2+ \left[-\frac{1\,166\,q}{147}+3\,\pi\,-\frac{4\,129\,i}{420}+6\,i\,\log\left(\frac{3}{2}\right)+18\,i\,\log \left(\frac{v}{v_0}\right)\right]\,v^3
\,,
\end{eqnarray}
\begin{eqnarray}\label{hatZ72w}
\hat{Z}_{72\omega_0}&=& 1-\frac{8\,q}{7}\,v-\frac{13\,619}{2\,142}\,v^2
\,,
\end{eqnarray}
\begin{eqnarray}\label{hatZ71w}
\hat{Z}_{71\omega_0}&=& 1-\frac{223}{34}\,v^2+ \left[-\frac{1\,262\,q}{147}+\pi -\frac{4\,129\,i}{1\,260}-2\,i\,\log  2 +6\,i\,\log \left(\frac{v}{v_0}\right)\right]\,v^3
\,,
\end{eqnarray}
\begin{eqnarray}\label{hatZ8evenmw}
\hat{Z}_{88\omega_0}= 1-\frac{3\,653}{342}\,v^2 \,,\qquad
\hat{Z}_{86\omega_0}= 1-\frac{353}{38}\,v^2 \,,\qquad
\hat{Z}_{84\omega_0}= 1-\frac{2\,837}{342}\,v^2 \,,\qquad
\hat{Z}_{82\omega_0}= 1-\frac{2\,633}{342}\,v^2 \,.
\end{eqnarray}
\begin{eqnarray}\label{hatZ8oddmw}
\hat{Z}_{87\omega_0}= 1-\frac{9\,q}{8}\,v \,,\qquad
\hat{Z}_{85\omega_0}= 1-\frac{9\,q}{8}\,v \,,\qquad
\hat{Z}_{83\omega_0}= 1-\frac{9\,q}{8}\,v \,,\qquad
\hat{Z}_{81\omega_0}= 1-\frac{9\,q}{8}\,v \,.
\end{eqnarray}
\end{widetext}

\section{Expressions of the $C_{\ell m}$'s modes for $ 4< \ell < 8$}
\label{AppendixE}

\begin{widetext}
\begin{subequations}
\begin{eqnarray}\label{hatC5m}
\hat{C}_{55}&=&\hat{Z}_{55\omega_0} -\frac{400\,q}{2\,457}\,v^5 ,\\
\hat{C}_{54}&=&\hat{Z}_{54\omega_0} +\frac{6\,q}{5}\,v-\frac{19\,213\,q}{2\,925}\,v^3 + \left[-\frac{332\,q^2}{125}+\frac{24\,\pi\,q}{5}-\frac{252\,i\,q}{25}+\frac{48}{5}\,i\,q \log  2 +\frac{144}{5}\,i\,q \log \left(\frac{v}{v_0}\right)\right]\,v^4 ,\\
\hat{C}_{53}&=&\hat{Z}_{53\omega_0} +\frac{64\,q}{25}\,v^3-\frac{8\,q^2}{15}\,v^4-\frac{20\,976\,q}{2\,275}\,v^5 ,\\
\hat{C}_{52}&=&\hat{Z}_{52\omega_0} +\frac{6\,q}{5}\,v-\frac{14\,017\,q}{2\,925}\,v^3+ \left[-\frac{458\,q^2}{125}+\frac{12\,\pi\,q}{5}-\frac{126\,i\,q}{25}+\frac{72}{5}\,i\,q \log \left(\frac{v}{v_0}\right)\right]\,v^4 ,\\
\hat{C}_{51}&=&\hat{Z}_{51\omega_0} +\frac{96\,q}{25}\,v^3-\frac{4\,q^2}{5}\,v^4-\frac{103\,312\,q}{8\,775}\,v^5 ,
\end{eqnarray}
\begin{eqnarray}\label{hatC6m}
&&\hat{C}_{66}=\hat{Z}_{66\omega_0} , \qquad\qquad\qquad\qquad\qquad\qquad
\hat{C}_{65}=\hat{Z}_{65\omega_0} + \frac{7\,q}{6}\,v-\frac{7\,043\,q}{882}\,v^3,\\
&&\hat{C}_{64}=\hat{Z}_{64\omega_0} + \frac{20\,q}{9}\,v^3-\frac{10\,q^2}{33}\,v^4, \qquad\quad\,
\hat{C}_{63}=\hat{Z}_{63\omega_0} + \frac{7\,q}{6}\,v-\frac{611\,q}{98}\,v^3,\\
&&\hat{C}_{62}=\hat{Z}_{62\omega_0} + \frac{32\,q}{9}\,v^3-\frac{16\,q^2}{33}\,v^4, \qquad\quad\,
\hat{C}_{61}=\hat{Z}_{61\omega_0} + \frac{7\,q}{6}\,v-\frac{4\,727\,q}{882}\,v^3,
\end{eqnarray}
\begin{eqnarray}\label{hatC7m}
\hat{C}_{77}=\hat{Z}_{77\omega_0} ,\qquad
\hat{C}_{76}=\hat{Z}_{76\omega_0} + \frac{8\,q}{7}\,v,\qquad
\hat{C}_{75}=\hat{Z}_{75\omega_0} + \frac{96\,q}{49}\,v^3,\qquad
\hat{C}_{74}=\hat{Z}_{74\omega_0} + \frac{8\,q}{7}\,v,\\
\hat{C}_{73}=\hat{Z}_{73\omega_0} + \frac{160\,q}{49}\,v^3,\qquad
\hat{C}_{72}=\hat{Z}_{72\omega_0} + \frac{8\,q}{7}\,v,\qquad
\hat{C}_{71}=\hat{Z}_{71\omega_0} + \frac{192\,q}{49}\,v^3,
\end{eqnarray}
\begin{eqnarray}\label{hatC8m}
\hat{C}_{87}=\hat{Z}_{87\omega_0} + \frac{9\,q}{8}\,v ,\qquad
\hat{C}_{85}=\hat{Z}_{85\omega_0} + \frac{9\,q}{8}\,v ,\qquad
\hat{C}_{83}=\hat{Z}_{83\omega_0} + \frac{9\,q}{8}\,v ,\qquad
\hat{C}_{81}=\hat{Z}_{81\omega_0} + \frac{9\,q}{8}\,v .
\end{eqnarray}
\end{subequations}
\end{widetext}

\section{Expressions of the $f_{\ell m}$'s modes for $ \ell > 4$}
\label{AppendixB}

\subsection{The odd-parity $f^L_{\ell m}$'s and even-parity $f_{\ell m}$'s}
\begin{widetext}
\begin{subequations}
\begin{eqnarray}
f_{55}&=&1-\frac{487}{78}\,v^2-\frac{10\,q}{3}\,v^3+\left(\frac{5\,q^2}{2}+\frac{50\,569}{6\,552}\right)\,v^4+\frac{1\,225\,q}{117}\,v^5 \,,\\
f^L_{54}&=&1-\frac{2\,908}{455}\,v^2-\frac{2\,q}{3}\,v^3+\left(2\,q^2+\frac{2\,168}{195}\right)\,v^4 \,,\\
f_{53}&=&1-\frac{125}{26}\,v^2-\frac{10\,q}{3}\,v^3+\left(\frac{5\,q^2}{2}+\frac{69\,359}{10\,920}\right)\,v^4+\frac{2\,191\,q}{195}\,v^5 \,,\\
f^L_{52}&=&1-\frac{2\,638}{455}\,v^2-\frac{2\,q}{3}\,v^3+\left(2\,q^2+\frac{15\,194}{1\,365}\right)\,v^4 \,,\\
f_{51}&=&1-\frac{319}{78}\,v^2-\frac{10\,q}{3}\,v^3+\left(\frac{5\,q^2}{2}+\frac{28\,859}{4\,680}\right)\,v^4+\frac{6\,797\,q}{585}\,v^5 \,,
\end{eqnarray}
\begin{eqnarray}
&f_{66}&=1-\frac{53}{7}\,v^2-4\,q\,v^3+\left(3\,q^2+\frac{133\,415}{9\,163}\right)\,v^4 \,,\qquad
f^L_{65}=1-\frac{185}{24}\,v^2-\frac{4\,q}{3}\,v^3 \,,\\
&f_{64}&=1-\frac{43}{7}\,v^2-4\,q\,v^3+\left(3\,q^2+\frac{312\,982}{27\,489}\right)\,v^4 \,,\qquad
f^L_{63}=1-\frac{169}{24}\,v^2-\frac{4\,q}{3}\,v^3 \,,\\
&f_{62}&=1-\frac{37}{7}\,v^2-4\,q\,v^3+\left(3\,q^2+\frac{1\,395\,521}{137\,445}\right)\,v^4 \,,\qquad
f^L_{61}=1-\frac{161}{24}\,v^2-\frac{4\,q}{3}\,v^3 \,,
\end{eqnarray}
\begin{eqnarray}
f_{77}=1-\frac{151}{17}\,v^2-\frac{14\,q}{3}\,v^3 \,,\quad
&f^L_{76}&=1-\frac{1\,072}{119}\,v^2 \,,\\
f_{75}=1-\frac{127}{17}\,v^2-\frac{14\,q}{3}\,v^3 \,,\quad
&f^L_{74}&=1-\frac{8\,878}{1\,071}\,v^2 \,,\\
f_{73}=1-\frac{111}{17}\,v^2-\frac{14\,q}{3}\,v^3 \,,\quad
&f^L_{72}&=1-\frac{8\,416}{1\,071}\,v^2 \,,\quad
f_{71}=1-\frac{103}{17}\,v^2-\frac{14\,q}{3}\,v^3 
\,,
\end{eqnarray}
\begin{eqnarray}
f_{88}=1-\frac{1\,741}{171}\,v^2 \,,\quad
&f^L_{87}&=1-\frac{3\,913}{380}\,v^2 \,,\quad
f_{86}=1-\frac{167}{19}\,v^2 \,,\quad
f^L_{85}=1-\frac{725}{76}\,v^2 \,,\\
f_{84}=1-\frac{1\,333}{171}\,v^2 \,,\quad
&f^L_{83}&=1-\frac{3\,433}{380}\,v^2 \,,\quad
f_{82}=1-\frac{1\,231}{171}\,v^2 \,,\quad
f^L_{81}=1-\frac{3\,337}{380}\,v^2 \,.
\end{eqnarray}
\end{subequations}
\end{widetext}

\subsection{The odd-parity $f_{\ell m}^H$'s}

\begin{widetext}
\begin{subequations}
\label{Hflm}
\begin{eqnarray}
f_{21}^H&=&1-\frac{3\,q}{2}\,v-\frac{3}{28}\,v^2-\frac{5\,q}{4}\,v^3+\left(3\,q^2-\frac{97}{126}\right)\,v^4-\frac{3\,q}{112}\left(28\,q^2+45\right)\,v^5
\nonumber\\
&&+ \left(\frac{75\,q^2}{14}-\frac{214\,\text{eulerlog}_1(v^2)}{105}+\frac{70\,479\,293}{11\,642\,400}\right)\,v^6+ \left(-\frac{535\,q^3}{168}+\frac{107}{35}\,q\,\text{eulerlog}_1(v^2)-\frac{12\,363\,787\,q}{1\,058\,400}\right)\,v^7
\nonumber\\
&&+ \left(\frac{107\,\text{eulerlog}_1(v^2)}{490}+\frac{5\,770\,262\,917}{1\,412\,611\,200}\right)v^8+ \left(\frac{10\,379\,\text{eulerlog}_1(v^2)}{6615}-\frac{23\,353\,414\,831}{13\,869\,273\,600}\right)\,v^{10} \,,
\end{eqnarray}
\begin{eqnarray}
f_{32}^H&=&1-\frac{74}{45}\,v^2-\frac{8\,q}{3}\,v^3+\left(2\,q^2-\frac{86}{55}\right)\,v^4-\frac{106\,q}{45}\,v^5+ \left(\frac{16\,q^2}{45}-\frac{104\,\text{eulerlog}_2(v^2)}{21}+\frac{96\,051\,082}{4\,729\,725}\right)\,v^6
\nonumber\\
&&+ \left(\frac{7\,696\,\text{eulerlog}_2(v^2)}{945}-\frac{708\,338\,174}{42\,567\,525}\right)\,v^8 \,,
\end{eqnarray}
\begin{eqnarray}
f_{43}^H&=&1-\frac{67}{22}\,v^2-\frac{10\,q}{3}\,v^3+\left(\frac{5\,q^2}{2}-\frac{1\,667}{3\,640}\right)\,v^4+\frac{7\,481\,q}{4\,620}\,v^5+ \left(\frac{11\,083\,164\,791}{277\,477\,200}-\frac{3\,142\,\text{eulerlog}_3(v^2)}{385}\right)\,v^6 \,, \\
f_{41}^H&=&1-\frac{169}{66}\,v^2-\frac{10\,q}{3}\,v^3+\left(\frac{5\,q^2}{2}+\frac{145\,021}{120\,120}\right)\,v^4+\left(\frac{89\,027\,q}{13\,860}-\frac{10\,q^3}{3}\right)\,v^5
\nonumber\\
&&+ \left(\frac{10\,765\,133\,231}{2\,497\,294\,800}-\frac{3\,142\,\text{eulerlog}_1(v^2)}{3\,465}\right)\,v^6 \,,
\end{eqnarray}
\begin{eqnarray}
f_{54}^H=1-\frac{1\,998}{455}\,v^2-4\,q\,v^3+\left(3\,q^2+\frac{3\,188}{1\,365}\right)\,v^4 \,, \qquad
f_{52}^H=1-\frac{1\,728}{455}\,v^2-4\,q\,v^3+\left(3\,q^2+\frac{4\,826}{1\,365}\right)\,v^4 \,,
\end{eqnarray}
\begin{eqnarray}
f_{65}^H=1-\frac{137}{24}\,v^2-\frac{14\,q}{3}\,v^3 \,, \qquad
f_{63}^H=1-\frac{121}{24}\,v^2-\frac{14\,q}{3}\,v^3 \,, \qquad
f_{61}^H=1-\frac{113\,v^2}{24}-\frac{14\,q}{3}\,v^3 \,,
\end{eqnarray}
\begin{eqnarray}
f_{76}^H=1-\frac{834}{119}\,v^2 \,, \qquad
f_{74}^H=1-\frac{6\,736}{1\,071}\,v^2 \,, \qquad
f_{72}^H=1-\frac{6\,274}{1\,071}\,v^2 \,.
\end{eqnarray}
\end{subequations}
\end{widetext}

\section{Expressions of the $\rho_{\ell m}$'s modes for $ \ell >  4$}
\label{AppendixC}

\subsection{The odd-parity $\rho^L_{\ell m}$'s and even-parity $\rho_{\ell m}$'s}
\begin{widetext}
\begin{subequations}
\begin{eqnarray}
\rho_{55}&=&1-\frac{487}{390} v^2-\frac{2\,q}{3}\,v^3+\left(\frac{q^2}{2}-\frac{3\,353\,747}{2\,129\,400}\right) v^4-\frac{241\,q}{195}\,v^5 \,,\\
\rho^L_{54}&=&1-\frac{2\,908}{2\,275} v^2-\frac{2\,q}{15}\,v^3+\left(\frac{2\,q^2}{5}-\frac{16\,213\,384}{15\,526\,875}\right) v^4 \,,\\
\rho_{53}&=&1-\frac{25}{26} v^2-\frac{2\,q}{3}\,v^3+\left(\frac{q^2}{2}-\frac{410\,833}{709\,800}\right) v^4-\frac{103\,q}{325}\,v^5 \,,\\
\rho^L_{52}&=&1-\frac{2\,638}{2\,275} v^2-\frac{2\,q}{15}\,v^3+\left(\frac{2\,q^2}{5}-\frac{7\,187\,914}{15\,526\,875}\right) v^4 \,,\\
\rho_{51}&=&1-\frac{319}{390} v^2-\frac{2\,q}{3}\,v^3+\left(\frac{q^2}{2}-\frac{31\,877}{304\,200}\right) v^4+\frac{139\,q}{975}\,v^5 \,,
\end{eqnarray}
\begin{eqnarray}
\rho_{66}=1-\frac{53}{42} v^2-\frac{2\,q}{3}\,v^3+\left(\frac{q^2}{2}-\frac{1\,025\,435}{659\,736}\right) v^4 \,,\qquad
&\rho^L_{65}&=1-\frac{185}{144} v^2-\frac{2\,q}{9}\,v^3 \,,\\
\rho_{64}=1-\frac{43}{42} v^2-\frac{2\,q}{3}\,v^3+\left(\frac{q^2}{2}-\frac{476\,887}{659\,736}\right) v^4 \,,\qquad
&\rho^L_{63}&=1-\frac{169}{144} v^2-\frac{2\,q}{9}\,v^3 \,,\\
\rho_{62}=1-\frac{37}{42} v^2-\frac{2\,q}{3}\,v^3+\left(\frac{q^2}{2}-\frac{817\,991}{3\,298\,680}\right) v^4 \,,\qquad
&\rho^L_{61}&=1-\frac{161}{144} v^2-\frac{2\,q}{9}\,v^3 \,,
\end{eqnarray}
\begin{eqnarray}
\rho_{77}=1-\frac{151}{119} v^2-\frac{2\,q}{3}\,v^3 \,,\quad
&\rho^L_{76}&=1-\frac{1072}{833} v^2 \,,\\
\rho_{75}=1-\frac{127}{119} v^2-\frac{2\,q}{3}\,v^3 \,,\quad
&\rho^L_{74}&=1-\frac{8878}{7497} v^2 \,,\\
\rho_{73}=1-\frac{111}{119} v^2-\frac{2\,q}{3}\,v^3 \,,\quad
&\rho^L_{72}&=1-\frac{8\,416}{7\,497} v^2 \,,\quad
\rho_{71}=1-\frac{103}{119} v^2-\frac{2\,q}{3}\,v^3 \,,
\end{eqnarray}
\begin{eqnarray}
\rho_{88}=1-\frac{1\,741}{1\,368} v^2 \,,\quad
&\rho^L_{87}&=1-\frac{3\,913}{3\,040} v^2 \,,\quad
\rho_{86}=1-\frac{167}{152} v^2 \,,\quad
\rho^L_{85}=1-\frac{725}{608} v^2 \,,\\
\rho_{84}=1-\frac{1\,333}{1\,368} v^2 \,,\quad
&\rho^L_{83}&=1-\frac{3\,433}{3\,040} v^2 \,,\quad
\rho_{82}=1-\frac{1\,231}{1\,368} v^2 \,,\quad
\rho^L_{81}=1-\frac{3\,337}{3\,040} v^2 \,.
\end{eqnarray}
\end{subequations}
\end{widetext}

\subsection{The odd-parity $\rho^H_{\ell m}$'s}

\begin{widetext}
\begin{subequations}\label{Hrholm}
\begin{eqnarray}
\rho_{21}^H&=&1-\frac{3\,q}{4}\,v-\frac{3}{224} \left(21\,q^2+4\right)\,v^2-\frac{1}{896} \left(q\,\left(189\,q^2+596\right)\right)\,v^3+\left( -\frac{405\,q^4}{2\,048}+\frac{1\,767\,q^2}{1\,792}-\frac{21\,809}{56\,448} \right)\,v^4
\nonumber\\
&&-\left( \frac{1\,701\,q^5}{8\,192}-\frac{1\,191\,q^3}{7\,168}+\frac{69\,851\,q}{75\,264} \right)\,v^5+ \left(\frac{7\,839\,703\,541}{2\,607\,897\,600}-\frac{15\,309\,q^6}{65\,536}+\frac{4\,113\,q^4}{16\,384}+\frac{342\,289\,q^2}{200\,704}\right.\nonumber\\
&&\left.-\,\frac{107\,\text{eulerlog}_1(v^2)}{105}\right)\,v^6+ \left(-\frac{72\,171\,q^7}{262\,144}+\frac{19\,683\,q^5}{65\,536}+\frac{3\,131\,q^3}{344\,064}+\frac{107}{140}\,q\,\text{eulerlog}_1(v^2)-\frac{40\,609\,146\,713\,q}{10\,431\,590\,400}\right)\,v^7
\nonumber\\
&&+ \left(\frac{107\,\text{eulerlog}_1(v^2)}{1\,960}+\frac{48\,499\,995\,300\,301}{22\,782\,593\,433\,600}\right)\,v^8+\left( \frac{2\,333\,563\,\text{eulerlog}_1(v^2)}{5\,927\,040}+\frac{3\,762\,995\,064\,239}{8\,679\,083\,212\,800} \right)\,v^{10} \,,
\end{eqnarray}
\begin{eqnarray}
\rho_{32}^H&=&1-\frac{74}{135}\,v^2-\frac{8\,q}{9}\,v^3+\left(\frac{2\,q^2}{3}-\frac{164\,726}{200\,475}\right)\,v^4-\frac{2\,138\,q}{1\,215}\,v^5+ \left(\frac{8\,q^2}{135}-\frac{104\,\text{eulerlog}_2(v^2)}{63}+\frac{61\,271\,294\,666}{10\,343\,908\,575}\right)\,v^6
\nonumber\\
&&+ \left(\frac{7\,696\,\text{eulerlog}_2(v^2)}{8\,505}+\frac{1\,593\,740\,014\,406}{3\,072\,140\,846\,775}\right)\,v^8 \,,
\end{eqnarray}
\begin{eqnarray}
\rho_{43}^H&=&1-\frac{67}{88}\,v^2-\frac{5\,q}{6}\,v^3+\left(\frac{5\,q^2}{8}-\frac{6\,934\,313}{7\,047\,040}\right)\,v^4-\frac{13\,847\,q}{9\,240}\,v^5
\nonumber\\
&&+ \left(\frac{1\,597\,804\,689\,571}{195\,343\,948\,800}-\frac{1\,571\,\text{eulerlog}_3(v^2)}{770}\right)\,v^6 \,, \\
\rho_{41}^H&=&1-\frac{169}{264}\,v^2-\frac{5\,q}{6}\,v^3+\left(\frac{5\,q^2}{8}-\frac{2,204\,777}{7\,047\,040}\right)\,v^4+\left(\frac{151\,q}{27\,720}-\frac{5\,q^3}{6}\right)\,v^5
\nonumber\\
&&+ \left(\frac{1\,299\,523\,316\,251}{1\,758\,095\,539\,200}-\frac{1\,571\,\text{eulerlog}_1(v^2)}{6\,930}\right)\,v^6 \,,
\end{eqnarray}
\begin{eqnarray}
\rho_{54}^H&=&1-\frac{1\,998}{2\,275}\,v^2-\frac{4\,q}{5}\,v^3+\left(\frac{3\,q^2}{5}-\frac{16\,699\,324}{15\,526\,875}\right)\,v^4 \,, \\
\rho_{52}^H&=&1-\frac{1\,728}{2\,275}\,v^2-\frac{4\,q}{5}\,v^3+\left(\frac{3\,q^2}{5}-\frac{6\,936\,754}{15\,526\,875}\right)\,v^4 \,,
\end{eqnarray}
\begin{eqnarray}
\rho_{65}^H=1-\frac{137}{144}\,v^2-\frac{7\,q}{9}\,v^3 \,, \qquad
\rho_{63}^H=1-\frac{121}{144}\,v^2-\frac{7\,q}{9}\,v^3 \,, \qquad
\rho_{61}^H=1-\frac{113}{144}\,v^2-\frac{7\,q}{9}\,v^3 \,,
\end{eqnarray}
\begin{eqnarray}
\rho_{76}^H=1-\frac{834}{833}\,v^2 \,, \qquad
\rho_{74}^H=1-\frac{6\,736}{7\,497}\,v^2 \,, \qquad
\rho_{72}^H=1-\frac{6\,274}{7\,497}\,v^2 \,.
\end{eqnarray}
\end{subequations}
\end{widetext}

\section{Expressions of the $\delta_{\ell m}$'s modes for $4< \ell \leq 7$}
\label{AppendixF}

\begin{subequations}
\begin{eqnarray}
\delta_{55} = \frac{31}{42}\,v^3\,, \quad
\delta_{53} = \frac{31}{70}\,v^3\,, \quad
\delta_{51} = \frac{31}{210}\,v^3\,, \\
\delta_{54} = \frac{12\,q}{5}\,v^4+\frac{8}{15}\,v^3\,, \quad
\delta_{52} = \frac{6\,q}{5}\,v^4+\frac{4}{15}\,v^3\,,
\end{eqnarray}
\begin{eqnarray}
\delta_{66} = \frac{43}{70}\,v^3\,, \quad
&\delta_{64}& = \frac{43}{105}\,v^3\,, \quad
\delta_{62} = \frac{43}{210}\,v^3\,, \\
\delta_{65} = \frac{10}{21}\,v^3\,, \quad
&\delta_{63}& = \frac{2}{7}\,v^3\,, \quad
\delta_{61} = \frac{2}{21}\,v^3\,,
\end{eqnarray}
\begin{eqnarray}
\delta_{77} = \frac{19}{36}\,v^3\,, \quad
&\delta_{75}& = \frac{95}{252}\,v^3\,, \\
\delta_{73} = \frac{19}{84}\,v^3\,, \quad
&\delta_{71}& = \frac{19}{252}\,v^3\,.
\end{eqnarray}
\end{subequations}

\section{Multipole moments for generic $\ell$ and $m$}
\label{AppendixD}

In Refs.~\cite{Kidder2008,DINresum} the authors have computed the even- and odd-parity 1PN multipoles 
for generic $\ell$ and $m$. Those calculations were crucial in understanding the $\ell$-scaling of the $f_{\ell m}$'s,  
suggesting the introduction of the $\rho_{\ell m}$'s functions. 

In this Appendix, we calculate the 0.5PN spin terms in the 
odd-parity multipoles $\hat{h}^{(1)}_{\ell m}$ and the 1.5PN spin terms 
in the even-parity multipoles $\hat{h}^{(0)}_{\ell m}$. Just for completeness 
we also reproduce the 1PN nonspinning terms in the odd-parity multipoles
 $\hat{h}^{(1)}_{\ell m}$, already computed in Ref.~\cite{DINresum}.

Henceforth, we make use of the standard multi-index notation for
tensors of arbitrary rank, which are displayed as
\begin{equation}
T_L \equiv T_{i_1i_2...i_\ell}\,,
\end{equation}
where each index $i_1$ to $i_\ell$ runs from 1 to 3. We also employ the notation 
$T_{<L>}={\rm STF}_L[T_L]$ to denote the symmetric trace-free projection over the indices 
$i_1$ to $i_\ell$. For example we have
\begin{equation}
T_{<ij>} = \frac{1}{2}(T_{ij} + T_{ji}) - \frac{1}{3}\delta_{ij}\delta^{pq}T_{pq}\,.
\end{equation}
Repeated multi-indices imply summation over all corresponding indices, e.g.
\begin{equation}
T_LS^L \equiv T_{i_1i_2...i_\ell}S^{i_1i_2...i_\ell}\,.
\end{equation}
Reference~\cite{Kidder2008} computed the expression of the full waveform 
as an expansion in -2 spin-weighted spherical harmonics through 
the coefficients $U_{\ell m}$ and $V_{\ell m}$ as follows
\begin{equation}
h_{lm} = \frac{1}{\sqrt{2}R}\,\Big(U_{\ell m} - i V_{\ell m}\Big) \,,\label{hlmKidder}
\end{equation}
where
\begin{subequations}
\begin{eqnarray}
U_{\ell m} &=& \frac{16\pi}{(2\ell+1)!!}\,\sqrt{\frac{(\ell+1)(\ell+2)}{2\ell(\ell-1)}}\,U_L\,\mathcal{Y}_L^{\ell m \, \ast} \,,\\
V_{\ell m} &=& -\frac{32\pi \, \ell}{(2\ell+1)!!}\,\sqrt{\frac{(\ell+2)}{2\ell(\ell+1)(\ell-1)}}\,V_L\,\mathcal{Y}_L^{\ell m \, \ast} \,. \nonumber \\
\end{eqnarray}
\end{subequations}
The radiative moments $U_L$ and $V_L$ are the $l^{\rm th}$ time derivatives
of the multipole moments $I_L$ and $J_L$ respectively, as we 
neglect tail contributions for our purposes here. In terms of the
vector $\hat{\bm{r}}$ defined above Eqs.~\eqref{nlamL}, the quantity
$\mathcal{Y}_L^{\ell m}$ is defined as follows
\begin{equation}\label{calYdef}
Y^{\ell m} = \mathcal{Y}_L^{\ell m}\,\hat{r}_L \,.
\end{equation}
\subsection{Odd-parity 0.5PN spin  multipoles}

The odd-parity contributions to the waveforms are provided by the 
expansion coefficients $V_{\ell m}$, which in turn are determined by the 
current multipole moments $J_L$. In the circular orbital case, the nonspinning 
1PN current-multipole moment $J_L$ is given by~\cite{DINresum}
\begin{eqnarray}
J_L^{\rm NS} &=& (\nu\,M\, r^{\ell+1}\,\Omega) 
\bigg[K_1\,\hat{L}_N^{<i_\ell}\,n^{L-1>} \nonumber \\
&& + v^2\,K_2\, \hat{L}_N^{<i_\ell}\,n^{L-3} \lambda^{i_{\ell-2} \,i_{\ell-1}>}\bigg]\,, 
\end{eqnarray} 
where $\Omega$ is the orbital frequency, $v = (M \Omega)^{1/3}$ and 
\begin{equation}
\bm{n}=\frac{\bm{r}}{r}, \quad \bm{\hat{L}_\textrm{N}}=\frac{\bm{r}\times \bm{\dot{r}}}{\lvert \bm{r}\times \bm{\dot{r}} \rvert},\quad  \bm{\lambda}= \bm{\hat{L}_\textrm{N}} \times \bm{n}\,,
\end{equation}
and where
\begin{subequations} 
\begin{eqnarray}
K_1 &=& c_{\ell+1} + v^2\,\left\{-\frac{\nu}{2\ell} + \frac{2\ell+3}{2\ell}\,b_{\ell+1}  + 2\nu\frac{\ell+1}{\ell}\,b_{\ell-1}\right. \nonumber \\
&& \left. + \frac{1}{2}\left[\frac{\ell+1}{\ell} - \frac{(\ell-1)\,(\ell+4)}{(\ell+2)\,(2\ell+3)}\right]\,c_{\ell+3}\right\}\,,\nonumber \\
&& \, \\
K_2 &=& \frac{(\ell-1)\,(\ell-2)\,(\ell+4)}{2(\ell+2)\,(2\ell+3)}\,c_{\ell+3} \,, \\
b_\ell &=& X_2^\ell + (-)^\ell\,X_1^\ell\,, \\
c_\ell &=& X_2^{\ell-1} + (-)^\ell\,X_1^{\ell-1}\,,  
\label{cl}
\end{eqnarray} 
\end{subequations} 
where $c_\ell$ coincides with Eq.~(\ref{cleps}), and $X_{1,2} = m_{1,2}/M$. For circular orbits, we have 
\begin{subequations}
\begin{eqnarray}
\bm{n} &=& (\cos \phi_{\rm orb},
\sin\phi_{\rm orb},0)\,, \\
\bm{\lambda} &=& (-\sin \phi_{\rm orb}, \cos\phi_{\rm orb},0)\,, \\
\hat{\bm{L}}_N &=& (0,0,1)\,.
\end{eqnarray}
\end{subequations}
In terms of the vector $\hat{\bm{r}} = (\sin\theta\cos\phi,
\sin\theta\sin\phi,\cos\theta)$, the following expressions will prove very helpful below
\begin{subequations}\label{nlamL}
\begin{eqnarray}
\bm{n} &=& [\hat{\bm{r}}]_{\theta = \pi/2, \phi = \phi_{\rm orb}} \,, \\
\bm{\lambda} &=& [\partial_\phi \hat{\bm{r}}]_{\theta = \pi/2, \phi = \phi_{\rm orb}} \,, \\
\hat{\bm{L}}_N &=& -[\partial_\theta \hat{\bm{r}}]_{\theta = \pi/2, \phi = \phi_{\rm orb}} \,.
\end{eqnarray}
\end{subequations}
The 0.5PN-order contribution to $J_L$ that is linear in the spins is given by Ref.~\cite{BBF}
\begin{equation}\label{JLspin}
J_L^{\rm S} = \frac{(\ell+1)}{2}{\rm STF}_L \Bigg\{\sum_A S_A^{i_\ell}\,y_A^{L-1}\Bigg\}\,.
\end{equation}
To rewrite Eq.~\eqref{JLspin} in the center-of-mass frame, we use $\bm{y}_1 = X_2\,\bm{r}$ and $y_2 = -X_1\,\bm{r}$, which leads to the following
\begin{eqnarray}
J_L^{\rm S} &=& \frac{(\ell+1)}{2}\,\bigg[\big(X_2^{\ell-1}\,S_1^{<i_\ell} + (-)^{\ell-1}\, X_1^{\ell-1}\,S_2^{<i_\ell} \big) x^{L-1>}\bigg] \nn \\
&\equiv&  \nu\,M^2\, r^{\ell-1}\,\frac{(\ell+1)}{2}\,\hat{\Sigma}_{(\ell)}^{<i_\ell}\,n^{L-1>}\,,
\end{eqnarray}
where
\begin{eqnarray}
\hat{\bm{\Sigma}}_{(\ell)} &=& X_1\,X_2^{\ell-2}\,\bm{\chi}_1 + (-)^{\ell-1}\,X_2\, X_1^{\ell-2}\, \bm{\chi}_2 \,,
\label{sigma}
\end{eqnarray}
and we define $\bm{\chi}_1 = \bm{S}_1/m_1^2$ and $\bm{\chi}_2 = \bm{S}_2/m_2^2$. For nonprecessing binaries, we have 
$\hat{\bm{\Sigma}}_{(\ell)} = \hat{\Sigma}_{(\ell)}\,\hat{\bm{L}}_N$, and hence we can write down the total 1PN-order 
current multipole moment as
\begin{eqnarray}
  J_L &&=  (\nu\,M\, r^{\ell+1}\, \Omega) \,{\rm STF}_L \Bigg\{\hat{L}_N^{i_\ell}\, \bigg[K_1 n^{L-1} \nn \\
&& + v^2\, K_2\, n^{L-3}\,\lambda^{i_{\ell-2} \, i_{\ell-1}} + v\,\frac{(\ell+1)}{2}\,\hat{\Sigma}_{(\ell)}\,n^{L-1}\bigg]\Bigg\}\,.
\nn \\
\end{eqnarray}
Next, in order to compute the radiative coefficient $V_{\ell m}$, we first need $J_{\ell m} = J_L\,\mathcal{Y}_L^{\ell m}$. 
It is therefore useful to rewrite all vectors appearing in $J_L$ in terms of $\hat{\bm{r}}$ as follows
\begin{eqnarray}\label{JLtotal1}
J_L &=& (\nu\, M\, r^{\ell+1}\, \Omega)\, {\rm STF}_L 
\Bigg\{\partial_{\theta} n^{i_\ell} \,\bigg[K_1 \,n^{L-1} \nonumber \\
&& + v^2\,K_2 n^{L-3}\, 
\partial_\phi\,n^{i_{\ell-2}}\,\partial_\phi n^{i_{\ell-1}} \nonumber\\
&&+ v\,\frac{(\ell+1)}{2}\,\hat{\Sigma}_{(\ell)}\,
n^{L-1}\bigg]\Bigg\}_{\rm orb}\,, 
\end{eqnarray}
where the ``orb'' subscript is shorthand for evaluating the bracket at 
$\theta = \pi/2, \phi=\phi_{\rm orb}$. The purpose of this rewriting is to allow us to eventually make use of Eq.~\eqref{calYdef}, together with the following identities
\begin{subequations}\label{partialrids}
\begin{eqnarray}
\partial_\theta n^{<L>} &=& \ell\,(\partial_\theta n^{<i_\ell} )\,n^{L-1>} \\
\partial_\phi^2 n^{<L-1>} &=& (\ell-1)\,\Big\{(\ell-2)n^{<L-3} \,\partial_\phi n^{i_{\ell-2}}\, \partial_\phi n^{i_{\ell-1}>}\nonumber \\
&&  - \Big[n^{<i_{\ell-1}} - (\bm{n}\cdot\hat{\bm{L}}_N)\,\hat{L}_N^{<i_{\ell-1}}\Big]\,n^{L-2>} \Big\} \,. \nonumber \\
\end{eqnarray}
\end{subequations}
By substituting Eqs.~\eqref{partialrids} into Eq.~\eqref{JLtotal1}, the current multipole moments become
\begin{eqnarray}\label{JLtotal2}
&& J_L = (\nu\, M\, r^{\ell+1}\, \Omega) {\rm STF}_L \Bigg\{ \frac{K_1}{\ell} \partial_\theta n^L 
+ v^2 \frac{K_2}{\ell(\ell-2)} \partial_\theta n^L \nn \\
&& + v^2 \frac{K_2}{\ell(\ell-1)(\ell-2)} \partial_\theta \partial_\phi^2 n^L 
+ v^2\,\frac{(\ell+1)}{2\ell}\,\hat{\Sigma}_{(\ell)}\,\partial_\theta n^L \Bigg\}_{\rm orb}\,. \nn\\
\end{eqnarray}
Contracting Eq.~\eqref{JLtotal2} with $\mathcal{Y}_L^{\ell m\,\ast}$ then yields
\begin{widetext}
\begin{eqnarray}
J_{\ell m}  &=& \frac{c_{\ell+1}}{\ell}(\nu M r^{\ell+1} \Omega) \left[\partial_\theta Y_{\ell m}^\ast(\theta,\phi_{\rm orb})\right]_{\theta = \pi/2}\Bigg\{1 + v\,\frac{(\ell+1)}{2c_{\ell+1}}\,\hat{\Sigma}_{(\ell)} - v^2 \left[\frac{\nu}{2\ell} - \frac{2\ell+3}{2\ell}\frac{b_{\ell+1}}{c_{\ell+1}} \right. \nn \\
&& \left. - 2\nu \frac{\ell+1}{\ell} \frac{b_{\ell-1}}{c_{\ell+1}} + \frac{1}{2}\left(\frac{m^2(\ell+4)}{(\ell+2)(2\ell+3)} - \frac{\ell+1}{\ell}\right)\frac{c_{\ell+3}}{c_{\ell+1}}\right] \Bigg\}\,.\label{JLtotal3}
\end{eqnarray}
\end{widetext}
From the parity properties of associated Legendre polynomials, $J_{\ell m}$ is non-vanishing only if $\ell+m$ is odd. The next step consists of converting $r^{\ell+1}$ into an expansion in $v$ by means of Kepler's third law, 
\begin{equation}
r^{\ell+1} = \left ({M}{v^{-2}}\right )^{\ell+1}\,\left[1 - v^2\,(\ell+1)\,\left(1-\frac{\nu}{3}\right)\right] \,,
\end{equation}
and substituting it into Eq.~\eqref{JLtotal3} yields
\begin{widetext}
\begin{eqnarray}
J_{\ell m} &=& \frac{c_{\ell+1}}{\ell}\,\left({M}{v^{-2}}\right)^{\ell+1}\,\nu\,v^{3}\,
\left[\partial_\theta Y_{\ell m}^\ast(\theta,\phi_{\rm orb})\right]_{\theta = \pi/2}\,
\Bigg(1 + v\,\frac{(\ell+1)}{2c_{\ell+1}}\,\hat{\Sigma}_{(\ell)} 
- v^2\,\left\{(\ell+1)\left(1-\frac{\nu}{3}\right)+ \frac{\nu}{2\ell} - \frac{2\ell+3}{2\ell}\,\frac{b_{\ell+1}}{c_{\ell+1}} \right.  \nn\\
&& \left. - 2\nu\,\frac{\ell+1}{\ell}\,\frac{b_{\ell-1}}{c_{\ell+1}} + \frac{1}{2}\left[\frac{m^2(\ell+4)}{(\ell+2)(2\ell+3)} - \frac{\ell+1}{\ell}\right]\,\frac{c_{\ell+3}}{c_{\ell+1}}\right\} \Bigg)\,.
\end{eqnarray}
Taking $\ell$ time derivatives and multiplying by the appropriate normalization factor finally gives
\begin{eqnarray}
V_{\ell m} &=&-\frac{32\pi \, \ell}{(2\ell+1)!!}\,\sqrt{\frac{(\ell+2)}{2\ell(\ell+1)(\ell-1)}}\nu\, M\, (-im)^\ell\, v^{(\ell+1)}\,\frac{c_{\ell+1}}{\ell}\,\left[\partial_\theta Y_{\ell m}^\ast(\theta,\phi_{\rm orb})\right]_{\theta = \pi/2}\,\Bigg\{1 + v\,\frac{(\ell+1)}{2c_{\ell+1}}\,\hat{\Sigma}_{(\ell)} \nn \\
&& - v^2\,\left[(\ell+1)\,\left(1-\frac{\nu}{3}\right)+ \frac{\nu}{2\ell} - \frac{2\ell+3}{2\ell}\,\frac{b_{\ell+1}}{c_{\ell+1}} - 2\nu\,\frac{\ell+1}{\ell}\,\frac{b_{\ell-1}}{c_{\ell+1}} + \frac{1}{2}\,\left(\frac{m^2\,(\ell+4)}{(\ell+2)\,(2\ell+3)} - \frac{\ell+1}{\ell}\right)
\,\frac{c_{\ell+3}}{c_{\ell+1}}\right] \Bigg\}\,,\nonumber \\\\
\label{Vlmfac}
\end{eqnarray}
The overall factor in front of the bracket in Eq.~\eqref{Vlmfac} coincides with 
the Newtonian contribution as given by Eq.~\eqref{hlmNewt}, using Eqs.~\eqref{hlmKidder} and~\eqref{nlmodd}.
Hence by definition [see Eq.~\eqref{hlm}], we find
\begin{eqnarray}
\label{h1}
\hat{h}^{(1)}_{\ell m} &=& 1 + v\,\frac{(\ell+1)}{2 c_{\ell+1}}\,\hat{\Sigma}_{(\ell)} - v^2\,\left[(\ell+1)\left(1-\frac{\nu}{3}\right)+ \frac{\nu}{2\ell} - \frac{2\ell+3}{2\ell}\,\frac{b_{\ell+1}}{c_{\ell+1}} - 2\nu\,\frac{\ell+1}{\ell} \frac{b_{\ell-1}}{c_{\ell+1}} \right. \nn \\
&& \left. + \frac{1}{2}\left(\frac{m^2\,(\ell+4)}{(\ell+2)\,(2\ell+3)} - \frac{\ell+1}{\ell}\right)\,\frac{c_{\ell+3}}{c_{\ell+1}}\right]\,.
\end{eqnarray}
\end{widetext}
Again, the 1PN-order terms in the above equation were computed in Appendix A of Ref.~\cite{DINresum}.

Quite interestingly, we find that in the nonspinning test-particle limit ($m_2 \ll m_1, \chi_1 = 
|\bm{\chi}_1| = a\, m_1 \equiv q, \chi_2 = 0$), 
only the odd-parity mode $\ell = 2$ contains the 0.5PN  spin term, for all the other odd-parity 
modes the 0.5PN spin terms vanish. In fact, using Eqs.~(\ref{cl}), (\ref{sigma}) we find that if we set $\chi_2=0$, 
the 0.5PN  spin terms reduces to 
\begin{equation}
\hat{h}^{(1), \rm 0.5PN}_{\ell m} = -\frac{(\ell+1)\,m_1^2\,\chi_1}{2[m_1\,m_2 + (-1)^\ell\,m_1^\ell\,m_2^{2-\ell}]}\,v\,.
\end{equation}
If $\ell = 2$, then $\hat{h}^{(1), \rm 0.5PN}_{21}= -3/2 v\,q$ when $\nu \rightarrow 0$, 
while if $\ell >2$, we have  $\hat{h}^{(1), \rm 0.5PN}_{\ell m} 
\propto \nu\,q\,v$ and the latter goes to zero as $\nu \rightarrow 0$. The fact that the odd-parity modes 
with $\ell > 2$ vanish, is consistent with the -2 spin-weighted spherical $C_{\ell m}$'s computed in the main part of 
this paper. However, it is worth noticing that the odd-parity -2 spin-weighted spheroidal $Z_{\ell m}$'s {\it do} 
contain 0.5PN  spin terms. 

Moreover, for the case of finite symmetric mass-ratio $\nu$, we find that the 0.5PN  spin 
terms in Eq.~(\ref{h1}) coincide with what was derived in PN theory~\cite{ABFO}. 
The $\ell$-dependence of the 0.5PN spin term in Eq.~(\ref{h1}) varies 
depending on the binary mass ratio and the spin magnitudes. For example we find that for 
maximally spinning and aligned black holes ($\chi_1 = \chi_2 = 1$) if the masses are equal, the 0.5PN spin 
term in Eq.~(\ref{h1}) scales as $\ell$, but if the masses are unequal, it generally doesn't scale as $\ell$.

Finally, we derive the corresponding generic 0.5PN spin
  contributions to $f_{\ell m}^{(1)}$ and $\rho_{\ell m}^{(1)}$. Since
  we know that there is no quadratic spin contribution at
  1PN order in $f_{\ell m}^{(1)}$, we need to introduce a 
  1PN quadratic spin term in $\rho_{\ell m}^{(1)}$. Thus, the spin portions read
\begin{eqnarray}
f^{(1), \rm 0.5PN}_{\ell m} &=& \hat{h}^{(1), \rm 0.5PN}_{\ell m} \,, \nonumber \\
\rho^{(1), \rm 0.5PN}_{\ell m} &=& \frac{1}{\ell}\hat{h}^{(1), \rm 0.5PN}_{\ell m} \,, 
\nonumber\\
\rho^{(1), \rm 1PN}_{\ell m} &=& -\frac{\ell-1}{2\,\ell^2}\left(\hat{h}^{(1), \rm 0.5PN}_{\ell m}\right)^2 \,.
\end{eqnarray}

\subsection{Even-parity 1.5PN spin multipoles}

The 1.5PN spin contributions to the even-parity waveform come from
two distinct sources. The first is the 1.5PN spin mass multipole
moment $I_L^{\rm S}$, given by (in the center-of-mass frame, for
non-precessing, circular orbits)
\begin{equation}
I^{\rm S}_L = M^2\nu^2\frac{2\ell}{\ell+1}r^\ell \tilde{\Sigma}_{(l)}{\rm STF}_L\Bigg[\ell\Omega n^L + \frac{\ell-1}{r^2\Omega}n^{L-2}v^{i_{\ell-1}}v^{i_\ell} \Bigg]\,, 
\end{equation}
where
\begin{equation}
\tilde{\Sigma}_{(\ell)} = X_2^{\ell-2} \chi_1 + (-)^\ell X_1^{\ell-2}\chi_2\,.
\end{equation}  
Making use of the following identity which is valid for circular orbits
\begin{equation}
{\rm STF}_L\Bigg[\frac{(\ell-1)}{r^2} n^{L-2}v^{i_{\ell-1}}v^{i_\ell}\Bigg] = {\rm STF}_L\Bigg[\frac{1}{\ell}\frac{d^2}{dt^2}n^L + \Omega^2 n^L\Bigg]\,,
\end{equation}
we can rewrite $I^{\rm S}_L$ as follows
\begin{equation}\label{ISL}
I^{\rm S}_L =  M^2\nu^2\frac{2\ell}{\ell+1}r^\ell \tilde{\Sigma}_{(l)}{\rm STF}_L\Bigg[(\ell+1)\Omega n^L + \frac{1}{\ell\Omega}\frac{d^2}{dt^2}n^L \Bigg]\,.
\end{equation}
The second contribution comes from the Newtonian mass multipole
moments, in two different ways. First, since the coordinate
transformation that takes us from a generic frame to the
center-of-mass frame involves the spins at 1.5PN order, the
Newtonian mass multipole moments acquire a spin contribution when
re-expressed in the center-of-mass frame. Second, when we use Kepler's law at
1.5PN order to rewrite the orbital separation $r$ as an expansion in
$v = (M\Omega)^{1/3}$, spin terms are generated which contribute to
the 1.5PN spinning waveform. In a general frame, the Newtonian mass
multipole moments are given by
\begin{equation}
I^{\rm N}_L = {\rm STF}_L\,\Big[m_1\, y_1^L + m_2\,y_2^L\Big]\,.
\end{equation} 
The coordinate transformation to the center-of-mass frame is given
by~\footnote{Strictly speaking this transformation contains
  non-spinning 1PN contributions. We shall not write those explicitly
  here to keep formulas as light as possible, as we are only concerned
  with spinning terms relative to the leading-order Newtonian
  contribution.}
\begin{subequations}
\begin{eqnarray}
\bm{y}_1 &=& X_2\,\bm{r} + \frac{\nu}{M}\,\bm{v}\times\bm{\Delta}\,, \\
\bm{y}_2 &=& -X_1\,\bm{r} + \frac{\nu}{M}\,\bm{v}\times\bm{\Delta}\,.
\end{eqnarray}
\end{subequations}  
Therefore in the center-of-mass frame, the Newtonian mass multipole moments read
\begin{equation}\label{INLCM}
I^{\rm N}_L = M\,\nu\, c_\ell\, r^l\, n^{<L>} + \nu^2\, \ell\, c_{\ell-1}r^{l-1}\, n^{<L-1}(\bm{v}\times \Delta)^{i_\ell>}\,.
\end{equation}
For non-precessing, circular orbits, Eq.~\eqref{INLCM} may be rewritten as
\begin{equation}\label{INLCM2}
I^{\rm N}_L =  M\,\nu\, c_\ell\, r^l\, n^{<L>}\,\Bigg[1+\nu\frac{\ell\,c_{\ell-1}}{c_\ell}(X_2\,\chi_2 - X_1\,\chi_1)v^3\Bigg]\,.
\end{equation}

\pagebreak

Adding together both contributions~\eqref{ISL} and~\eqref{INLCM2},
contracting with $\mathcal{Y}_{\ell m}^{\ast}$ and finally taking $\ell$ 
time derivatives as well as multiplying by the appropriate overall factor, 
we arrive at the following expression for the even-parity radiative moment
\begin{widetext}
\begin{eqnarray}\label{Ulm}
U_{\ell m} &=& \frac{16\pi}{(2\ell+1)!!}\,\sqrt{\frac{(\ell+1)(\ell+2)}{2\ell(\ell-1)}}M\nu c_\ell(-im\Omega)^\ell r^\ell Y_{\ell m}^\ast(\pi/2,\phi_{\rm orb})\Bigg[1+\nu\frac{\ell c_{\ell-1}}{c_\ell}(X_2\chi_2 - X_1\chi_1)v^3 \nn \\
&& +\frac{\nu}{c_{\ell}}\left(\frac{2\ell}{\ell+1}\right)\left(\ell+1-\frac{m^2}{\ell}\right)\tilde{\Sigma}_{(\ell)}v^3\Bigg]\,.
\end{eqnarray}
\end{widetext} 
The overall factor in front of the bracket in Eq.~\eqref{Ulm} coincides with 
the Newtonian contribution as given by Eq.~\eqref{hlmNewt}, using Eqs.~\eqref{hlmKidder} and~\eqref{nlmeven}.
Next we use Kepler's third law to replace the orbital separation $r$
by the following expansion in $v$. Aagain, we do not write the 1PN
  order non-spinning contributions explicitly here to keep formulas
  short.

\ 

\begin{equation}\label{Kepler}
r=M\,v^{-2}\, \Bigg\{1 + \left[\frac{2}{3}(X_1^2\,\chi_1 + X_2^2\,\chi_2) + \nu\,(\chi_1+\chi_2)\right]\,v^3\Bigg\}^{-1}\,.
\end{equation}
Substituting~\eqref{Kepler} into~\eqref{Ulm}, we can finally isolate the 1.5PN spin contribution 
to the even-parity waveform as
\begin{widetext}
\begin{equation}\label{hath0lm}
\hat{h}_{\ell m}^{(0),1.5{\rm PN}} = \Bigg\{-\ell\,\left[\frac{2}{3}\,(X_1^2\,\chi_1 + X_2^2\,\chi_2) + \nu\,(\chi_1+\chi_2)\right] +\nu\,\frac{\ell c_{\ell-1}}{c_\ell}(X_2\,\chi_2 - X_1\,\chi_1)+\frac{\nu}{c_{\ell}}\left(\frac{2\ell}{\ell+1}\right)\left(\ell+1-\frac{m^2}{\ell}\right)\tilde{\Sigma}_{(\ell)}\Bigg\}\,v^3\,.
\end{equation}
\end{widetext}
In the non-spinning test particle limit, Eq.~\eqref{hath0lm} simply reduces to
\begin{equation}
\hat{h}_{\ell m}^{(0),1.5{\rm PN}} \rightarrow -\frac{2\ell}{3}\,q\, v^3\,,
\end{equation} 
thus, it scales as $\ell$. Finally, we derive the corresponding generic 1.5PN spin 
contribution to $f_{\ell m}^{(0)}$ and $\rho_{\ell m}^{(0)}$ and they
read
\begin{eqnarray}
f_{\ell m}^{(0),1.5{\rm PN}} &=& \hat{h}_{\ell m}^{(0),1.5{\rm PN}}\,, \nonumber\\
\rho_{\ell m}^{(0),1.5{\rm PN}} &=& \frac{1}{\ell}f_{\ell m}^{(0),1.5{\rm PN}} \,.
\end{eqnarray}
Therefore, the generic $f_{\ell m}^{(0)}$ and $\rho_{\ell m}^{(0)}$
expressions through 1.5PN are given by the above equation combined
with the 1PN nonspinning result given in Eq. (A15) of
Ref.~\cite{DINresum} (note that there is no 1.5PN nonspinning
contribution to $f_{\ell m}^{(0)}$ or $\rho_{\ell m}^{(0)}$.


%
\end{document}